\documentclass[twocolumn]{aastex631}

\usepackage{graphicx}
\usepackage{amssymb}
\usepackage{amsmath}

\accepted{May 14, 2022}

\shorttitle{How the stellar populations of passive central galaxies depend on stellar and halo mass}
\shortauthors{Oyarz\'un et al.}
\graphicspath{{./}{figures/}}
\begin{document}
	
	\title{SDSS-IV MaNGA: How the stellar populations of passive central galaxies \\depend on stellar and halo mass}

	\email{goyarzun@ucsc.edu}
	
	\author{Grecco A. Oyarz\'un}
	\affiliation{Astronomy Department, University of California, Santa Cruz, CA 95064, USA}
	
	\author{Kevin Bundy}
	\affiliation{University of California Observatories - Lick Observatory, University of California, Santa Cruz, CA 95064, USA}
	
	\author{Kyle B. Westfall}
	\affiliation{University of California Observatories - Lick Observatory, University of California, Santa Cruz, CA 95064, USA}
	
	\author{Jeremy L. Tinker}
	\affiliation{Center for Cosmology and Particle Physics, Department of Physics, New York University, New York, USA, 10003}
	
	\author{Francesco Belfiore}
	\affiliation{INAF – Osservatorio Astrofisico di Arcetri, Largo E. Fermi 5, I-50157, Firenze, Italy}
	
	\author{Maria Argudo-Fern{\'a}ndez}
	\affiliation{Instituto de F{\'i}sica, Pontificia Universidad Cat{\'o}lica de Valpara{\'i}so, Casilla 4059, Valpara{\'i}so, Chile}
	
	\author{Zheng Zheng}
	\affiliation{National Astronomical Observatories, Chinese Academy of Sciences, A20 Datun Road, Beijing, China}
	\affiliation{CAS Key Laboratory of FAST, NAOC, Chinese Academy of Sciences}
	
	\author{Charlie Conroy}
	\affiliation{Center for Astrophysics \textbar\ Harvard \& Smithsonian, 60 Garden Street, Cambridge, MA 02138, USA}
	
	\author{Karen L. Masters}
	\affiliation{Departments of Physics and Astronomy, Haverford College, 370 Lancaster Avenue, Haverford, PA 19041, USA}
	
	\author{David Wake}
	\affiliation{Department of Physics, University of North Carolina Asheville, One University Heights, Asheville, NC 28804, USA}	
	
	\author{David R. Law}
	\affiliation{Space Telescope Science Institute, 3700 San Martin Drive, Baltimore, MD 21218, USA}
	
	\author{Richard M. McDermid}
	\affiliation{MQAAAstro Research Centre, Macquarie University, NSW 2109, Australia}
	
	\author{Alfonso Arag\'on-Salamanca}
	\affiliation{School of Physics \& Astronomy, University of Nottingham, University Park, Nottingham, NG7 2RD, UK}
	
	\author{Taniya Parikh}
	\affiliation{Max-Planck-Institut f\"ur extraterrestrische Physik, Giessenbachstrasse 1, 85748 Garching bei M\"unchen, Germany}	
	
	\author{Renbin Yan}
	\affiliation{Department of Physics and Astronomy, University of Kentucky, 505 Rose St., Lexington, KY 40506-0055, USA}
	
	\author{Matthew Bershady}
	\affiliation{University of Wisconsin - Madison, Department of Astronomy, 475
		N. Charter Street, Madison, WI 53706-1582, USA}
	\affiliation{South African Astronomical Observatory, PO Box 9, Observatory
		7935, Cape Town, South Africa}
	\affiliation{Department of Astronomy, University of Cape Town, Private Bag X3,
		Rondebosch 7701, South Africa}
	
	\author{Sebasti\'an F. S\'anchez}
	\affiliation{Instituto de Astronom\'ia, Universidad Nacional Aut\'onoma de M\'exico, A. P. 70-264, C.P. 04510}
	
	\author{Brett H. Andrews}
	\affiliation{PITT PACC, Department of Physics and Astronomy, University of Pittsburgh, Pittsburgh, PA 15260, USA}	
	
	\author{Jos\'e G. Fern\'andez-Trincado}
	\affiliation{Instituto de Astronom\'ia, Universidad Cat\'olica del Norte, Av. Angamos 0610, Antofagasta, Chile}
	
	\author{Richard R. Lane}
	\affiliation{Centro de Investigaci\'on en Astronom\'ia, Universidad Bernardo O Higgins, Avenida Viel 1497, Santiago, Chile}
	
	\author{D. Bizyaev}
	\affiliation{Apache Point Observatory, P.O. Box 59, Sunspot, NM 88349, USA}
	
	\author{Nicholas Fraser Boardman}
	\affiliation{School of Physics and Astronomy, University of St Andrews, North Haugh, St Andrews KY16 9SS, UK}
	
	\author{Ivan Lacerna}
	\affiliation{Instituto de Astronom\'ia y Ciencias Planetarias, Universidad de Atacama, Copayapu 485, Copiap\'o, Chile}
	\affiliation{Millennium Institute of Astrophysics, Nuncio Monsenor Sotero Sanz 100, Of. 104, Providencia, Santiago, Chile}
	
	\author{J. R. Brownstein}
	\affiliation{Department of Physics and Astronomy, University of Utah, 115 S. 1400 E., Salt Lake City, UT 84112, USA}
	
	\author{Niv Drory}
	\affiliation{McDonald Observatory, The University of Texas at Austin, 1 University Station, Austin, TX 78712, USA}
	
	\author{Kai Zhang}
	\affiliation{Lawrence Berkeley National Laboratory, 1 Cyclotron Road, Berkeley, CA 94720, USA}
	
	
	
	\begin{abstract}
		
		We analyze spatially resolved and co-added SDSS-IV MaNGA spectra with signal-to-noise $\sim 100$ from 2200 passive central galaxies ($z \sim 0.05$) to understand how central galaxy assembly depends on stellar mass (M$_*$) and halo mass (M$_h$).
		We control for systematic errors in M$_h$ by employing a new group catalog from \citet{tinker2020a,tinker2020b} and the widely-used \citet{yang2007} catalog. At fixed M$_*$, the strength of several stellar absorption features varies systematically with M$_h$. Completely model-free, this is one of the first indications that the stellar populations of centrals with identical M$_*$ are affected by the properties of their host halos. To interpret these variations, we applied full spectral fitting with the code \texttt{alf}. At fixed M$_*$, centrals in more massive halos are older, show lower [Fe/H], and have higher [Mg/Fe] with $3.5\sigma$ confidence. We conclude that halos not only dictate how much M$_*$ galaxies assemble, but also modulate their chemical enrichment histories. Turning to our analysis at fixed M$_h$, high-M$_*$ centrals are older, show lower [Fe/H], and have higher [Mg/Fe] for M$_h>10^{12}h^{-1}$M$_{\odot}$ with confidence $>4\sigma$. While massive passive galaxies are thought to form early and rapidly, our results are among the first to distinguish these trends at fixed M$_h$. They suggest that high-M$_*$ centrals experienced unique \textit{early} formation histories, either through enhanced collapse and gas fueling, or because their halos were early-forming and highly concentrated, a possible signal of galaxy assembly bias.
	\end{abstract}
	% Estimates from the \citet{yang2007} and \citet{tinker2020b} group catalogs are compared to control for systematic errors in halo mass (M$_h$) estimation. 

	
	
	\section{Introduction} \label{1}
	
	The historical debate over spheroidal galaxy formation pitted an \textit{in-situ} process (so-called ``monolithic collapse,'' e.g. \citealt{eggen1962,larson1974,arimoto-yoshii1987,bressan1994}) against an \textit{ex-situ} one (so-called ``hierarchical assembly,'' e.g. \citealt{toomre1977,white-rees1978}).  More recently, this debate has been recast in the form of a proposed ``two-phase'' scenario that incorporates elements from both formation pathways (\citealt{oser2010,oser2012,johansson2012}).  The question now is what physical processes, both \textit{in-situ} and \textit{ex-situ}, dominate and at what redshifts?  
	% To gain insight with available observables, we want to distinguish the stellar populations at different radial scales and estimate the properties of the dark matter halo environment on the largest scales.
	
	The two-phase formation scenario found particular motivation and success in explaining why,  
	% At low redshift, the effective radii ($R_e$) of massive (stellar mass $>10^{11.5}$M$_{\odot}$) quenched galaxies typically exceed 5 kpc (e.g. \citealt{vanderwel2014}). However, galaxies of these stellar masses (M$_*$) at $z>2$ are two to three times smaller in radius (\citealt{toft2007,cimatti2008,buitrago2008,vandokkum2010}). 
	since $z\sim 2$, the M$_*$ of passive spheroidal galaxies has apparently increased by only a factor of two (\citealt{toft2007,cimatti2008,buitrago2008,vandokkum2010}), while their effective radii ($R_e$) have increased by a factor of three to six (\citealt{daddi2005,trujillo2006b,trujillo2006a,trujillo2007,zirm2007,vanderwel2008,vandokkum2008,damjanov2009,cassata2010,cassata2011b,vanderwel2014}). 
	After an initial phase that forms the 
	``red nuggets" observed at $z \sim 2$ (\citealt{vandokkum2009,newman2010,damjanov2011,whitaker2012b,dekel-burkert2014}), the second evolutionary phase involves stellar accretion through minor mergers which preferentially adds \textit{ex-situ} stars to the outskirts (e.g. \citealt{zolotov2009,hopkins2010,tissera2013,tissera2014,cooper2015,rodriguez-gomez2016}). It has been shown that such accretion efficiently increases $R_e$ while keeping M$_*$ roughly constant (e.g. \citealt{bezanson2009,barro2013,cappellari2013b,wellons2015}).
	%  These ideas are at the basis of the current cosmological picture for structure formation and evolution, in which massive systems accrete stellar envelopes from satellite galaxies (\citealt{oser2010,oser2012,johansson2012}).
	
	Direct support of this picture comes from studies of the surface brightness profiles of massive nearby galaxies that almost always feature multiple components (e.g. \citealt{huang2013a}). Their inner parts ($R<1$ kpc) are very compact and populate the same region of the mass-size plane as massive $z>1$ galaxies (\citealt{huang2013b}). In contrast, their outer envelopes can be quite extended ($R>10$ kpc) and are consistent with being built through minor mergers (\citealt{huang2013b}). These outer envelopes can also show greater ellipticity than the inner parts, potentially reflecting the orbital properties of accreted satellites (\citealt{huang2018}).
	
	In \citet{oyarzun2019}, we applied a different test of the two phase formation scenario by searching for the predicted flattening of the stellar metallicity profile that is expected from the accretion of lower mass galaxies  (e.g. \citealt{cook2016,taylor2017}).  
	% Another way of testing the two phase formation scenario is through the analysis of stellar populations, as minor mergers are predicted to flatten the stellar metallicity profiles of galaxies (e.g. \citealt{cook2016}). \citet{oyarzun2019} searched for evidence of this feature in early-type galaxies (ETGs) from
	Using a sample of early-type galaxies (ETGs) from the MaNGA survey (\citealt{bundy2015}), we detected a flattening beyond $R_e$ in the otherwise declining metallicity profiles. This flattening grows more prominent with increasing M$_*$,
	% The metallicity profiles of M$_*<10^{11}$M$_{\odot}$ ETGs fall steeply with radius, whereas 
	especially for ETGs with M$_*>10^{11}$M$_{\odot}$. This observation not only suggests that massive ETGs assembled their outskirts through minor mergers, but also that their \textit{ex-situ} stellar mass fraction is higher, in agreement with theoretical predictions (\citealt{rodriguez-gomez2016}).
	
	% In this picture presented by simulations and observations, environment plays a major role in galaxy formation. This motivates searching for signatures of environment-driven growth at fixed M$_*$. \citet{wetzel2012} revealed that central and satellite galaxies in SDSS (\citealt{york2000,gunn2006}) show significant differences in their assembly history at fixed M$_*$. On average, satellite galaxies are older and are more chemically enriched than centrals of the same M$_*$ (\citealt{pasquali2010}). In order to understand what physical mechanisms are driving these differences, radial information is key (e.g. \citealt{huang2013a,huang2013b,huang2018,oyarzun2019}). Thus, the natural step is to replicate these analyses with spatially resolved surveys. 
	
	With mergers and stellar accretion driving the second phase of spheroidal galaxy evolution, it is natural to search for a link between passive centrals and their host halo environments, given that those environments determine the central galaxy's merger history (e.g. \citealt{hopkins2010b}).
	Unfortunately, searches for signatures of environment-driven growth in spatially resolved surveys have been so far inconclusive. \citet{santucci2020} compared the stellar population gradients of central and satellite galaxies in the SAMI survey (\citealt{allen2015}) at fixed M$_*$ and found no significant differences. In initial efforts with MaNGA data, \citet{zheng2017,goddard2017a,goddard2017b} studied the stellar age and metallicity gradients of massive galaxies and found no significant correlation with the local environment or different proxies for the large scale structure. The consensus of recent work is that at fixed M$_*$, the stellar populations of massive galaxies within $\sim$1 $R_e$ are largely determined by their \textit{in-situ} formation histories rather than their environment (\citealt{peng2012,greene2015,goddard2017b,scott2017,contini2019,bluck2020}).
	
	In this paper, we return to the important question of \textit{in-situ} versus \textit{ex-situ} evolution in passive galaxies.  By adopting the perspective of the stellar-to-halo mass relation, as defined for central galaxies  (SHMR; \citealt{moster2013}), we re-frame the question as a search for secondary correlations in the stellar populations of passive centrals as a function of halo mass (M$_h$) at fixed stellar mass (M$_*$), as well as correlations with M$_*$ at fixed M$_h$.  The first characterization allows us to revisit the role of environment, as expressed by M$_h$, in modulating galaxy formation at fixed M$_*$.  The second lets us ask, to what extent does M$_h$ determine the fate of a central galaxy?
	
	% The question of how environment modulates galaxy formation leads us to ask: if we fix the initial host halo environment, how diverse is the range of possible galaxies that can form and what mechanisms drive that diversity? In other words, we are interested in the specific question: To what extent does halo mass determine the fate of a central galaxy?
	
	This latter question is significant because theoretical models ultimately tie galaxy properties to their dark matter halos.  Deviations, especially in the intrinsic
	% We know that halo mass has a strong impact on the amount of stellar mass central galaxies can form. The deeper potential wells of more massive halos efficiently limit the impact of feedback (e.g. \citealt{booth-schaye2010}), leading to the formation of more massive systems (\citealt{matteucci1994}) at the peak of star-formation efficiency (M$_h\sim 10^{12}$M$_{\odot}$; \citealt{behroozi2013,girelli2020}). Mergers and stellar accretion are believed to lead to further growth (\citealt{oser2010,oser2012,johansson2012}). As a result of these mechanisms, M$_h$ and M$_*$ follow a strong correlation known as the stellar-to-halo-mass relation of central galaxies. 
	scatter of the SHMR, have garnered a lot of interest and motivated work on ``galaxy assembly bias,'' the possible existence of correlations between galaxy and secondary halo properties at fixed M$_h$ (e.g. \citealt{zentner2014,wechsler2018,xu2020}). According to the theory, galaxy luminosity correlates with halo formation time and concentration at fixed M$_h$ (\citealt{croton2007,matthee2017,kulier2019,xu2020}). More concentrated halos, for example, facilitate the formation of deeper potential wells that may promote a more rapid assembly of M$_*$ (\citealt{booth-schaye2010,matthee2017,kulier2019}).  Gas accretion and star formation should commence earlier in halos that formed early for their M$_h$ (\citealt{kulier2019}). 
	% This prediction can be tested with our dataset by exploring correlations between stellar age and M$_*$ at fixed M$_h$. Another prediction by \citealt{kulier2019} is that the star formation timescales of centrals are longer in highly concentrated halos. This scenario can be tested by exploring correlations between chemical abundances and M$_*$ at fixed M$_h$. 
	Yet, it remains unclear whether these predicted differences in halo and galaxy assembly history have an impact on the nature of the stellar populations across the SHMR as observed today.

	To make progress on these questions and begin to delineate subtle secondary correlations between M$_*$ and M$_h$, we construct an updated catalog of 2200 passive centrals drawn from the MaNGA survey.  Previous work has emphasized the importance of $\sim$1\% level or better precision in measuring stellar age and abundances in ETG spectra (e.g. \citealt{conroy2014}).  Galaxy-integrated spectra in our sample can exceed signal-to-noise ($S/N$) $\sim 100$ per galaxy, making MaNGA the premiere data set for co-added spectral analyses of nearby galaxies at this level of precision.  The $S/N$ from stacking all single-fiber ETG spectra in the SDSS \textsc{main} Galaxy Survey (\citealt{york2000,gunn2006,dr12}) would be a factor 0.6 times lower than the equivalent from the MaNGA stack.  Equally important, the MaNGA data allow for simultaneous spatially-resolved measurements, allowing for consistency checks across radial bins.  Our high $S/N$ sample allows us to first detect subtle spectral differences in a model-free manner and then employ sophisticated full spectral fitting codes like \texttt{Prospector} and \texttt{alf} for the interpretation (\citealt{leja2017,johnson2020,conroy2012a,conroy2018}).  These codes model the non-linear response of spectral features (\citealt{conroy2013}) to infer the age, abundance of various elements, and initial mass function (IMF) of old stellar systems ($\gtrsim$1 Gyr), while also accounting for uncertainties in stellar evolution.
	
	% K Bundy (2021-04-13) - Stacked S/N after stacking all ETGs
	%  SDSS:  sqrt(5e6*0.8*0.1)*10 = 2,190
	%. MaNGA: sqrt(1e4*0.8*0.15)*100 = 3,464
	
	% Though we address sample limitations by using one of the best samples available, searches for signatures of environment-driven growth face further challenges. One of them is stellar population characterization. The spectral response to changes in stellar population parameters is highly non-linear (e.g. \citealt{conroy2018}). The spectral imprints left by stellar age, metallicity, element abundances, and IMF are also quite degenerate. As a result, modeling the stellar components in galaxies is very difficult (\citealt{conroy2013}). We cannot neglect these challenges, but we can assess their impact by directly comparing galaxy spectra before analyzing stellar population fitting outputs.
	
	% In this work, we tackle these difficulties by searching for signatures of environment-driven evolution directly in the spectra. As we will show, we find significant spectral differences that correlate with M$_h$ at fixed M$_*$. In order to interpret this differences, we turn to stellar population fitting with the program \texttt{alf} (\citealt{conroy2012a,conroy2018}). The code \texttt{alf} allows us to fit for the age, abundance of various elements, and initial mass function (IMF) of old stellar systems ($\gtrsim$1 Gyr), while also accounting for uncertainties in stellar evolution.
	
	We also pay close attention to the impact of systematic errors on our results from potential biases in the M$_h$ estimates. We are unfortunately limited in this study to halo estimates from group finding algorithms, which must first distinguish centrals from satellites and then require total M$_*$ measurements of all group members (e.g. \citealt{yang2005,yang2007}). 
	% However, total M$_*$ are highly uncertain in large photometric surveys (e.g. \citealt{bundy2017}), as the outer light profiles of galaxies can be missed due to shallow surface brightness limits (\citealt{bernardi2013,huang2018}). In addition, most group finding algorithms are not calibrated on observations, but on mock galaxy catalogs instead (see \citealt{tinker2020a} for an example). As a result, algorithms 
	Group catalogs often fail to reproduce the fractions of red and blue satellites, the dependence of the stellar-to-halo mass relation for centrals (SHMR) on galaxy color, and correlations between M$_h$ and secondary galaxy properties (\citealt{tinker2020a}). 	
	% In this paper, we lessen the impact of systematics in group finding algorithms by utilizing the catalog by
	We address these concerns by utilizing the new SDSS halo catalog in \citet{tinker2020a,tinker2020b}, which exploits deep photometry from the DESI Legacy Imaging Survey (\citealt{dey2019}).
	% which allows them to constrain halo mass better than with SDSS photometry. In addition,
	\citet{tinker2020a} also implement a group finding algorithm that is calibrated on observations of color-dependent galaxy clustering and estimates of the total satellite luminosity. As a result, their catalog better reproduces the color-dependent satellite fraction of galaxies and improves on the purity and completeness of central galaxy samples. In this work, we use the M$_h$ estimates by \citet{tinker2020a,tinker2020b} and compare them against those by \citet{yang2005,yang2007}, allowing us to assess how sensitive our results are to systematics in halo catalogs.   
	
	% By addressing limitations in signal-to-noise, approach to stellar population characterization, and measure of halo mass, in this work we can ask how the formation histories of central galaxies depend on M$_*$ and M$_h$. To do so, we construct a sample of over 1800 passive central galaxies from the MaNGA survey and group them according to their M$_*$ and M$_h$. We construct high signal-to-noise stacked spectra to show that the strength of various spectral absorption features depend on M$_*$, M$_h$, and radius. We associate these spectral differences to stellar population variations with \texttt{alf}, which we use to inform our understanding of how passive galaxies form. We also discuss how our findings might be affected by the biases, limitations, and assumptions of different group catalogs.
	
	% This motivates the second question we ask in this paper: are the stellar populations of central galaxies consistent with galaxy assembly bias predictions? To answer this question, we use \texttt{alf} to constrain the stellar populations of passive central galaxies as a function of M$_*$ and M$_h$. We will show that, at fixed M$_h$, high-M$_*$ centrals have stellar populations that are older, have higher [Fe/H], and have greater [$\alpha$/Fe] than low-M$_*$ centrals. We will discuss several scenarios to explain our findings, including galaxy assembly bias and differences in accretion history.
	
	This paper is structured as follows. In Section \ref{2}, we introduce our dataset and sample of passive central galaxies. In Section \ref{3}, we describe our treatment of the spectra and the stellar population fitting process. In Section \ref{4}, we show the results from direct spectral comparison and stellar population fitting. We interpret our findings in Section \ref{5} and summarize in Section \ref{6}. Stellar masses throughout were obtained assuming a \citet{kroupa2001} IMF. The halo masses used in this work are reported in units of $h^{-1}$M$_{\odot}$. For all other physical quantities, this work adopts $H_0=70$ km s$^{-1}$Mpc$^{-1}$. All magnitudes are reported in the AB system (\citealt{oke1983}).

	\section{Dataset}
	\label{2}
	
	\subsection{The MaNGA survey}
	\label{2.1}
	
	The MaNGA survey (\citealt{bundy2015,yan2016b}) is part of the now complete fourth generation of SDSS (\citealt{york2000,gunn2006,blanton2017,dr15}) and obtained spatially resolved spectra for more than ten thousand nearby galaxies ($z<0.15$). By means of integral field unit spectroscopy (IFS; \citealt{drory2015,law2015}), every galaxy was observed with fiber bundles with diameters varying between 12\farcs 5 and 32\farcs 5 and composed of 19 to 127 fibers. The resulting radial coverage reaches between 1.5$R_e$ and 2.5$R_e$ for most targets (\citealt{wake2017}). The spectra cover the wavelength range 3600-10300 \AA \ at a resolution of R$\sim$2000 (\citealt{smee2013}). The reduced spectra have a median spectral resolution of $\sigma$=72 km s$^{-1}$.
	
	All MaNGA data used in this work was reduced by the Data Reduction Pipeline (DRP; \citealt{law2016,law2021,yan2016a}). The data cubes typically reach a 10$\sigma$ continuum surface brightness of 23.5 mag arcsec$^{-2}$, and their astrometry is measured to be accurate to 0\farcs1 (\citealt{law2016}). De-projected distances, stellar kinematics, and spectral index maps were calculated by the MaNGA Data Analysis Pipeline (DAP; \citealt{belfiore2019,westfall2019}). This work also used Marvin (\citealt{cherinka2017}), the tool specially designed for access and handling of MaNGA data\footnote{\href{https://www.sdss.org/dr15/manga/marvin/}{https://www.sdss.org/dr15/manga/marvin/}}.
	
	Effective radii ($R_e$) for all MaNGA galaxies are publicly available as part of the NASA-Sloan Atlas\footnote{\href{http://nsatlas.org}{http://nsatlas.org}}(NSA). These $R_e$ were determined using an elliptical Petrosian analysis of the $r$-band image from the NSA, using the detection and deblending technique described in \citet{blanton2011}.
	
	\subsection{Selection of passive galaxies}
	\label{2.2}
	
	This paper is based on the MaNGA Product Launch 11 (MPL-11) dataset, which consists of observations for over 10,000 MaNGA targets (see Table 1 in \citealt{law2021} for reference on the various release versions). To select passive galaxies, we used estimates of the spatially integrated specific star-formation rate (sSFR) of MaNGA galaxies derived as part of the pipeline for the pipe3D Value Added Catalog for DR17\footnote{\href{https://www.sdss.org/dr17/manga/manga-data/manga-pipe3d-value-added-catalog/}{https://www.sdss.org/dr17/manga/manga-data/manga-pipe3d-value-added-catalog/}}. These sSFRs are based on measurements of the H$\alpha$ equivalent width that were corrected by dust attenuation using the Balmer decrement (\citealt{sanchez2016}). We defined our sample of passive galaxies by setting the criterion $\log{(\mbox{sSFR})}<-11.5$ M$_{\odot} yr^{-1}$.%, which is commonly used to distinguish between ionization due to a smooth background of hot evolved stars and that are due to star formation and AGN (\citealt{cidfernandes2011,belfiore2016,sanchez2020}). No morphological selection criteria were implemented.
	
	Our approach yielded a subset with 3957 passive galaxies, of which 2217 are identified as centrals in the catalog by \citet{tinker2020a,tinker2020b} and 952 in the \citet{yang2007,wang2016} catalog. Details on the environmental classification are presented in Section \ref{2.4}. %According to the Galaxy Zoo     VAC\footnote{\href{https://www.sdss.org/dr15/data_access/value-added-catalogs/?vac_id=manga-morphologies-from-galaxy-zoo}{https://www.sdss.org/dr15/data\_access/value-added-catalogs/?vac\_id=manga-morphologies-from-galaxy-zoo}} (\citealt{willett2013,hart2016}), the morphology for 55\% of our centrals can be qualified as ``smooth"  (i.e., spheroids), whereas 45\% ``have features or a disk".
	
	\begin{figure*}
		\centering
		\includegraphics[width=7in]{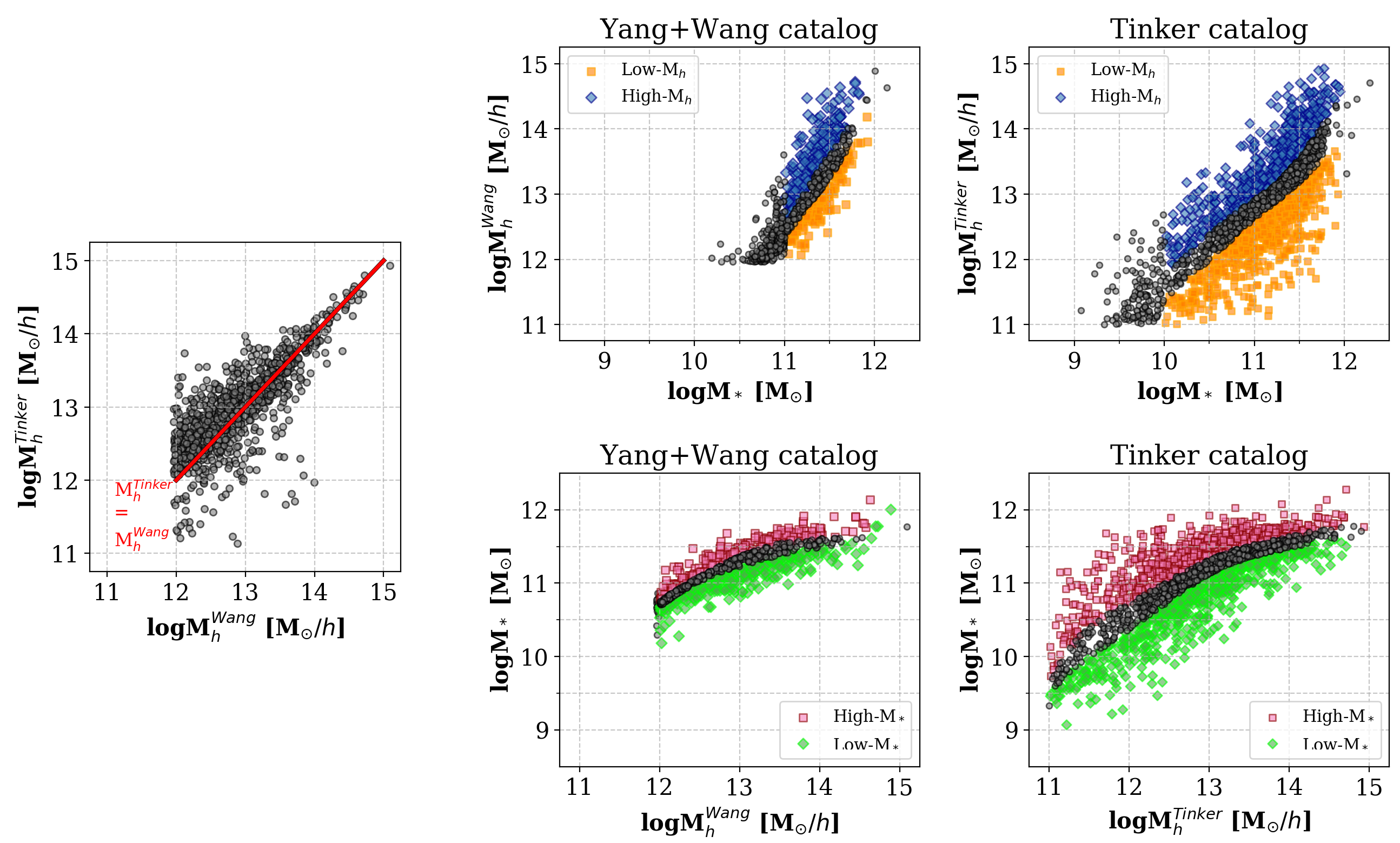}
		\caption{Comparison between the M$_h$ measured by \citet{wang2016} and \citet{tinker2020b} for our passive centrals. The M$_h^{\mbox{\tiny Wang}}=$M$_h^{\mbox{\tiny Tinker}}$ line is shown in red. \textbf{Right:} Stellar-to-halo mass relation for passive centrals in the two catalogs. We define our subsamples based on the independent property and stellar-to-halo mass ratio. At fixed M$_*$ (upper panels), high-M$_h$ centrals are shown in blue and low-M$_h$ centrals in orange. At fixed M$_h$ (lower panels), high-M$_*$ centrals are shown in red and low-M$_*$ in green. Galaxies in regions of the SHMR with narrow dynamic range were not included in the analysis (gray). We also excluded galaxies between the 33rd and 66th percentiles in M$_*$-to-M$_h$ ratio.}
		\label{fig3}
	\end{figure*}	
	
	\begin{deluxetable*}{|c|c|c|c|c|c|c|}
		\renewcommand*{\arraystretch}{1.2}
		\centering
		\tablecaption{Number of central galaxies per sample}
		\tabletypesize{\footnotesize}
		\tablewidth{0pt}
		\tablehead{
			\colhead{M$_*$[M$_{\odot}$]=} & \colhead{$10^{10}-10^{10.5}$\tablenotemark{*}} & \colhead{$10^{10.5}-10^{11}$\tablenotemark{*}} & \colhead{$10^{11}-10^{11.5}$}  & \colhead{$10^{11.5}-10^{12}$} & Out of range & \colhead{Total}}
		%\colnumbers
		\startdata
		\label{table1}
		Yang+Wang & 7 & 267 & 523 & 153 & 2 & 952\\	
		Yang+Wang low-M$_h$ & 0 & 0 & 174 & 48 & - & 222\\	
		Yang+Wang high-M$_h$ & 0 & 0 & 172 & 51 & - & 223\\		
		\hline		
		Tinker & 211 & 514 & 948 & 412 & 132 & 2217 \\
		Tinker low-M$_h$ &  70 & 188 & 317 & 128 & - & 703\\	
		Tinker high-M$_h$ & 70 & 143 & 353 & 148 & - & 714 \\			
		\hline
		\colhead{} \\
		\hline
		\hline
		\colhead{M$_h$[M$_{\odot}/h$]=} & \colhead{$10^{11}-10^{12}$\tablenotemark{*}} & \colhead{$10^{12}-10^{13}$}  & \colhead{$10^{13}-10^{14}$} & \colhead{$10^{14}-10^{15}$} & Out of range & \colhead{Total} \\
		\hline
		Yang+Wang & 27 & 517 & 358 & 49 & 1 & 952\\
		Yang+Wang low-M$_*$ & 0 & 170 & 120 & 19 & - & 309\\	
		Yang+Wang high-M$_*$ & 0 & 172 & 113 & 15 & - & 300\\
		\hline		
		Tinker & 329 & 888 & 860 & 139 & 1 & 2217\\
		Tinker low-M$_*$ & 109 & 297 & 265 & 45 & - & 716\\	
		Tinker high-M$_*$ & 112 & 310 & 301 & 50 & - & 773\\	
		\enddata
		\tablenotetext{*}{These bins were not considered as a result of the narrow dynamic ranges in M$_h$ or M$_*$ (Figure \ref{fig3}).}
	\end{deluxetable*}
	
	\begin{figure}
		\centering
		\includegraphics[width=3.2in]{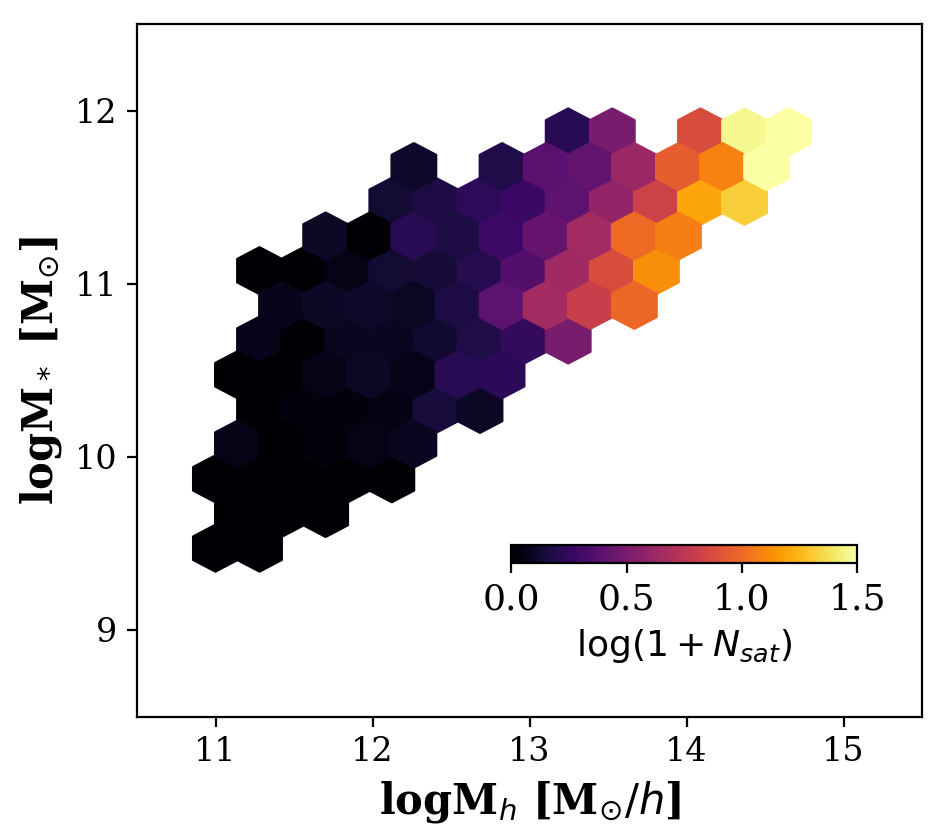}
		\caption{Number of satellite galaxies per halo as a function of M$_h$ and M$_*$ in the \citet{tinker2020a,tinker2020b} catalog. At fixed M$_*$, centrals in high-M$_h$ halos tend to have more satellites than centrals in low-M$_h$ halos.}
		\label{fig3a}
	\end{figure}

	\subsection{Stellar masses}
	\label{2.3}
	
	%To estimate the M$_*$ of our galaxies, we used r-band luminosities from the DESI Legacy Imaging Survey (DLIS; \citealt{dey2019}). This ongoing survey targets to provide deep imaging for 14000 deg$^2$ of the extragalactic sky in optical and mid-infrared bands. In the r-band, DLIS targets a 5$\sigma$ detection limit of $r>23$. In this work, we use the r-band luminosities (L$_r^{\mbox{\tiny DLIS}}$) reported in the group catalog by \citet{tinker2020a,tinker2020b}, which correspond to the DLIS DR6 and DR7. The luminosities in this catalog cover $\sim 75\%$ of the SDSS footprint (\citealt{tinker2020b}) and are publicly available\footnote{\href{https://galaxygroupfinder.net}{https://galaxygroupfinder.net}}.
	
	%To convert from L$_r^{\mbox{\tiny DLIS}}$ to M$_*^{\mbox{\tiny DLIS}}$, we first estimated spectroscopic stellar masses for every galaxy as probed by MaNGA (M$_*^{\mbox{\tiny MaNGA}}$). 
	
	To estimate the M$_*$ of our galaxies, we first co-added the MaNGA spectra within the $R_e$ of every galaxy (measured on r-band imaging from SDSS). Before co-addition, we shifted every spectrum back to the rest-frame using the stellar systemic velocity ($v_*$) maps calculated by the DAP. Then, we estimated the mass within 1$R_e$ by running the stellar population fitting code \texttt{Prospector} (\citealt{leja2017,johnson2020}) on the co-added spectrum. Our runs adopted the MILES stellar library (\citealt{sanchez-blazquez2006}), MIST isochrones (\citealt{dotter2016,choi2016}), and \citet{kroupa2001} IMF. Further details on our \texttt{Prospector} runs are presented in Section \ref{3.4}. 
	
	We then assumed the total spectroscopic stellar mass of every galaxy to be 
	\begin{gather}
	\mbox{M}_*^{total}= 2\mbox{M}_*^{R_e} \times 10^{-0.15}, 
	\end{gather}
	where $\mbox{M}_*^{R_e}$ is the spectroscopic stellar mass within the effective radius measured with \texttt{Prospector}. Since it has been found that half-mass radii are smaller than half-light radii (\citealt{garcia-benito2017}), our M$_*^{total}$ may be overestimated. Yet, we do not expect any biases to arise from this definition, since differences between half-mass and half-light radii have not been found to correlate with stellar mass (\citealt{szomoru2013}). Any overestimation of our stellar masses was corrected by implementing an offset of 0.15 dex (see the multiplicative term in the equation).  This value was obtained by measuring the offset between our $2\mbox{M}_*^{R_e}$ and the stellar masses measured through k-correction fits to the Sersic fluxes in the NSA (\citealt{blanton2007}). For the rest of the paper, we will simply refer to M$_*^{total}$ as M$_*$.

	\subsection{Yang+Wang halo masses}
	\label{2.4}	
	
	For our first characterization of environment, we used the MPL-9 version of the Galaxy Environment for MaNGA Value Added Catalog\footnote{\href{https://www.sdss.org/dr15/data\_access/value-added-catalogs/?vac\_id=gema-vac-galaxy-environment-for-manga-value-added-catalog}{https://www.sdss.org/dr15/data\_access/value-added-catalogs/?vac\_id=gema-vac-galaxy-environment-for-manga-value-added-catalog}} (GEMA-VAC; \citealt{argudo-fernandez2015}). We used this catalog to identify central galaxies and retrieve estimates of their halo masses. The environmental classification and halo mass entries were computed by cross-matching MaNGA MPL-9 with the \citealt{yang2007} group catalog for SDSS (see also \citealt{yang2005}). The halo masses computed by \citealt{yang2007} assumed a WMAP3 cosmology (\citealt{spergel2007}). In this work, we use the values in the GEMA-VAC that were updated to a WMAP5 cosmology (\citealt{dunkley2009}) as part of the work by \citet{wang2009,wang2012,wang2016}. 
	
	The group catalog by \citet{yang2007} was computed on the SDSS New York University Value-Added Galaxy Catalog (NYU-VAGC; \citealt{blanton2005b}) based on SDSS DR4 (\citealt{adelman2006}). To compute the \citet{yang2007} catalog, a series of steps were iterated until convergence was achieved. First, clustering analysis in redshift space was used to find potential cluster centers and groups. In the second step, group luminosities (L$_{19.5}$) were computed as the combined luminosity of all group members with $M_r-5\log{h}\leqslant-19.5$ (hence the subscript in L$_{19.5}$). Dark matter halo masses, sizes, and velocity dispersions were then estimated. In particular, tentative M$_h$ were assigned according to the L$_{19.5}$-M$_h$ relation measured in the previous iteration, which assumes a one-to-one correspondence between L$_{19.5}$ and M$_h$. Finally, membership probabilities in redshift space around group centers were estimated. This allowed for group memberships to update. The process from the second to the final step was repeated until no further changes in group memberships were observed. After convergence, final M$_h$ were derived through abundance matching using the \citet{vandenbosch2007} halo mass function.  
	
	Of the 3957 passive galaxies in our sample, 952 are centrals with halo mass measurements in the GEMA-VAC (see Figure \ref{fig3}). The SHMR using the stellar masses from Section \ref{2.3} and the halo masses from the Yang+Wang catalog is plotted in Figure \ref{fig3}. Details on the number of galaxies as a function of M$_*$ and M$_h$ are shown in Table \ref{table1}.
	
	\subsection{Tinker halo masses}
	\label{2.5}	
	
	Group finding algorithms, like the one described in Section \ref{2.4}, are affected by several issues. By not breaking down galaxy samples into star-forming and quiescent subsamples, they can fail to reproduce the fraction of quenched satellite galaxies and mis-estimate by an order of magnitude M$_h$ (\citealt{campbell2015}). Some of these shortcomings can be tackled by calibrating the free parameters of the group finder with real data instead of mock catalogs. In this paper, we work with the group finder by \citet{tinker2020a}, which uses observations of color-dependent galaxy clustering and total satellite luminosity for calibration.
	
	In the self-calibrating halo-based galaxy group finder by \citet{tinker2020a}, the probability of a galaxy being a satellite depends on galaxy type and luminosity. This dependence is quantified by 14 different parameters that are calibrated until the best-fitting model is found. First, a starting value for the parameters is adopted and the group finder is run until the fraction of red and blue satellites match the input dataset. The assigned groups and halo masses are then used to populate the Bolshoi-Planck simulation (\citealt{klypin2016}) and predict galaxy clustering and the total satellite luminosity. These predictions are compared to observational measurements to quantify the adequacy of the model. The process is then repeated until the best-fitting model is found.
	
	This algorithm was applied to SDSS galaxies in \citet{tinker2020b}. Compared to the dataset available to \citet{yang2005,yang2007}, \citet{tinker2020b} had access to deeper photometry from the DESI Legacy Imaging Survey (DLIS; \citealt{dey2019}). This is quite important, since good accounting of group and galaxy M$_*$ is key to properly constraining M$_h$ (\citealt{bernardi2013,wechsler2018}). As a result, the SDSS group catalog by \citet{tinker2020b} is an improvement in both dataset and algorithm.
	
	Despite improvements, the approach by \citet{tinker2020a} still has limitations. Like all group finding algorithms, it is susceptible to central galaxy mis-identification. It also assumes that the amount of light in satellite galaxies is a function of halo mass only and its implementation on SDSS data fails to match the clustering of faint quiescent galaxies (\citealt{tinker2020b}). There is also room for further freedom in how the algorithm fits the data, in particular for taking into account secondary correlations between galaxy and halo properties (\citealt{tinker2020b}).	
	
	The self-calibrating halo-based galaxy group finder applied to SDSS is publicly available\footnote{\href{https://galaxygroupfinder.net}{https://galaxygroupfinder.net}}. We downloaded the catalog and cross-matched with our sample of quenched systems. We selected all galaxies with satellite probabilities lower than 0.1 and obtained a sample with 2217 passive central galaxies (see Table \ref{table1}). The M$_h$ measured by Tinker and Yang+Wang are compared in the left panel of Figure \ref{fig3}. The right panel shows the SHMR using our M$_*$ and M$_h^{\mbox{\tiny Tinker}}$, which extends down to M$_*\sim10^{10}$M$_{\odot}$ (M$_h^{\mbox{\tiny Tinker}}\sim10^{11}h^{-1}$M$_{\odot}$). Figure \ref{fig3a} shows that the average number of satellites per halo monotonically increases with M$_h^{\mbox{\tiny Tinker}}$.
	
	\section{Methodology}
	\label{3}
	%\textbf{In this section, we describe our methodology to \textbf{co-add spectra for every galaxy} and the stellar population fitting process with Firefly, pPXF, and \texttt{Prospector}.}
	%\textbf{In this paper, we explore the existence of radial trends in two ETG scaling relations that involve metallicity (hereafter log$Z/Z_{\odot}$): the log$\sigma_*$-log$Z/Z_{\odot}$ and logM$_*$-log$Z/Z_{\odot}$ relations. In this section, we first introduce our scheme to define radial bins. Then, we detail the Bayesian framework that we use in order to constrain the parameters of the scaling relations.}
	
	\subsection{Sample definitions}
	\label{3.1}	
	
	The correlation between the M$_*$ and M$_h$ of central galaxies has significant scatter, some of which is thought to be intrinsic (e.g. \citealt{xu2020}). This would imply that the stellar masses of central galaxies are not uniquely determined by the mass of their halos, but also by secondary properties (e.g. \citealt{zentner2014}). In consequence, the SHMR is a very useful tool for testing the mechanisms driving the galaxy-halo connection (e.g. \citealt{leauthaud2017}). The purpose of this work is to probe these mechanisms through the analysis of galaxy stellar populations. 
	
	Figure \ref{fig3} shows the SHMR for our sample of centrals according to both environmental catalogs. We computed the 33rd and 66th percentiles in M$_*$-to-M$_h$ ratio as a function of M$_*$ to define two subsamples. High-M$_h$ centrals are those that reside in high M$_h$ for their M$_*$ and are shown in blue. Low-M$_h$ centrals, on the other hand, reside in low M$_h$ halos for their M$_*$ and are shown in orange. In equation form,
	
	\begin{align}
	\mbox{High-M$_h$ sample:  } \frac{\mbox{M$_*$}}{\mbox{M$_h$}}(\mbox{M$_*$})<\left[ \frac{\mbox{M$_*$}}{\mbox{M$_h$}}(\mbox{M$_*$}) \right]_{.33} \\
	\mbox{Low-M$_h$ sample:  } \frac{\mbox{M$_*$}}{\mbox{M$_h$}}(\mbox{M$_*$})>\left[ \frac{\mbox{M$_*$}}{\mbox{M$_h$}}(\mbox{M$_*$}) \right]_{.66}.
	\end{align}
	
	To quantify correlations at fixed M$_h$, we also computed the 33rd and 66th percentiles in M$_*$-to-M$_h$ ratio as a function of M$_h$. High-M$_*$ centrals have high M$_*$-to-M$_h$ ratios, whereas low-M$_*$ centrals have low M$_*$-to-M$_h$ ratios. Put in an equation,
	
	\begin{align}
	\mbox{High-M$_*$ sample:  } \frac{\mbox{M$_*$}}{\mbox{M$_h$}}(\mbox{M$_h$})>\left[ \frac{\mbox{M$_*$}}{\mbox{M$_h$}}(\mbox{M$_h$}) \right]_{.66} \\
	\mbox{Low-M$_*$ sample:  } \frac{\mbox{M$_*$}}{\mbox{M$_h$}}(\mbox{M$_h$})<\left[ \frac{\mbox{M$_*$}}{\mbox{M$_h$}}(\mbox{M$_h$}) \right]_{.33}.
	\end{align}
	
	Galaxies between the 33rd and 66th percentiles in M$_*$-to-M$_h$ ratio were not used and are shown as gray points in Figure \ref{fig3}. Since the distribution of centrals in M$_*$-to-M$_h$ ratio as a function of M$_*$ or M$_h$ is rather flat, no subsample is biased as a result of uncertainties in M$_*$ or M$_h$. Note that the sample membership of a given galaxy depends on the environmental catalog, since our classification scheme is based on M$_h$. The number of galaxies in every M$_*$ and M$_h$ bin for the two environmental catalogs is presented in Table \ref{table1}. Due to the narrow dynamic range in M$_h$ with the Yang+Wang catalog for M$_*<10^{11}$M$_{\odot}$, the corresponding bins were not included in the analysis (see Table \ref{table1}). The average number of satellites per halo (N$_{sat}$) for every subsample is plotted in Figure \ref{fig3a}. At fixed M$_*$, N$_{sat}$ correlates with M$_h$. %At fixed M$_h$, N$_{sat}$ anti-correlates with M$_*$. 
	
	\subsection{Co-addition and stacking of spectra}
	\label{3.2}
	The spectral precision needed by our stellar population analysis requires high signal-to-noise (S/N$\gtrsim50$) spectra (e.g. \citealt{conroy2014}). Thus, we first co-added the MaNGA spectra within every galaxy following a radial binning scheme. We then computed the median of these co-additions to generate galaxy stacks for every subsample. The co-addition and stacking steps are described below.
	
	For every galaxy, we associated galactocentric distances to all spaxels by retrieving elliptical polar radii ($R$) from the DAP that account for the axis ratio of every object. We binned all the spaxels within every galaxy into the three annuli $R=[0, 0.5]$, $[0.5, 1]$, and $[1, 1.5]$ $R_e$. After masking sky line residuals and spectra outside the wavelength range (3700\AA, 9200\AA) in the observed frame, the co-added spectra in every bin were co-added. To do this, we shifted every spectrum back to the rest-frame using the stellar systemic velocity ($v_*$) maps computed by the DAP with a Voronoi binning scheme that aims for a minimum signal-to-noise ratio of 10 per bin. We did not convolve the spectra to a common $\sigma_*$ prior to stacking. We also masked all spaxels that were flagged as unusable by the DRP and DAP.
	
	After co-addition, we ran pPXF (\citealt{cappellari-emsellem2004,cappellari2017}) with the MILES Single Stellar Population (SSP) library (\citealt{vazdekis2010}) on the resulting spectra to measure the co-added $v_*$ and $\sigma_*$. Co-added $v_*$ showed values $v_*\lesssim 1$ km/s, indicating that spectra were properly shifted back to the rest-frame. We also ran \texttt{Prospector} (details in Section \ref{3.4}) to measure stellar mass surface density profiles for every galaxy.
	
	After binning in M$_*$ and M$_h$ (see Table \ref{table1}), we stacked the spectra across every subsample in each of the four radial annuli. For stacking, all co-added spectra were convolved to $\sigma_*=350\ km s^{-1}$ and median normalized. This value is motivated by the maximum observed dispersion, and it mitigates line-strength variations caused by different Doppler broadening. Stacks were obtained by computing the median at each wavelength after removing the continuum. Errors on the stacks were quantified through Monte Carlo simulations of the stacking process that took into account the propagated errors. Two stacked spectra are shown in Figure \ref{fig4} after all emission lines were masked. The dependence of the stacks on whether masking is performed before or after stacking is minimal.
	
	Appendix \ref{appendix1} shows stacked spectra at multiple radii and M$_h$. We typically reach $S/N>200$ at the centers and $S/N>100$ in the outskirts. We find the spectral $S/N$ of the stacks to show dependence on both the individual $S/N$ and the total number of spectra. As a result, the highest M$_*$ and M$_h$ stacks show the lowest spectral $S/N$ of all bins by factors of $\sim 2$.

	\subsection{Evidence for environmental differences}
	\label{3.3}	
\begin{figure*}
		\centering
		\includegraphics[width=7in]{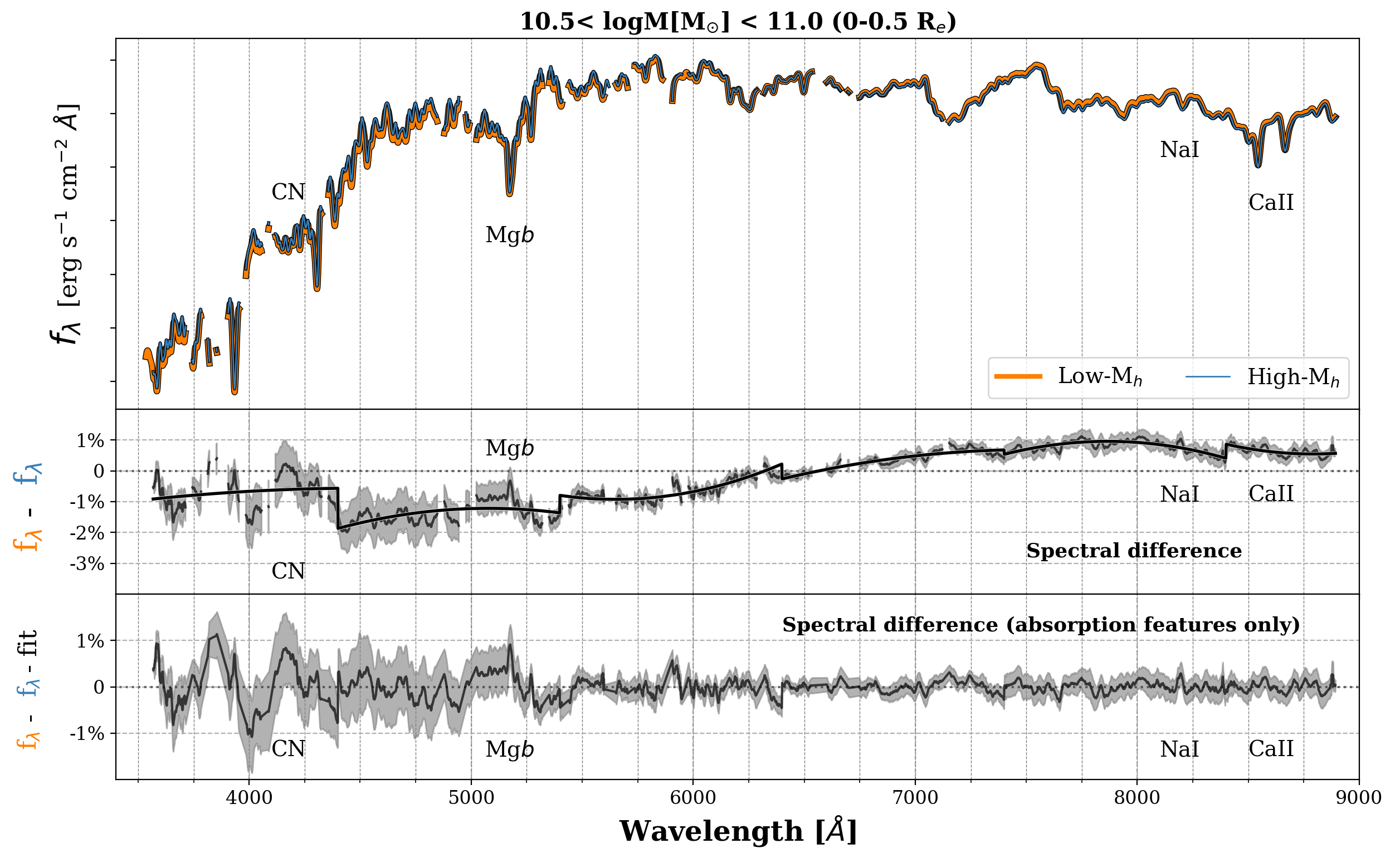}
		\caption{\textbf{Top:} Stacked spectra of central galaxies at fixed M$_*$ that reside in low-M$_h$ halos (orange) and high-M$_h$ halos (blue). The data represent stacks within the central 0.5$R_e$ for M$_*=10^{10.5}-10^{11}$M$_{\odot}$ centrals. \textbf{Middle:} The high-M$_h$ stack subtracted from the low-M$_h$ stack with a fit plotted in black. Gray shades show the error on this difference. \textbf{Bottom:} Result of subtracting the fit from the spectral difference. This highlights variations in several absorption features, some of which help to break degeneracies between various stellar population parameters. The two subsets reveal significant differences in both spectral shape and absorption features. Requiring no model assumptions, this figure demonstrates that M$_h$ has an impact on the stellar populations of passive central galaxies with same M$_*$.}
		\label{fig4}
	\end{figure*}

	The top panel of Figure \ref{fig4} compares the stacked spectra at the centers of high- and low-M$_h$ centrals with M$_*=10^{10}-10^{10.5}$M$_{\odot}$. With M$_*$ held constant (see Section \ref{3.1}), any differences between the spectra in this comparison can be interpreted as environmental signatures. In this subsection, we describe our method to finding and highlighting these signatures. 
	
	We started by subtracting high-M$_h$ stacks from low-M$_h$ counterparts of the same M$_*$. An example of the resulting spectral difference is shown in the middle panel of Figure \ref{fig4}. Then, to highlight variations in spectral features, we subtracted fits from the spectral differences. The result is shown in the bottom panel of Figure \ref{fig4}. Note how the spectral difference captures variations in both spectral shape and features. As we will show in Section \ref{4}, we find some significant environmental signatures at different M$_*$ and radii. This is evidence that M$_h$ has an impact on the stellar populations of passive central galaxies at fixed M$_*$. 
	
	For context, some abundance sensitive features are labeled in this figure. Difference at these locations highlight the fact that environmental signatures can manifest because of differences in stellar age, metallicity, element abundances, and the IMF of central galaxies (\citealt{conroy2018}). To inform our interpretation of these differences, we turned to stellar population fitting codes \texttt{Prospector} and \texttt{alf}.
	
	\begin{figure*}[htp]
		\centering
		\includegraphics[width=7in]{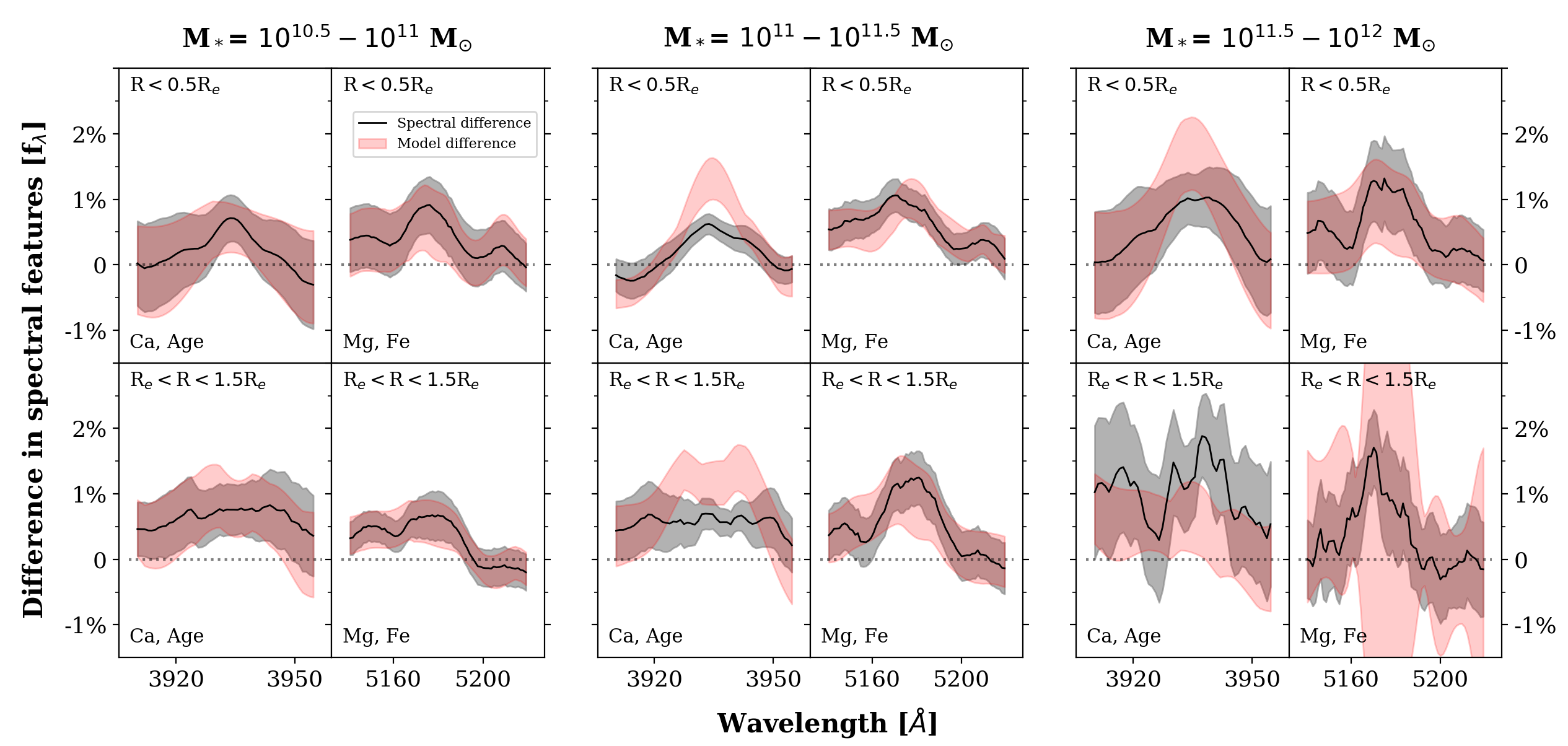}
		\caption{Differences in spectral features between low- and high-M$_h$ centrals of the same M$_*$. Positive difference indicates stronger absorption in high-M$_h$ galaxies, with the 1$\sigma$ error plotted in gray. Different panels show different spectral features for which we detect significant differences that are systematic with M$_*$ and radius. Stellar mass increases from left to right, galactocentric distance from top to bottom. Annotated are the stellar population parameters that dominate the strength of each feature. The red shading represents the 1$\sigma$ posterior distribution of the best-fitting model spectra produced by \texttt{alf}. The \texttt{alf} posteriors reproduce the observed spectral differences well, although some model mismatch at the 1\% level is apparent.}
		\label{fig4b}
	\end{figure*}	
	
	\subsection{Stellar mass surface density profiles with Prospector}
	\label{3.4}
	
	We used the stellar population fitting code \texttt{Prospector}\footnote{\href{https://github.com/bd-j/prospector/blob/master/doc/index.rst}{https://github.com/bd-j/prospector/blob/master/doc/index.rst}} (\citealt{leja2017,johnson2020}) on the co-added spectra to estimate stellar masses (Section \ref{2.3}) and stellar mass surface density ($\Sigma_*$) profiles for all galaxies.
	
	\texttt{Prospector} samples the posterior distribution for a variety of stellar population parameters and star formation history (SFH) prescriptions defined by the user. In this code, stellar population synthesis is handled by the code FSPS\footnote{\href{https://github.com/cconroy20/fsps}{https://github.com/cconroy20/fsps}}(\citealt{conroy2009,conroy-gunn2010}). Our runs used the MILES stellar library (\citealt{sanchez-blazquez2006}), MIST isochrones (\citealt{dotter2016,choi2016}), and Kroupa IMF (\citealt{kroupa2001}) as inputs. We adopted a non-parametric SFH with a continuity prior, which emphasizes smooth SFHs over time (\citealt{leja2019}). As in \citet{leja2019}, we used the following time bins:
	\begin{align}
	\nonumber
	0< &\mbox{ }t <30 \mbox{ Myr} \\
	\nonumber
	30 \mbox{ Myr}< &\mbox{ }t <100 \mbox{ Myr} \\
	\nonumber
	100 \mbox{ Myr}<&\mbox{ }t <330 \mbox{ Myr} \\
	\nonumber
	330 \mbox{ Myr}< &\mbox{ }t <1.1 \mbox{ Gyr} \\
	\nonumber
	1.1 \mbox{ Gyr}< &\mbox{ }t <3.6 \mbox{ Gyr} \\
	\nonumber
	3.6 \mbox{ Gyr}< &\mbox{ }t <11.7 \mbox{ Gyr} \\
	11.7 \mbox{ Gyr}< &\mbox{ }t <13.7 \mbox{ Gyr}
	\end{align}
	
	Our parameter space also included the optical depth of dust in the V-band (\citealt{kriek-conroy2013}), stellar mass, stellar velocity dispersion, and mass-weighted stellar ages and metallicities. To derive the posterior distributions, we used the Dynamic Nested Sampling package dynesty\footnote{\href{https://github.com/joshspeagle/dynesty/blob/master/docs/source/index.rst}{https://github.com/joshspeagle/dynesty/blob/master/docs \\ /source/index.rst}} (\citealt{speagle2019}).
	
	\subsection{Stellar ages and element abundances with \texttt{alf}}
	\label{3.5}
	We used the program \texttt{alf} to characterize the stellar populations in more detail. This code fits the absorption line optical-near infrared spectrum of old ($\gtrsim$1 Gyr) stellar systems (\citealt{conroy2012a,conroy2018}). Unlike \texttt{Prospector}, \texttt{alf} allows us to fit for the abundances of multiple elements and the IMF. This comes at the cost of computation times that are longer by a factor of 100. 
	\texttt{alf} is based on the MIST isochrones (\citealt{dotter2016,choi2016}) and the empirical stellar libraries by \citet{sanchez-blazquez2006} and \citet{villaume2017}. Full spectral variations induced by deviations from the solar abundance pattern are quantified in the theoretical response functions (\citealt{conroy2018,kurucz2018}). These allow \texttt{alf} to sample a multivariate posterior that includes the abundances of 19 elements (including C, N, O, Mg, and Fe).
	
	We ran \texttt{alf} in ``full" mode, which fits for a two-component SFH, stellar velocity dispersion, IMF, and the abundances of 19 elements. We adopted a triple power law IMF with two free parameters. Power laws were fit in the ranges 0.08-0.5M$_{\odot}$ and 0.5-1M$_{\odot}$. The IMF slope was set to -2.35 in the range 1-100 M$_{\odot}$ (\citealt{salpeter1955}). Posterior sampling was performed with emcee (\citealt{emcee}). We found our runs to fully converge under the default configuration, which is 1024 walkers, 10$^{4}$ burn-in steps, and 100-step chains. For this setup, \texttt{alf} took $\sim$100 CPU hours per spectrum to run.
	
	With \texttt{alf}, we recover stellar velocity dispersions within 1\% of the input value ($\sigma_*=350$ km s$^{-1}$). We derived mass-weighted stellar ages by mass-weighting the posteriors of the two-component SFHs. The stellar ages measured by \texttt{Prospector} and \texttt{alf} show agreements within 2 Gyr. The most significant age differences between high- and low-M$_h$ centrals are found by both codes. More subtle differences are only detected by \texttt{alf} (more details in Section \ref{4}). 
	
	%To estimate [Mg/Fe], we computed the average between the enhancement of O, Mg, Si, Ca, and Ti (e.g. \citealt{kirby2008}). Conversions from [X/H] to [X/Fe] assumed the correction factors in \citet{schiavon2007}. The abundances measured by \texttt{alf} cannot be directly compared to the stellar metallicity reported by \texttt{Prospector}, since the latter only fits a scaled solar abundance pattern.
	Following the approach of Lick indices, we use [Mg/Fe] as a proxy for [$\alpha$/Fe] throughout(\citealt{johansson2012b}, see \citealt{kirby2008} for other elements that can be used as tracers for [$\alpha$/Fe]). Conversions from [X/H] to [X/Fe] assumed the correction factors in \citet{schiavon2007}. The abundances measured by \texttt{alf} cannot be directly compared to the stellar metallicity reported by \texttt{Prospector}, since the latter only fits a scaled solar abundance pattern.	
	
	For estimating uncertainties on the fitted parameters, we adopted a three-step method that was iterated over 5 times. First, we bootstrapped the galaxies selected in the sample assignment step from Section \ref{3.1}. Then, we computed the stacked spectra for the corresponding bootstrapped galaxies as in Section \ref{3.2}. Lastly, we fit each stacked spectrum with \texttt{alf}. This resulted in 5 posterior distributions for every subsample in each stellar and halo mass bin. The 5 distributions were then folded in, such that the final posteriors account for uncertainties in both methodology and modeling. Some of the model fits and posterior distributions are shown in Appendix \ref{appendix1}.

	\begin{figure*}[htp]
		\centering
		\includegraphics[width=6.8in]{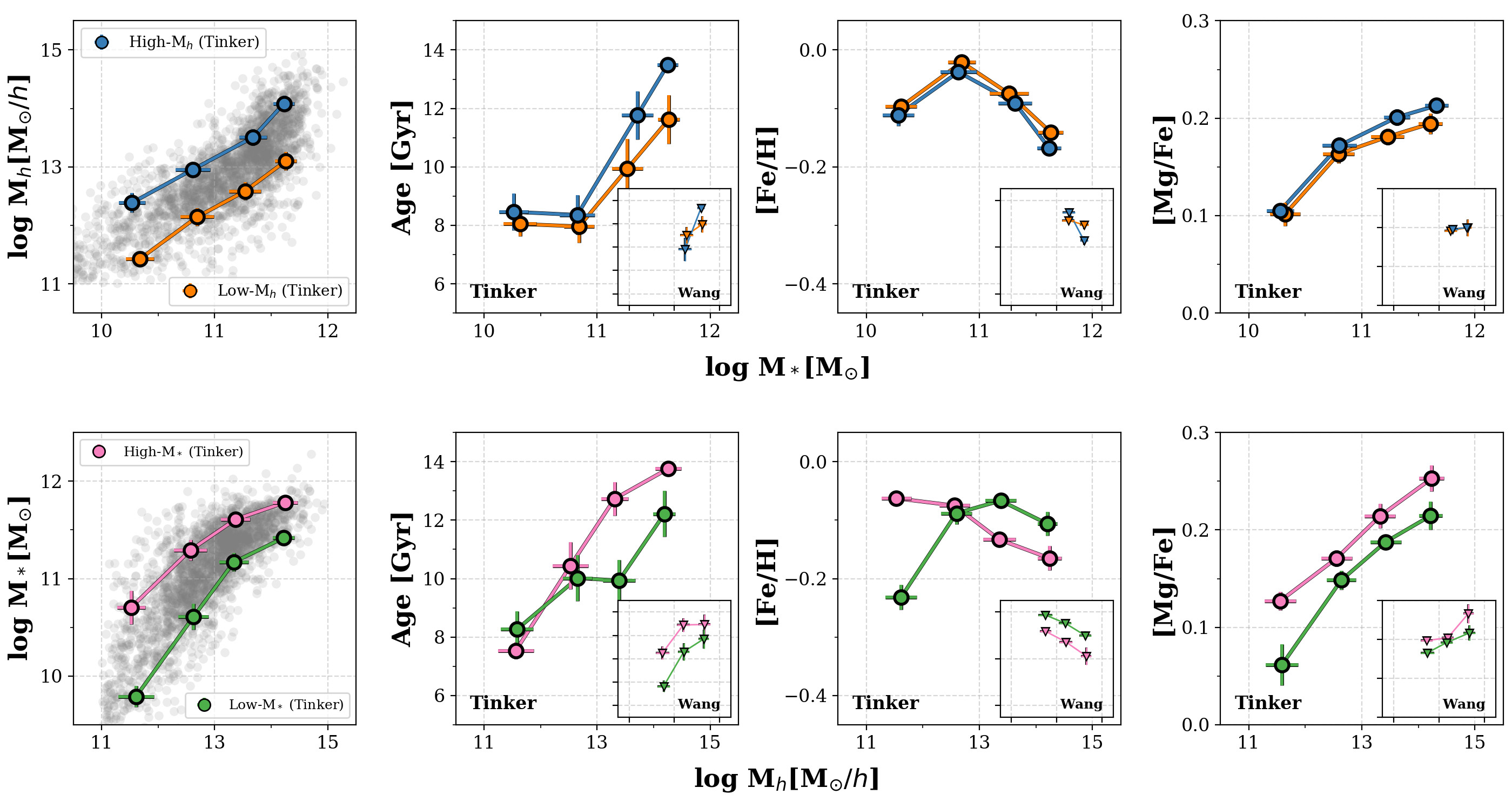}
		\caption{Stellar population parameters of passive central galaxies within $1.5R_e$ as a function of M$_*$ (top) and M$_h$ (bottom). Measurements that used the Tinker catalog for sample assignment are plotted in circles. Measurements that used the Yang+Wang catalog are shown in triangles. The stellar populations of passive centrals better correlate with M$_*$ than with M$_h$. \textbf{Top:} High-M$_h$ centrals are older (3.6$\sigma$), have lower [Fe/H] (3.6$\sigma$), and show higher [Mg/Fe] (3.4$\sigma$) than low-M$_h$ centrals. \textbf{Bottom:} For M$_h>10^{12}h^{-1}$M$_{\odot}$, the stellar populations of high-M$_*$ centrals are older (4.4$\sigma$), have lower [Fe/H] (5$\sigma$), and have greater [Mg/Fe] (4.4$\sigma$) than those of low-M$_*$ counterparts.}
		\label{fig9}
		\label{fig:integrated}
	\end{figure*}	
	
	\begin{figure*}[htp]
		\centering
		\includegraphics[width=7in]{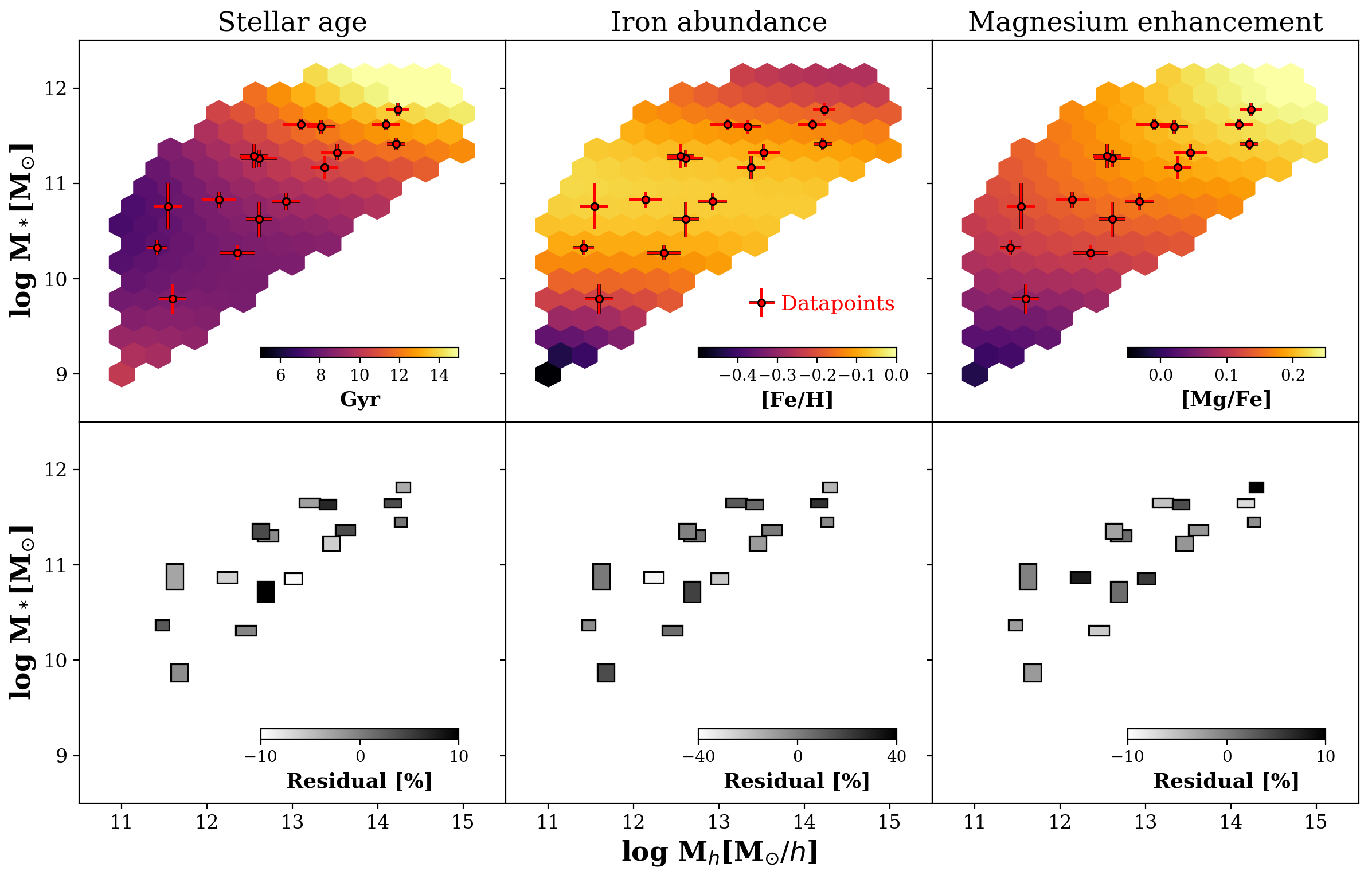}
		\caption{\textbf{Top: }The SHMR of passive centrals colored by stellar age, [Fe/H], and [Mg/Fe]. Values were derived by fitting two-dimension polynomials in M$_*$-M$_h$ space to the measurements from Figure \ref{fig9} that adopted M$_h^{\mbox{\tiny Tinker}}$ (red datapoints). All parameters depend on both M$_*$ and M$_h$. \textbf{Bottom: }Residuals relative to the median value of the parameter in the SHMR.}
		\label{fig9a}
	\end{figure*}
	
	\section{Results}
	\label{4}
	\subsection{Empirical spectral differences with environment}
	\label{4.1}
	Using the methods in Section \ref{3.3}, we quantified differences between the spectra of centrals in high- and low-M$_h$ halos. Our findings are plotted in Figure \ref{fig4b}. The M$_*$ of the compared galaxies increases from left to right, with the top row displaying results for an inner bin of galactocentric radius and the bottom row for an outer bin. Shaded contours show 1$\sigma$ errors, which fully account for uncertainties in sample assignment and stacking. Absorption features that are deeper in low-M$_h$ galaxies are negative whereas features that are stronger in high-M$_h$ galaxies are positive. We listed the stellar parameters that drive feature changes as quantified in the response functions by \citet{conroy2018}.
	
	At all M$_*$ and radii, high-M$_h$ centrals show stronger Mg$\textit{b}$ 5172\AA\ absorption (\citealt{faber1976,faber1985}). Apart from being dominated by magnesium abundance, this spectral feature is also sensitive to iron enrichment. Another significant set of spectral differences are found around 4000\AA, which tend to be stronger in high-M$_h$ centrals. These features are in a Ca-sensitive spectral region, as evidenced by the widely used CaHK, Ca4227, and Ca4455 spectral indices (e.g. \citealt{worthey1994,tripicco-bell1995}). Apart from being sensitive to the Calcium abundance, features at these wavelengths are also sensitive to stellar age and overall metallicity. 
	
	The detection of these differences demonstrates that halos have an impact on the stellar populations of passive central galaxies that is secondary to the correlation between M$_h$ and M$_*$. We now turn to our results from stellar population fitting to attempt to interpret these results.
	
	\subsection{Integrated measurements}
	\label{4.2}
	To derive the mean stellar population properties within the galaxy, we averaged the radial profiles derived with \texttt{alf} (Section \ref{3.5}). The results are shown with M$_*$ and M$_h$ as the controlling variable in the top and bottom panels of Figure \ref{fig9}, respectively. Measurements that used the Tinker catalog are plotted in circles, whereas those that used the Yang+Wang catalog are shown in the insets as triangles.
	
	To first order, it is well known that stellar age and metallicity increase with the M$_*$ or central velocity dispersion of galaxies (\citealt{faber-jackson1976,cidfernandes2005,gallazzi2005,thomas2005,thomas2010,gonzalez-delgado2014,mcdermid2015}). While the top row of Figure \ref{fig9} confirms this trend with age for the passive central galaxies in our study, we see that [Fe/H] decreases with M$_*$. In the literature, the word metallicity refers to a weighted average of the abundance of various elements (i.e., a rescaling of the solar abundance pattern), whereas the [Fe/H] measurements that we derived with \texttt{alf} map the abundance of iron only. The decrease in [Fe/H] with M$_*$ for M$_*>10^{11}$M$_{\odot}$ is likely the result of how the formation timescales of galaxies become more rapid as M$_*$ increases.
	
	The plot in the top row, second column compares the stellar ages of centrals in low- and high-M$_h$ at fixed M$_*$. With the Tinker catalog, we find low-M$_h$ centrals to be younger than high-M$_h$ counterparts by $\sim 1-2$ Gyr. Differences between the [Fe/H] of low- and high-M$_h$ centrals are also present, with high-M$_h$ centrals showing lower [Fe/H] by $\lesssim 0.05$ dex at all M$_*$. High-M$_h$ centrals also show slightly greater [Mg/Fe], especially for M$_*>10^{11}$M$_{\odot}$. Figure \ref{fig9} shows that the deeper absorption seen in high M$_h$ centrals (Figure \ref{fig4b}) is primary a result of their older ages and higher magnesium-enhancement. 
	
	We can assign confidence levels to these results by computing the Bayes factor. As an example, we can consider the following two hypotheses: either high-M$_h$ centrals are older or low-M$_h$ centrals are older. The Bayes factor is just the ratio between the marginalized likelihoods of the models since we are interested in adopting flat, uninformative priors. In equation form, the Bayes factor is equal to
	
	\begin{gather}
	\mathcal{B} = \prod_{\scriptsize \mbox{M}_*}{B_i}= \prod_{\scriptsize \mbox{M}_*} \frac{\mbox{I\kern-0.15em P}(\mbox{age}_{\mbox{\scriptsize high-M}_h} - \mbox{age}_{\mbox{\scriptsize low-M}_h}>0)}{ \mbox{I\kern-0.15em P}(\mbox{age}_{\mbox{\scriptsize high-M}_h} - \mbox{age}_{\mbox{\scriptsize low-M}_h}<0)}.
	\end{gather}
	
	Figure \ref{fig9} indicates that for the first M$_*$ bin, the stellar ages of the two subsamples are quite similar, and therefore $\mathcal{B}_1\sim 1$. On the other hand, high-M$_h$ are significantly older in the highest M$_*$ bin, and hence the associated Bayes factor is $\mathcal{B}_4\sim 100$. If we assume that the two hypotheses cover all possible model options (i.e. that at least one of the subsamples is older), we can impose that the sum of the model probabilities is equal to one (e.g. \citealt{oyarzun2017}). With this method, we conclude that high M$_h$ are older, have lower [Fe/H], and feature higher [Mg/Fe] than low-M$_h$ centrals with $3.6\sigma$, $3.6\sigma$, and $3.4\sigma$ confidence levels, respectively.
	
	We can now invert this analysis and use M$_h$ as the controlling variable. The results are shown in the bottom row of Figure \ref{fig9}, and indicate clear secondary behavior in the stellar population of galaxies with different values of M$_*$ at fixed M$_h$. We see that high-M$_*$ centrals are nearly always older. For M$_h>10^{12}h^{-1}$M$_{\odot}$, this difference can exceed 2 Gyr, has a significance of 4.4$\sigma$, and is clear in both halo catalogs. Differences in [Fe/H] show more complicated behavior that is dependent on M$_h$. For M$_h<10^{12}h^{-1}$M$_{\odot}$, high-M$_*$ centrals show higher [Fe/H] by as much as $0.2$ dex (8$\sigma$). For M$_h>10^{12}h^{-1}$M$_{\odot}$, the difference reverses and [Fe/H] decreases with M$_h$ (5$\sigma$). Significant differences are also observed in [Mg/Fe], with high-M$_*$ centrals showing greater [Mg/Fe] at all M$_h$ (5.6$\sigma$) and for M$_h>10^{12}h^{-1}$M$_{\odot}$ (4.4$\sigma$) with both catalogs.
	
	To help visualize the results across this multi-dimensional space, Figure \ref{fig9a} shows the SHMR colored by each of the stellar population properties we considered. This figure was made by fitting a two-dimensional, second-degree polynomial in M$_*$-M$_h$ space to the results from Figure \ref{fig9}. Note how stellar age and [Mg/Fe] not only vary with M$_*$, but also with M$_h$. On the other hand, [Fe/H] depends almost exclusively on M$_*$.
	
	%In summary, M$_*$ emerges as the variable that captures how the the stars in passive galaxies assemble. Minor secondary correlations with M$_h$ at fixed M$_*$ are not driven by residual correlations with M$_*$, as our approach to sample assignment is designed to define subsamples with the same M$_*$ distributions (Section \ref{2.1}). On the other hand, differences at fixed M$_h$ are of greater magnitude. High-M$_*$ centrals have older stellar populations, lower [Fe/H], and greater [Mg/Fe] than low-M$_*$ counterparts at all M$_h>10^{12}$M$_{\odot}$. Taken together, these results indicate that high-M$_*$ centrals formed earlier and in shorter timescales. Further interpretation of these findings is presented in Section \ref{5}.
	
	%We should note that these parameter variations are driven by trends we observe in the data. The distinct spectral differences between our subsamples are reproduced by our best fit models (see Figure \ref{fig4b}). This is the case at all radial scales, as we will show in our analysis of the stellar population profiles.
	
	\subsection{The stellar population profiles of central galaxies}
	
\begin{figure*}[htp]
		\centering
		\includegraphics[width=7in]{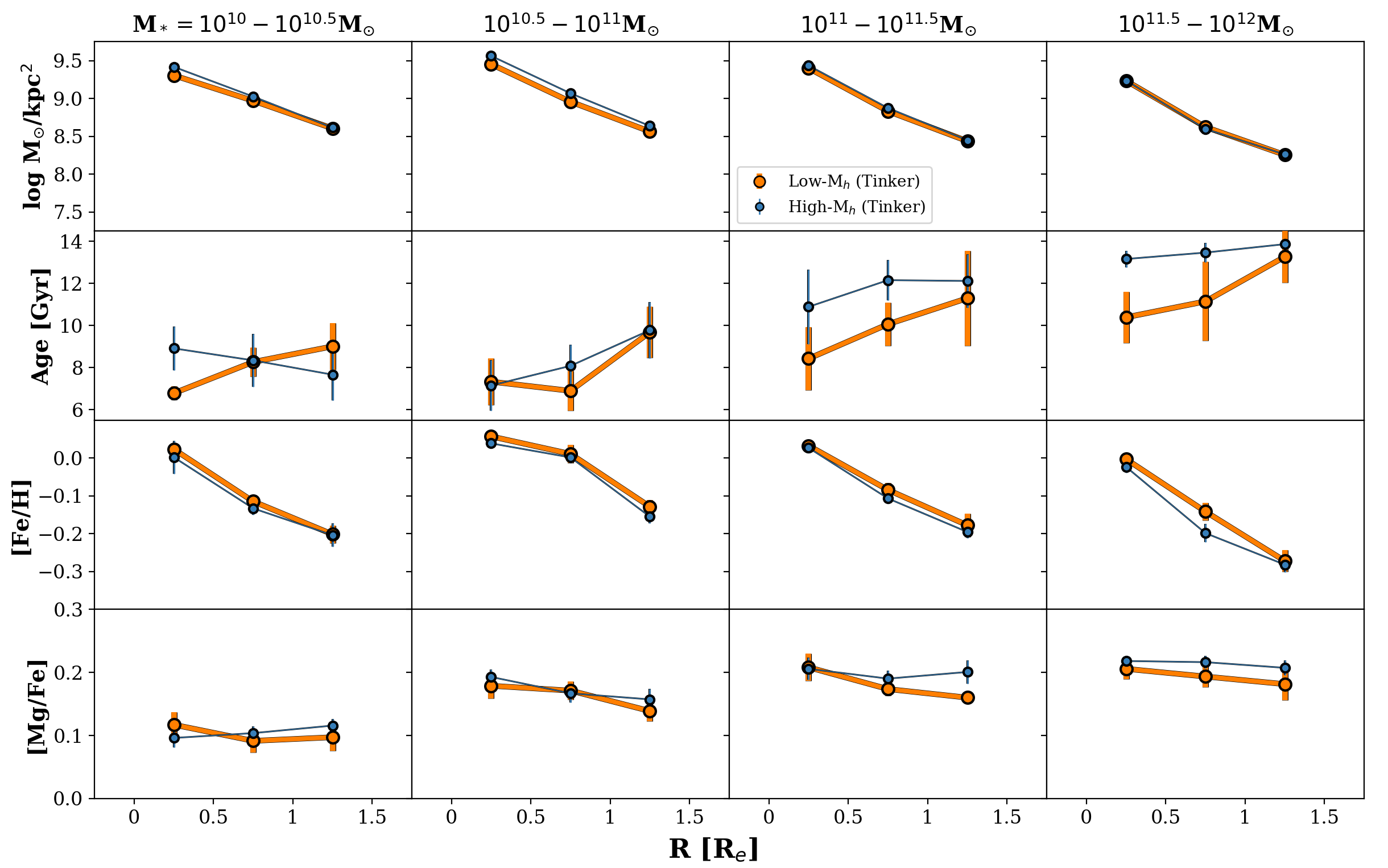}
		\caption{Stellar population profiles of low-M$_h$ (orange) and high-M$_h$ (blue) centrals at fixed M$_*$ with the Tinker catalog. From top to bottom, shown are stellar mass surface density, stellar age, [Fe/H], and [Mg/Fe]. Stellar mass increases from left to right. Low-M$_h$ centrals have younger stellar populations at all radii for M$_*>10^{11}$M$_{\odot}$ and higher [Fe/H] at all radii and all M$_*$ than high-M$_h$ centrals.}
		\label{fig5}
	\end{figure*}

\begin{figure*}[htp]
		\centering
		\includegraphics[width=7in]{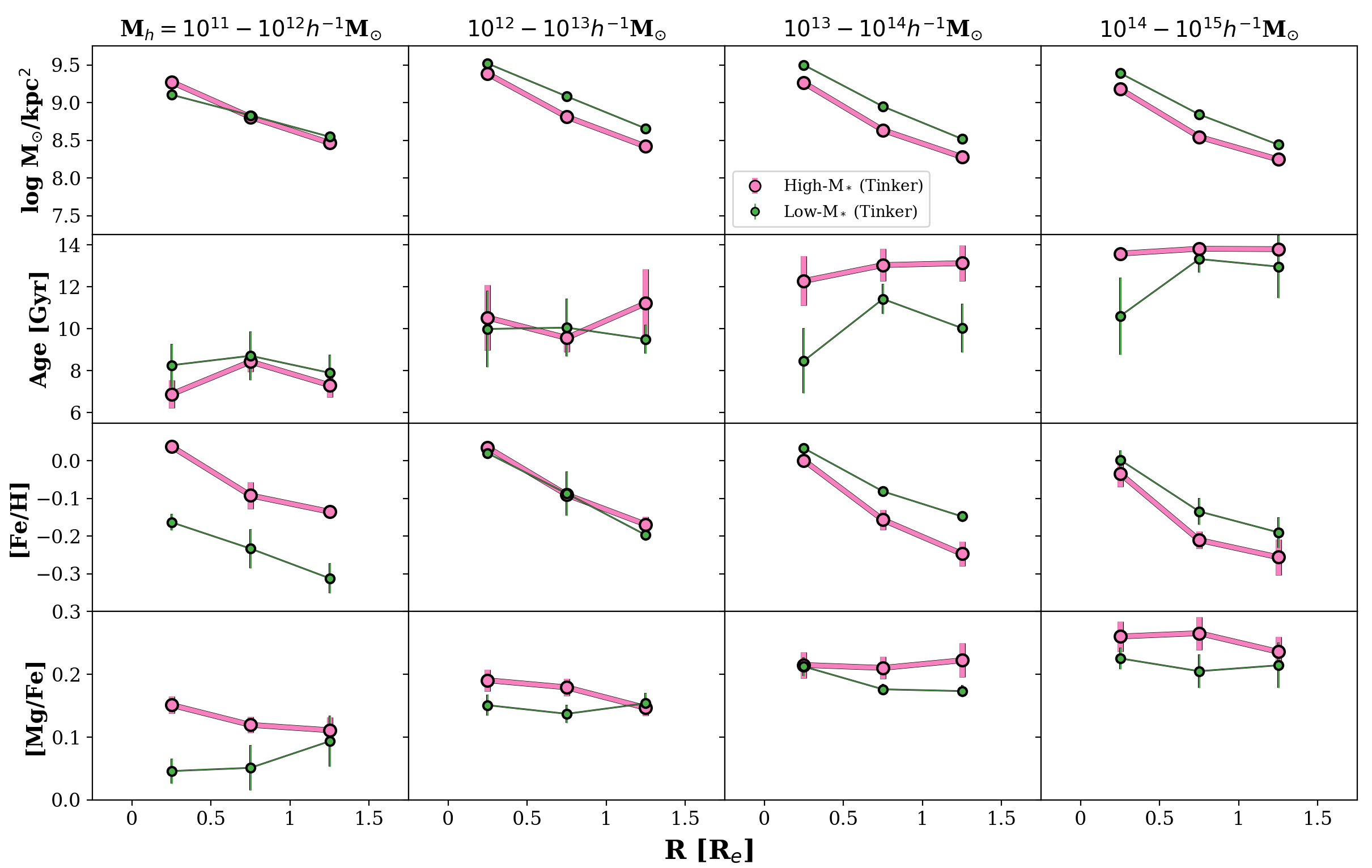}
		\caption{Stellar population profiles of high-M$_*$ (red) and low-M$_*$ (green) centrals at fixed M$_h$ with the Tinker catalog. From top to bottom, shown are stellar mass surface density, stellar age, [Fe/H], and [Mg/Fe]. Stellar mass increases from left to right. Despite their lower $\Sigma_*$, high-M$_*$ are more massive due to their much larger $R_e$. For M$_h>10^{12}h^{-1}$M$_{\odot}$, high-M$_*$ centrals have stellar populations that are older, have lower [Fe/H], and show greater [Mg/Fe] than low-M$_*$ centrals at most radii.}
		\label{fig6}
	\end{figure*}

	We now look at the radial dependence of our derived measurements. The stellar population profiles at fixed M$_*$ are shown in Figure \ref{fig5}, with M$_*$ increasing from left to right. High-M$_h$ profiles are shown in blue and low-M$_h$ profiles in orange. The comparison at fixed M$_h$ is shown in Figure \ref{fig6}, with high-M$_*$ centrals in magenta and low-M$_*$ centrals in green. The findings reported in this section apply to both catalogs, albeit only measurements using the Tinker catalog are shown to keep the figures simple.
	
	%The $\Sigma_*$ profiles are compared in the top row. $\Sigma_*$ decreases as M$_*$, indicating that the stellar content in massive galaxies is more diffusely distributed (\citealt{kormendy1977}). Despite their lower $\Sigma_*$, these galaxies have more M$_*$ as a result of their larger $R_e$ (\citealt{daddi2005,vanderwel2014}). At fixed M$_*$, the profiles of high- and low-M$_h$ centrals are very similar in normalization and shape. The similarity in shape is expected, since differences with M$_h$ in the $\Sigma_*$ profiles of centrals only become apparent beyond 100 kpc (\citealt{huang2020}). 
	
	In general, the stellar population profiles confirm the differences in normalization that we reported in Figure \ref{fig9} and Section \ref{4.2}. We find stellar population differences to be rather constant with radius, indicating that no significant differences in profile shape are apparent between the subsamples. The stellar age and [Mg/Fe] profiles all tend to be flat, while the [Fe/H] profiles all fall with radius, as reported in previous work (e.g. \citealt{greene2015,vandokkum2017,alton2018,parikh2018,parikh2019,parikh2021,zheng2019,lacerna2020}).
	
	The fact that parameter differences are mostly constant with radius indicates that the results in the integrated properties are not driven by outlier radial bins. This could  have been more problematic in the outskirts, where the spectral $S/N$ is a factor of two lower than at the centers (see Appendix \ref{appendix1}). The profiles would also reveal if any trends in the integrated measurements are driven by standout physical behavior. For example, recent central star formation could also lower the mass-weighted stellar ages at the center, but we see no evidence for this either. 
	
	We therefore conclude that any variations in the shape of the stellar population profiles within 1.5$R_e$ must be very subtle. Processes like radial migration could contribute to ``wash-out" any subtle differences that might have been imparted by past events in the assembly history (\citealt{minchev2012,el-badry2016}). Moreover, the galactocentric distances probed in this paper are just inward of the radii at which the stellar metallicity profiles of nearby galaxies start to show flattening due to minor mergers and stellar accretion from satellite galaxies (\citealt{oyarzun2019}).

	\section{Discussion}
	\label{5}
	
	% \textbf{WAKE: Could your results simply be that the current stellar mass of these passive galaxies is a better indicator of the conditions of the halos that they occupied when they formed the bulk of their stars 6-11 Gyrs ago than whatever the group catalogs are measuring now. That could be to do with the overall halo mass growth varying a lot from variable accretion since the galaxy formed, but that growth having much less of an effect on the center of the halo. Or it could just be that the halo mass in these group catalogs is so much more badly measured than the stellar mass. Or a combination. Do you have a sense of the relative contribution to the scatter in the SHMR from the error in the group catalog halo mass compared to the intrinsic scatter?}
	
	% \subsection{How the stellar populations of passive centrals depend on Halo Mass at fixed Stellar Mass}
	\subsection{How halo mass modulates the stellar populations of passive centrals at fixed M$_*$}
	
	\label{5.1}
	\label{fix_Mstar}
	
	In the two phase scenario for galaxy evolution, the \textit{in-situ} formation of central galaxies is followed by a phase of \textit{ex-situ} growth through stellar accretion (\citealt{oser2010,oser2012,johansson2012,moster2013,furlong2017}). Recent observations of nearby massive galaxies provide support for this secondary phase. The stellar density profiles of massive galaxies at low redshift have faint, extended stellar envelopes that could have originated from late time stellar accretion (\citealt{huang2013a,huang2013b,huang2018}), as also identified locally in the Milky Way (e.g. \citealt{fernandez-trincado2019,fernandez-trincado2020}). Massive nearby galaxies also show flat stellar metallicity profiles beyond the $R_e$, as it would be expected if their outskirts assembled through minor mergers (\citealt{oyarzun2019}).
	
	In \citet{oyarzun2019}, we showed that this transition in the shape of the stellar metallicity profiles becomes prominent in galaxies with stellar mass greater than M$_*=10^{11}$M$_{\odot}$. Our estimates of the \textit{ex-situ} stellar mass fraction at $R_e$ are consistent with zero for M$_*<10^{11}$M$_{\odot}$ and unity for M$_*>10^{11}$M$_{\odot}$. %In this work, results from Figure \ref{fig9} also indicate a phase transition at M$_*=10^{11}$M$_{\odot}$, with \textcolor{red}{[Fe/H] peaking at these M$_*$.}
	A phase transition around M$_*=10^{11}$M$_{\odot}$ also emerges in theoretical predictions. The baryon-to-star conversion efficiency is believed to peak around M$_*=10^{10.5}$M$_{\odot}$ (e.g. \citealt{behroozi2013,girelli2020}), thus creating an inflection point in the SHMR (e.g. \citealt{moster2010,posti-fall2021}). As M$_*$ increases, the conversion efficiency decreases and mergers grow in importance (\citealt{rodriguez-gomez2016,rodriguez-puebla2017}).  To first order, stellar mass that would have formed in the central galaxy at lower masses becomes increasingly locked up in satellites and the ``intra-group'' medium of host halos at larger masses.  We can use these insights to inform our interpretation of Figures \ref{fig9} and \ref{fig5}.
	
	Focusing on lower mass galaxies with M$_* < 10^{11}$M$_{\odot}$, we see that centrals in low-M$_h$ halos have higher [Fe/H] within $1.5R_e$ than centrals with the same M$_*$ in larger halos. The same trend was recovered by \citet{greene2015} in the MASSIVE survey (\citealt{greene2013}) (though we should note that the opposite result was found by \citet{labarbera2014} and \citet{rosani2018}; more in Section \ref{sec:halo_errors}). This result can be interpreted two ways. First, it might imply that central galaxies in high-M$_h$ halos more efficiently retained their gas throughout their star formation episodes. This could have led to rapid star formation, therefore enhancing their [Mg/Fe], which is precisely what we observe in \ref{fig9}. This Mg-enhancement in high-M$_h$ centrals was also detected by \citet{scholz-diaz2022}, who also characterized the stellar populations of central galaxies, but with single-fiber SDSS spectroscopy.
	
	Alternatively, we know by definition that at fixed M$_*$, low-M$_h$ centrals also have higher M$_*$-to-M$_h$ ratios and lower numbers of satellites (see Figure \ref{fig3a}).  Perhaps our results are simply a sign of more extended central star formation histories at lower halo masses where lower virial temperatures and fewer satellites allow for longer periods of cold flow accretion (e.g. \citealt{dekel2006,zu2016}).
	
	Before turning to the interpretation of our results toward M$_*>10^{11}$M$_{\odot}$, we should emphasize that our sample binning and spectral $S/N$ requirements in this work prevent us from studying the flattening of metallicity profiles in the low surface brightness outskirts of massive galaxies that were the subject of \citet{oyarzun2019}.  The profiles in Figure \ref{fig5} are mostly limited to the inner regions of galaxies, within $\sim 1.5R_e$.  That said, we see in this work that M$_h$ has an impact on the formation of passive galaxies, even at the centers of M$_*>10^{11}$M$_{\odot}$ centrals. Though subtle, these signatures of halo-modulated evolution are significantly detected, as evidenced in Figures \ref{fig4} and \ref{fig4b}.
	
	For M$_*>10^{11}$M$_{\odot}$, the stars in high-M$_h$ halo centrals are older than their counterparts in low-M$_h$ halos. At these large M$_*$, high-M$_h$ centrals also have lower [Fe/H] and higher [Mg/Fe] at all radii. The fact that these differences are present at the centers and show little radial variation points to differences in \textit{in-situ} formation, as opposed to minor merger accretion at larger radii. In line with our interpretation at lower M$_*$, a possible explanation is that low-M$_h$ centrals continued forming stars over longer timescales.  A later onset of quenching would yield younger stellar ages, higher [Fe/H], and lower [Mg/Fe] as measured today. %It is interesting to speculate on whether the apparent lack of an age difference at lower masses (M$_*<10^{11}$M$_{\odot}$) provides additional physical insight or reflects challenges in modeling 30\% differences in mean age among 8-12 Gyr old populations.  We note that when we bin our samples using the Yang+Wang M$_h$ catalog, we see discrepant results among high-mass galaxies, making firm conclusions here difficult.
	% However, the absence of any differences in the [Mg/Fe] profiles are in tension with this scenario.
	
	Given the more active merger history of high-M$_*$ galaxies, it is worth exploring the possibility that a fraction of the stars at the centers of M$_*>10^{11}$M$_{\odot}$ centrals was accreted. To lower the [Fe/H] abundance and increase the stellar age in high-M$_h$ centrals, accreted stars would have to be metal poor and old. \citet{oyarzun2019} argued that signatures of minor mergers are produced by the accretion of stellar envelopes that are more metal poor than the inner regions of the central, giving credibility to this explanation. However, explaining why accreted stellar populations would be old is more challenging. The stellar ages of satellite galaxies rarely exceed 10 Gyr, even in halos as massive as M$_h>10^{14}h^{-1}$M$_{\odot}$ (\citealt{pasquali2010}). Yet, the stellar ages of high-M$_h$ centrals can be as old as 12 Gyr for M$_*=10^{11.5}$M$_{\odot}$. 
	
	% Recent observations have found that centrals in high-M$_h$ halos not only have more M$_*$ at the centers, but also in the outskirts (\citealt{huang2020}). 
	
	Recently, \citealt{huang2020} exploited deep, wide-field imaging to measure individual early-type galaxy surface brightness profiles to $R \sim$ 150 kpc for a sample large enough that precise M$_h$ estimates could be derived through galaxy-galaxy weak lensing.  They found that, as a function of their host halo M$_h$, central galaxies not only contain more M$_*$ in their centers, but in their distant outskirts as well.  If dry minor mergers are required to build those outskirts, does this mean that larger halos must be effective in suppressing star-formation in both their central galaxies \textit{and} the within-the-satellite population that will later merger onto those centrals?
	% Would this scenario enable more minor merging that then built the outskirts of the central galaxy? 
	Further exploration of this question requires a direct comparison between the stellar population profiles of central and satellite galaxies, a subject we will return to in future work. 
	
	% \subsection{Influence of systematic errors}	
	
	% Through stacking, we have shown that the spectra of big and \textcolor{red}{low-M$_h$} centrals at fixed M$_*$ are different. However, these signatures of halo-driven evolution are only at the 1-2\% level. Furthermore, we struggled to conclusively associate these signatures to stellar population variations, except for differences in [Fe/H] (see Figure \ref{fig5}). This is likely a combination of the difficulties faced by stellar population fitting, M$_h$ estimation, and signal-to-noise.
	
	% Limitations in sample size and signal-to-noise also affect our constraining power. Though the MaNGA dataset is now available in its totality to the collaboration, this work made use of an earlier data release that consists of 80\% of the whole sample. Future work taking advantage of the full dataset might be able to constrain the stellar population profiles more accurately and thus more conclusively determine the astrophysical mechanisms driving the differences in stellar population parameters with M$_h$ at fixed M$_*$. 
	
	% \subsection{Galaxy Assembly Bias and the Stellar-to-Halo Mass Relation of Centrals}
	
	\subsection{How M$_*$ drives evolution within dark matter halos of identical mass}
	
	\label{5.2}
	
	So far, we have discussed the influence of the host dark matter halo mass (M$_h$) as a secondary, modulating variable in the formation and evolution of passive central galaxies at fixed M$_*$.  This approach follows a long line of literature seeking to understand the role of ``environment'' after controlling for luminosity or stellar mass (e.g. \citealt{dressler1980,kauffmann2004,peng2010b}).
	
	Our theoretical understanding of galaxy formation, however, begins by assuming an underlying distribution of evolving dark matter halos and then seeks to build physical models on top (see \citealt{benson2000,moster2013,somerville2015}). Acknowledging our imperfect ability in this paper to measure dark matter halos observationally (see Section \ref{sec:halo_errors}), we nonetheless turn to a theory-minded perspective, in which halo mass is the primary variable.  By studying trends with M$_*$ in bins of fixed M$_h$, we gain insight into the range of evolution that occurs within halos of fixed mass today.
	
	% fixed M$_h$ for the rest of the discussion. First, we will discuss what mechanisms are responsible for the scatter of the SHMR in simulations. Then, we will use this information to interpret our findings and better understand the connection between central galaxies and their host halos.
	
	% Preferred cosmological models predict the formation of galaxies to be closely linked to the evolution of dark matter halos (e.g. \citealt{wechsler2018}). As a result, M$_h$ has been a key ingredient in galaxy formation models for over two decades. Early implementations that used M$_h$ as the main parameter were able to reproduce some of the observable properties of galaxies (e.g. \citealt{benson2000}). Though more recent models better reproduce observations, some still resort to abundance matching in M$_h$- M$_*$ space (e.g. \citealt{moster2013}).
	
	% \subsection{On the assembly of passive central galaxies}	
	% \label{5.3}
	
	Considering stellar age, [Fe/H], and [Mg/Fe] in nearly every mass bin, the bottom row of Figure \ref{fig6} shows that the stellar populations of central galaxies at fixed M$_h$ depend strongly on stellar mass.  On the one hand, this result seems to be a familiar expression of how stellar populations depend on the luminosity of early-type galaxies (see \citealt{renzini2006} for a review). But when we remember that our trends are seen at fixed M$_h$, they are perhaps more surprising.  In two halos of identical mass today, central galaxies with different M$_*$ have markedly different formation histories.
	
	We discuss two physical interpretations of this result before concluding the discussion with an examination of potential observational biases.
	
	\subsubsection{Varying conditions of early formation}
	
	Our first interpretation follows in the spirit of ``monolithic collapse,'' namely that the early conditions ($z \sim 4$) for gas accretion and mergers determine how the vast majority of stars in early type galaxies form.  Rapid, intense formation leads to a greater stellar mass content (\citealt{vandokkum2009,newman2010,damjanov2011,whitaker2012b,dekel-burkert2014,zolotov2015}). The number of stars formed in this early phase may be therefore largely independent of the progenitor halo properties, let alone the halo's final mass at $z=0$.  
	
	In this picture, certain halos would happen to host the conditions needed for gas-rich mergers which promote the early formation of massive central galaxies. These events would have to rapidly exhaust gas supplies to produce old and chemically enriched stellar populations today.  Unfortunately, testing this scenario with stellar population profiles is challenging because the imprint of gas-rich mergers is hard to predict and sensitive to the initial conditions of the encounter (\citealt{kobayashi2004}).
	
	We note, however, that at all M$_h$, high-M$_*$ centrals are larger in size and show higher $\Sigma_*$ [M$_{\odot}$/kpc] within $\sim$1 $R_e$ compared to their low-M$_*$ counterparts (see Figure \ref{fig6}).  Their initial ``collapse'' may have driven up such large central gas densities that the resulting deep potential wells limited the impact of feedback, driving runaway growth in M$_*$  (e.g. \citealt{matteucci1994,wellons2015b}).  Such a period of collapse might leave kinematic and morphological signatures. Gas-rich major mergers might preserve the spin of the merger remnant, for example, giving rise to compact disks at $z \sim 2$ (e.g. \citealt{vanderwel2011}) and so-called fast rotators (e.g. \citealt{graham2018}). On the other hand, a more extended formation period followed by dry mergers might produce slow rotator galaxies (\citealt{naab2014}). 
	% Going back to galaxy assembly bias: would halo formation time bias leave any imprints in the kinematics of central galaxies? What about signatures in their morphology? 
	
	In a sense, this general picture aligns with Section \ref{fix_Mstar} in that stellar mass, and not the dark matter halo, emerges as the dominant variable that controls how the bulk of stars in passive centrals form. This would seem antithetical to current theoretical models, however, and inconsistent with a variety of observational studies emphasizing the importance of M$_h$, or at least proxies thereof (e.g. $\Sigma_{\rm 1 kpc}$ and $\sigma_*$) in driving galaxy properties (Figures \ref{fig4}, \ref{fig4b} of this paper and e.g. \citealt{franx2008,wake2012,chen2020,estrada-carpenter2020}).  
	
	% .  The future growth of the halo can follow rather different assembly histories, resulting in a 

	% We should also consider alternative scenarios for our results. The assembly of central galaxies at late times ($z<2$) is thought to be dominated by the accretion of stellar envelopes from satellite galaxies (\citealt{oser2010,oser2012,johansson2012,moster2013,furlong2017}). These accretion processes have been shown to efficiently increase the sizes and M$_*$ of passive galaxies (e.g. \citealt{bezanson2009,hopkins2010,barro2013,cappellari2013b,wellons2015}) and leave imprints in their stellar population profiles (\citealt{cook2016,taylor2017,oyarzun2019}). Can the differences between the stellar populations of overmassive and undermassive centrals be driven by differences in their late-time assembly? 
	
	% The larger sizes, higher M$_*$, older ages, and greater chemical enrichment of overmassive centrals could be the consequence of the accretion of old, metal rich, compact systems known as red nuggets (\citealt{vandokkum2009,newman2010,damjanov2011,whitaker2012b}). Though plausible, there are two problems with this explanation. Minor merger signatures grow in intensity with M$_h$ and radius (\citealt{rodriguez-gomez2016,oyarzun2019}), but the stellar population differences between overmassive and undermassive centrals do not. If anything, the magnitude of the difference in iron abundance decreases with M$_h$ (third row in Figure \ref{fig6}). Also in contrast with this scenario, differences in [Mg/Fe] tend to disappear at large radii (fourth row in Figure \ref{fig6}).

	\subsubsection{Galaxy Assembly Bias}
	
	If a galaxy's growth history is ultimately tied to its dark matter halo, then Figure \ref{fig:integrated} tells us that other halo parameters, beyond M$_h$, must play a role. These might include halo formation time and halo clustering (e.g. \citealt{gao2005,wechsler2006}), secondary properties which invoke the idea of ``halo assembly bias.'' By definition, this term encapsulates all correlations between halo clustering and the assembly histories of halos at fixed M$_h$ (\citealt{mao2017,mansfield2020}). For example, dark matter halos that assembled early tend to be more strongly clustered than counterparts that assembled at lower redshift (\citealt{gao2005,contreras2019}), especially for M$_h<10^{13}h^{-1}$M$_{\odot}$ (\citealt{li2008}). 
	
	The strong statistical relationship between halos and the galaxies they host (e.g. \citealt{leauthaud2017}) has motivated predictions for how halo assembly bias might impact galaxy formation.  \citet{croton2007} used semi-analytic models to predict that halo formation time should correlate with galaxy luminosity at fixed M$_h$. In a halo mass bin of M$_h=10^{14}-10^{15.5}h^{-1}$M$_{\odot}$, they found that bright centrals formed in those halos that had mostly assembled by $z\sim1$, whereas fainter centrals were associated with halos that assembled later ($z\sim 0.5$).  \citet{zehavi2018} expanded on this result, predicting a dependence of central M$_*$ on halo formation time at fixed M$_h$.   
	
	An implied secondary dependence of stellar mass growth on halo age has also been studied with semi-empirical models.  \citealt{bradshaw2020}, for example, report factor of $\sim$3 differences in ``central'' M$_*$ between the 20\% youngest and oldest halos at fixed halo mass above M$_h>10^{13}$M$_{\odot}$ as modeled in the UniverseMachine (UM; \citealt{behroozi2019}).  This difference in M$_*$ and age (of approximately 1--2 Gyr) is similar to our results in the two bottom-left panels of Figure \ref{fig:integrated}.
	% Halo age can account for $\sim$0.25 dex of the scatter in the SHMR for M$_h>10^{13}$M$_{\odot}$.
	
	% Of great importance is also the intrinsic scatter of the SHMR, since it contains further clues about how galaxies and halos co-evolve (\citealt{xu2020}).
	
	A galaxy assembly bias signal is also predicted in cosmological hydrodynamic simulations like EAGLE (\citealt{crain2015,schaye2015,matthee2017,kulier2019}) and Illustris (\citealt{vogelsberger2014a,nelson2015,xu2020}), where it is driven by the fact that halos in dense regions not only collapse earlier but do so with higher halo concentrations (e.g. \citealt{wechsler2002,zhao2009,correa2015,hearin2016}).  That yields deeper potential wells (\citealt{matthee2017}) and more efficient central star formation (e.g. \citealt{booth-schaye2010}).  For example,  \citet{matthee2017} found an order-of-magnitude variation in central M$_*$ at fixed halo mass below M$_h\lesssim10^{12}$M$_{\odot}$.  This variation correlates strongly with halo assembly time, which spans $z= $0.6--3 (see their Figure 7).
	
	In IllustrisTNG, galaxy assembly bias is strongly imprinted in galaxy observables. At fixed M$_h$, central galaxies that are dispersion dominated, redder, larger, and have higher M$_*$ are more strongly clustered, especially for M$_h<10^{13}h^{-1}$M$_{\odot}$ (\citealt{montero-dorta2020}). In addition, centrals in IllustrisTNG with high M$_*$-to-M$_h$ ratios exhausted their gas reservoirs earlier, thus quenching at higher redshifts (\citealt{montero-dorta2021}).
	
	It is tempting to map these predictions to our observational results. At fixed M$_h$, high-M$_*$ centrals have stellar populations that are older, have higher [Mg/Fe], and show lower [Fe/H], revealing that they formed earlier and more rapidly. In the galaxy assembly bias scenario, this is a consequence of galaxy formation in older, highly concentrated halos. This explanation was also proposed by \citet{scholz-diaz2022} to explain why the same trends for stellar age and [Mg/Fe] are present in single-fiber spectra from SDSS.
	
	Before we consider observational biases on this conclusion (Section \ref{sec:halo_errors}), it is worth noting that there is disagreement among theoretical predictions for the detailed behavior of galaxy assembly bias. In the UM analysis, the secondary correlation with halo formation time is strongest for M$_h\gtrsim10^{14}$M$_{\odot}$ (\citealt{bradshaw2020}; see also \citealt{croton2007}). But \citet{matthee2017} find that halo formation time in EAGLE has no effect above M$_h\gtrsim 10^{12}$M$_{\odot}$, instead strengthening at lower masses (see also \citealt{kulier2019}). Meanwhile, \citet{zehavi2018} and \citet{xu2020} find a correlation between halo age and M$_*$ that peaks around M$_h\sim10^{12}h^{-1}$M$_{\odot}$ and mildly decreases in significance toward M$_h\gtrsim10^{13.5}h^{-1}$M$_{\odot}$. Uncertainties in late-time growth explain some of the discrepancy. In hydrodynamical simulations, the stochasticity of late mergers can wash out formation-time bias at high-M$_h$ (\citealt{matthee2017}). The opposite happens in the UM because the bias here is driven by accreted stellar populations (\citealt{bradshaw2020}). Halo age primarily influences the stellar mass that was accreted through mergers rather than the populations formed \textit{in-situ}. 
	
	Finally, we note that the more rapid formation of high-M$_*$ centrals, as inferred from the [Mg/Fe] and [Fe/H] measurements, is in agreement with the analysis by \citet{montero-dorta2021} in IllustrisTNG. However, it is in contrast with how high-M$_*$ centrals assemble in EAGLE, where they grow by forming stars over long timescales (\citealt{kulier2019}). As \citet{contreras2021} concluded in their comparison of different galaxy assembly bias models, the amplitude and behavior of the signal is strongly model-dependent.

	\subsubsection{Systematic errors and assembly bias signal}
	\label{sec:halo_errors}
	
	% Observations of assembly bias
	Many studies have searched for signatures of galaxy assembly bias and tentative detections have been reported, including 
	% , the correlation between properties that show halo assembly bias and galaxy properties at fixed M$_h$ (e.g. \citealt{zentner2014,wechsler2018,xu2020}). Some tentative detections of galaxy assembly bias include 
	correlations between halo concentration and the occupation fraction of centrals (\citealt{zentner2019,lehmann2017}), spatial clustering of galaxies (e.g. \citealt{berlind2006,lacerna2014,montero-dorta2017}), and galactic ``conformity'' (i.e., similarity in the physical properties of galaxies within a halo; \citealt{weinmann2006,hearin2016b,berti2017,calderon2018,mansfield2020}).  Unfortunately, systematic errors in the distinction between centrals and satellites, halo mass estimates, and other problems have called into question many of these results (see \citealt{campbell2015,lin2016,tinker2018,wechsler2018}).
	
	Much of the concern in our work centers on the group catalogs we use to infer the host dark matter halo properties.  Indeed, we show explicitly that the choice of group catalog can affect our results.  For example, the Yang+Wang and Tinker catalogs disagree as to whether low-M$_h$ centrals have higher or lower [Fe/H] than their high-M$_h$ counterparts (see Figure \ref{fig9}). These discrepancies are also present in other work. \citet{labarbera2014} and \citet{rosani2018} used the \citet{yang2007} catalog to find that high-M$_h$ centrals have younger and more iron enriched populations than low-M$_h$ centrals. The opposite was found by \citet{greene2015} with the catalog by \citet{crook2007}.  
	
	While we favor the new group catalogs from \citet{tinker2020a,tinker2020b}, which are more sophisticated, self-consistent, and robust thanks to deeper imaging, systematic errors in M$_h$ may significantly impact our conclusions. This is in large part because the M$_h$ estimates \textit{depend}, at least in part, on the M$_*$ of the central galaxy.  In addition to counting satellite luminosity, the Tinker catalog also employs color and $r$-band light concentration when assigning M$_h$ to an associated central galaxy.  Our results indicate that high-M$_*$ centrals have higher central $\Sigma_*$, the opposite of the expected bias with concentration.  But they do have fewer satellites (apparent in Figure \ref{fig3a}).  This is expected because the central and satellite M$_*$ values must roughly add  to a constant at fixed M$_h$.  It is possible that the history of satellite accretion drives the stellar population trends we see, independently of the halo assembly history.
	
	Finally, the identified sample of ``central'' galaxies (which is roughly identical in the Tinker and Yang+Wang catalogs) is both incomplete and contaminated (by actual satellite galaxies) in ways that likely depend on M$_*$ and inferred M$_h$.  Disentangling these effects requires substantial mock observations that are beyond the scope of this paper.
	
	Instead, in future work we can make progress by measuring observables associated with our subsamples that are independent of galaxy luminosity (i.e., independent of M$_*$).  For example, the asssembly bias interpretation would be strengthened by a detection of different large-scale density signals on 10 Mpc scales for high- versus low-M$_*$ galaxies at fixed M$_h$.  Likewise, we can exploit new, more accurate proxies for M$_h$ \citep[e.g.][]{bradshaw2020}, eventually aiming to derive M$_h$ estimates from stacked weak lensing.
	
	% since it improves on some of the limitations faced by older group finding algorithms. By implementing color dependent parameters in the calibration process and taking advantage of deep DLIS photometry, they better account for how M$_h$ depends on galaxy color and total satellite luminosity.
	
	% We can summarize our conclusions as follows. Overmassive centrals are larger, older, and have higher [Mg/Fe] and greater [Fe/H] than undermassive counterparts of the same M$_h$. Our preferred explanation for these results is that overmassive centrals formed in deeper potential wells, where gas accretion is triggered at high redshift and the impact of feedback is limited. This fits very well with predictions of galaxy assembly bias arguing that overmassive centrals assemble in early forming, highly concentrated halos. An alternative explanation for these results would be that overmassive galaxies grew in M$_*$ and size through the accretion of old, metal rich systems.
	
	\section{Summary}
	\label{6}
	
	We constructed a sample of over 2200 passive central galaxies from the MaNGA survey to study how their assembly histories depend on M$_*$ and M$_h$. We constrained the stellar populations of our galaxies to high precision through spectral stacking and characterization with the codes \texttt{Prospector} and \texttt{alf}. We also control for systematics in M$_h$ estimation by comparing the outputs from the group catalogs by \citet{yang2007,wang2016} and \citet{tinker2020a,tinker2020b}. Our findings can be summarized as follows:
	
	1. At fixed M$_*$, there are significant differences in the spectra of passive centrals as a function of M$_h$. These differences are present at all M$_*$ and we detect them at all radii ($R<1.5R_e$). With no modeling involved, this shows that host halos have an impact on the formation of central galaxies. 
	
	2. To associate these spectral differences to stellar population variations, we turned to fitting with \texttt{Prospector} and \texttt{alf}. At fixed M$_*$, centrals in more massive halos show older stellar ages, lower [Fe/H], and higher [Mg/Fe]. A possible explanation for this result is that centrals in massive halos more efficiently retained their gas, allowing for early and rapid formation. Alternatively, the fewer satellites in low-M$_h$ centrals could allow for longer periods of cold flow accretion onto the central galaxy, resulting in more extended star formation histories.
	
	3. At fixed M$_h$, centrals with high-M$_*$ have older stellar populations and formed in shorter timescales (low [Fe/H] and high [Mg/Fe]) than centrals with low-M$_*$. At first glance, this result might be expected given how the stellar populations of early-type galaxies depend on M$_*$. However, our results are among the first to distinguish these evolutionary trends at fixed M$_h$. We propose two different scenarios to explain these results:
	
	\textit{Varying conditions of early formation:} at fixed M$_h$, centrals that undergo gas-rich mergers can fuel rapid, intense star-formation episodes followed by runaway growth in M$_*$. This process of ``enhanced collapse" leads to the formation of old, $\alpha$-enhanced stellar populations.
	
	\textit{Galaxy assembly bias:} according to theory, central galaxies with high M$_*$-to-M$_h$ ratios assembled in early-forming, highly concentrated halos. Their gravitational potentials lead to early star-formation, efficient metal retention, and rapid exhaustion of their gas reservoirs.
	
	4. Though we use two group catalogs in our analysis, we are still sensitive to sample contamination and systematic errors in M$_h$ estimation. Future work can improve on these difficulties by measuring observables that are independent of galaxy luminosity or by exploiting more accurate proxies for M$_h$.
	
	\medskip
	
	%\begin{acknowledgments}
		We would like to thank the Referee and Statistics Editor for the helpful comments and suggestions. We would also like to thank everyone at UC Santa Cruz involved in the installation and maintenance of the supercomputer Graymalkin that was used to run \texttt{alf} on MaNGA spectra. This work made use of GNU Parallel (\citealt{tange2018}). GO acknowledges support from the Regents' Fellowship from the University of California, Santa Cruz. KB was supported by the UC-MEXUS-CONACYT Grant. J.G.F-T gratefully acknowledges the grant support provided by Proyecto Fondecyt Iniciaci\'on No. 11220340, and also from ANID Concurso de Fomento a la Vinculaci\'on Internacional para Instituciones de Investigaci\'on Regionales (Modalidad corta duraci\'on) Proyecto No. FOVI210020, and from the Joint Committee ESO-Government of Chile 2021 (ORP 023/2021). MAF acknowledges support by the FONDECYT iniciaci\'on project 11200107. This research made use of Marvin, a core Python package and web framework for MaNGA data, developed by Brian Cherinka, Jos\'e S\'anchez-Gallego, and Brett Andrews (MaNGA Collaboration, 2018). Funding for the Sloan Digital Sky Survey IV has been provided by the Alfred P. Sloan Foundation, the U.S. Department of Energy Office of Science, and the Participating Institutions. SDSS acknowledges support and resources from the Center for High-Performance Computing at the University of Utah. The SDSS web site is www.sdss.org. SDSS is managed by the Astrophysical Research Consortium for the Participating Institutions of the SDSS Collaboration including the Brazilian Participation Group, the Carnegie Institution for Science, Carnegie Mellon University, the Chilean Participation Group, the French Participation Group, Harvard-Smithsonian Center for Astrophysics, Instituto de Astrof\'isica de Canarias, The Johns Hopkins University, Kavli Institute for the Physics and Mathematics of the Universe (IPMU) / University of Tokyo, Lawrence Berkeley National Laboratory, Leibniz Institut f\"ur Astrophysik Potsdam (AIP), Max-Planck-Institut f\"ur Astronomie (MPIA Heidelberg), Max-Planck-Institut f\"ur Astrophysik (MPA Garching), Max-Planck-Institut f\"ur Extraterrestrische Physik (MPE), National Astronomical Observatories of China, New Mexico State University, New York University, University of Notre Dame, Observat\'orio Nacional / MCTI, The Ohio State University, Pennsylvania State University, Shanghai Astronomical Observatory, United Kingdom Participation Group, Universidad Nacional Aut\'onoma de M\'exico, University of Arizona, University of Colorado Boulder, University of Oxford, University of Portsmouth, University of Utah, University of Virginia, University of Washington, University of Wisconsin, Vanderbilt University, and Yale University.
	%\end{acknowledgments}
	
	%\newpage
	
	%\medskip
	
	\appendix
	\section{Model fits and posterior distributions}
	\label{appendix1}
\begin{figure*}[htp]
		\centering
		\gridline{\fig{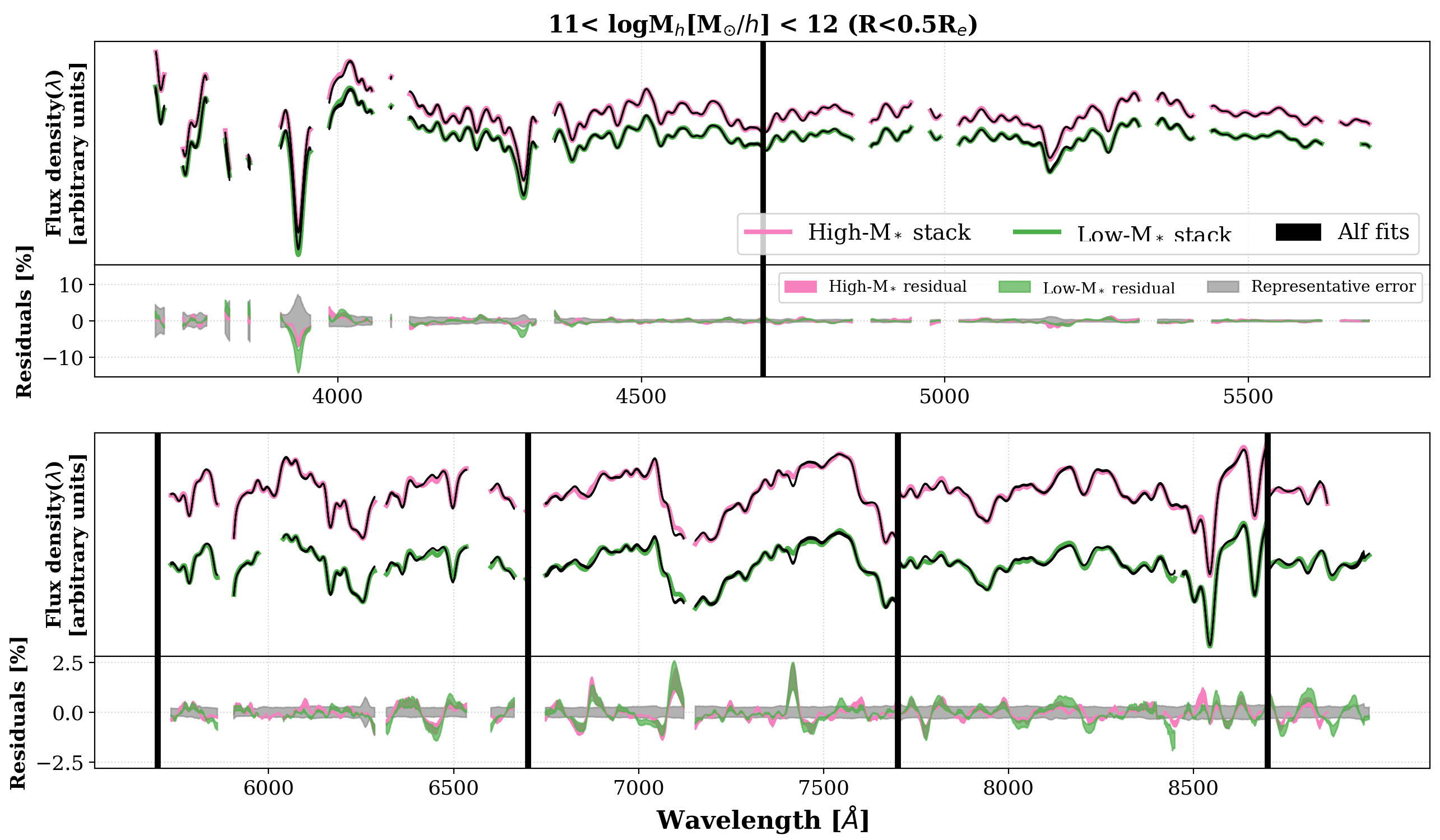}{\textwidth}{}}
		\gridline{\fig{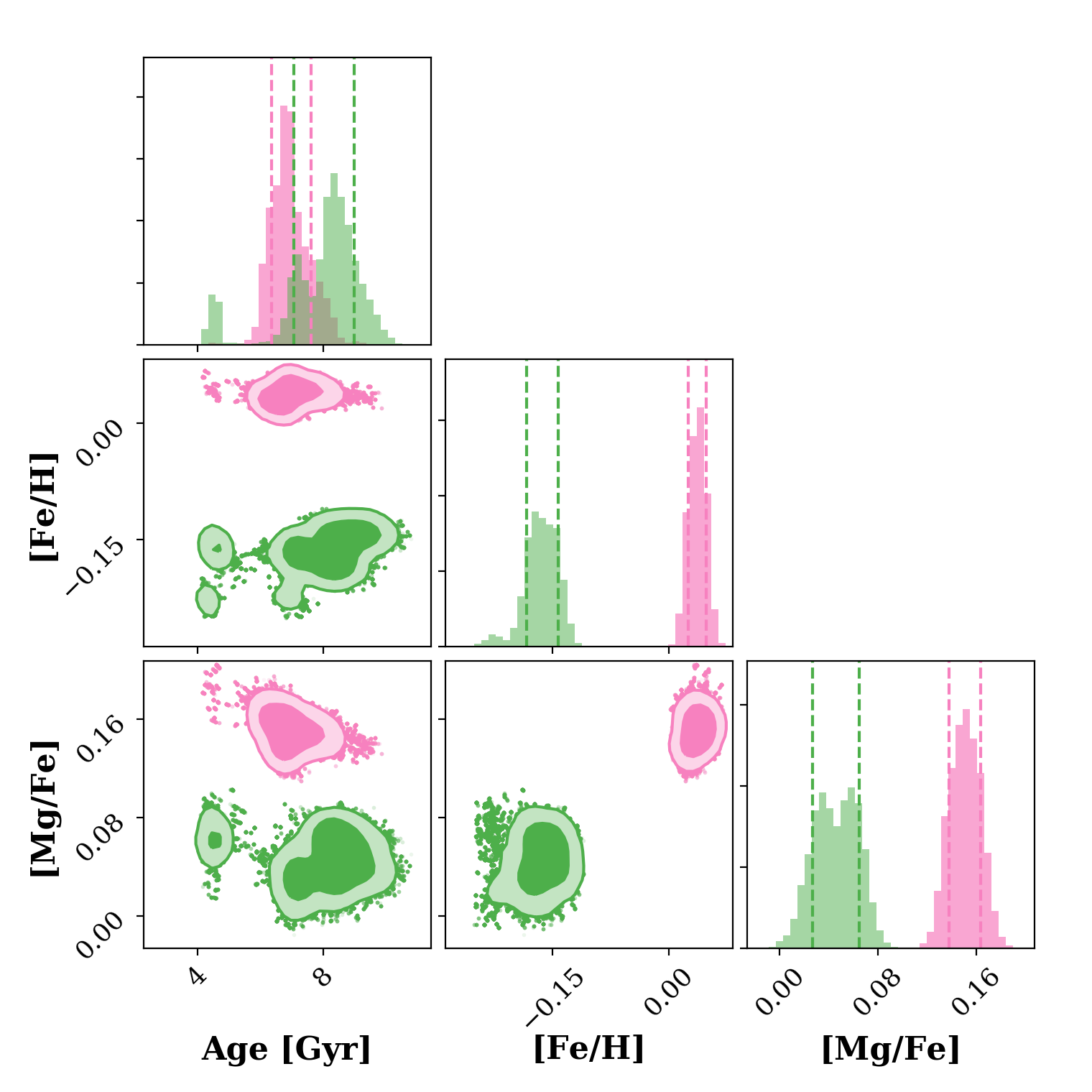}{.6\textwidth}{}}
		\caption{\textbf{Top:} Best fits and residuals for the $11<$logM$_h$[M$_{\odot}/h$]$<12$ bin ($R<0.5R_e$). \textbf{Bottom:} Posterior distributions. Details in Appendix {\ref{appendix1}}.}
		\label{figa1}
	\end{figure*}

\begin{figure*}[htp]
		\centering
		\gridline{\fig{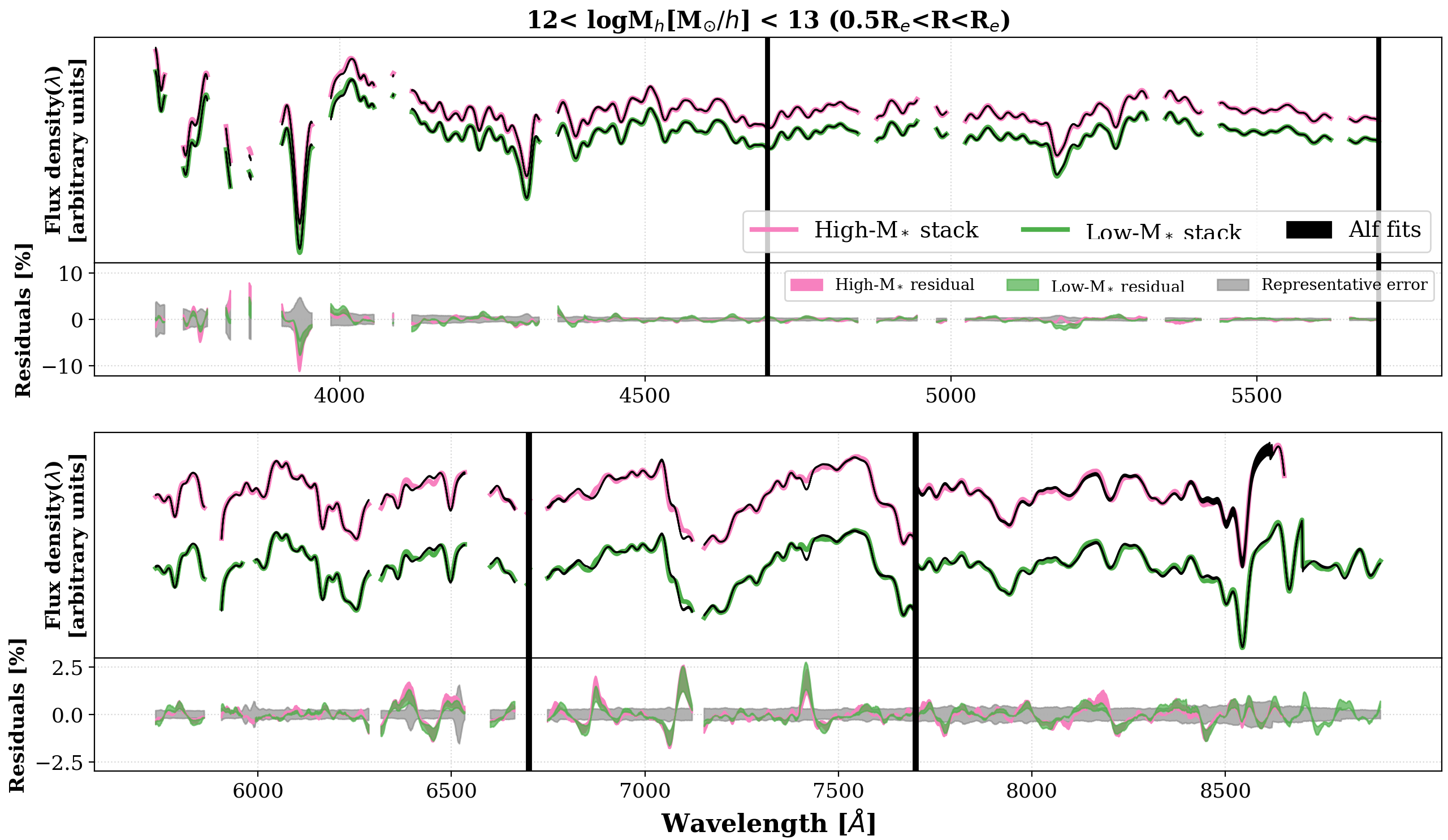}{\textwidth}{}}
		\gridline{\fig{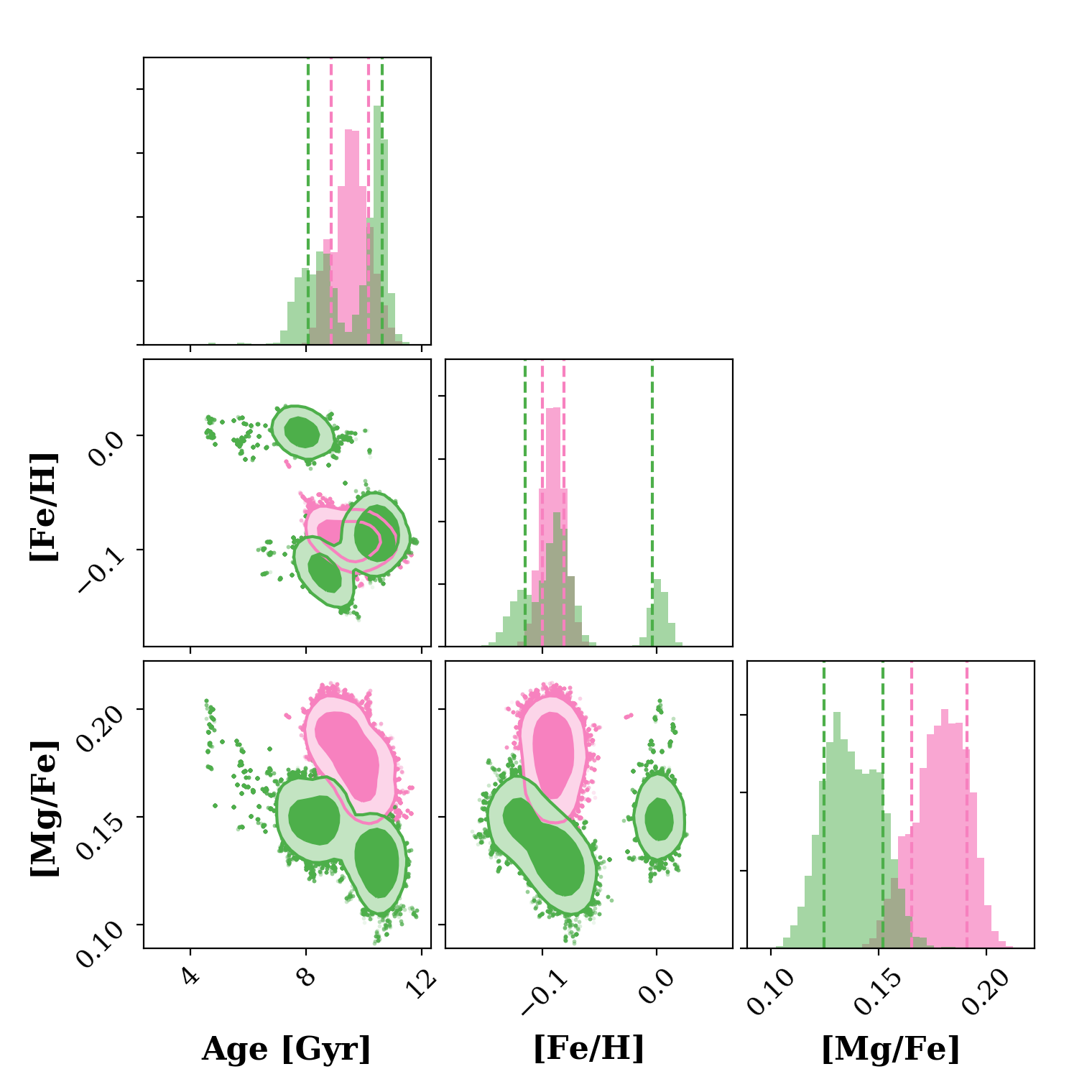}{.6\textwidth}{}}
		\caption{\textbf{Top:} Best fits and residuals for the $12<$logM$_h$[M$_{\odot}/h$]$<13$ bin (0.5$R_e<R<R_e$). \textbf{Bottom:} Posterior distributions. Details in Appendix {\ref{appendix1}}.}
		\label{figa2}
	\end{figure*}

\begin{figure*}[htp]
		\centering
		\gridline{\fig{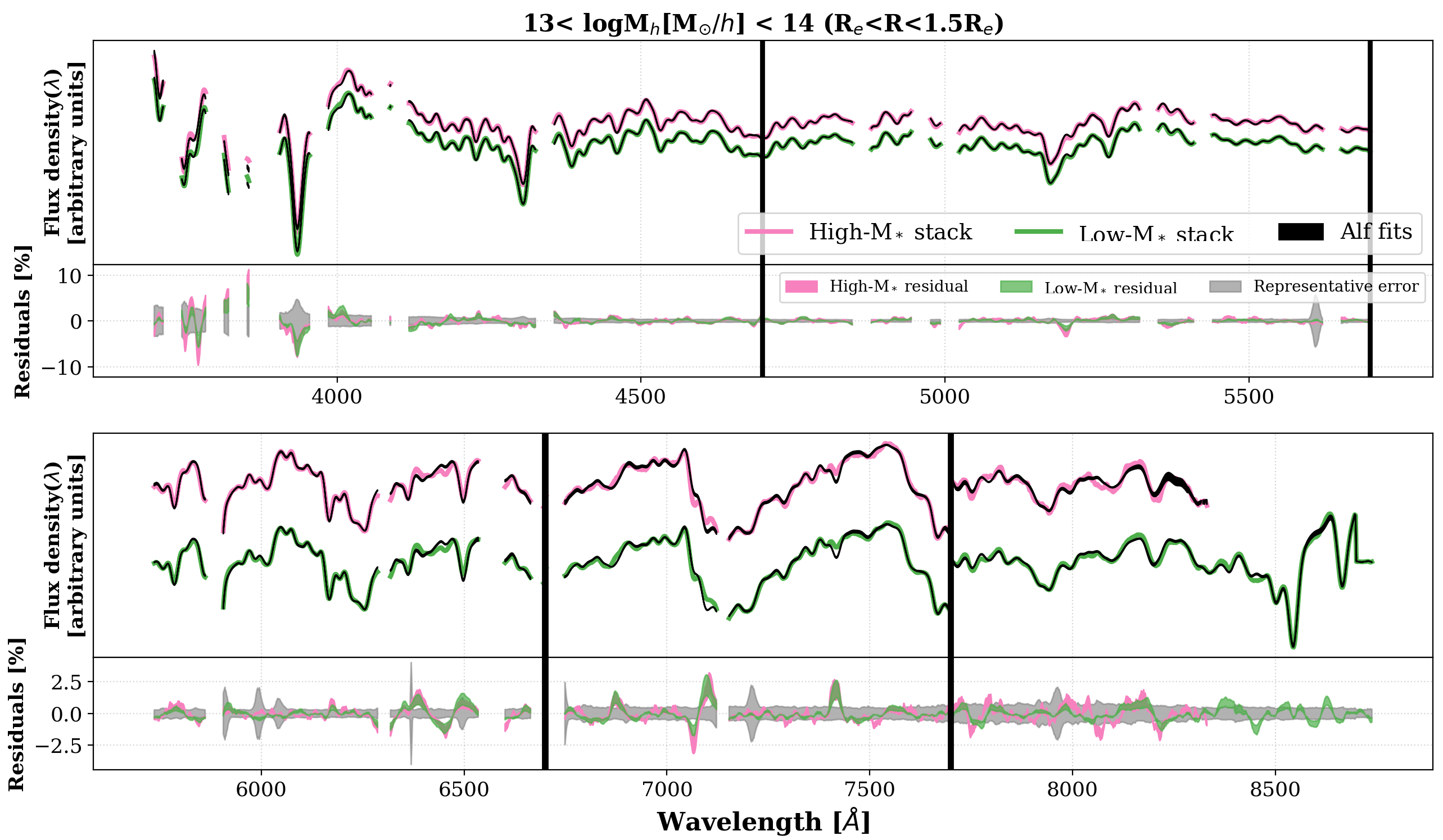}{\textwidth}{}}
		\gridline{\fig{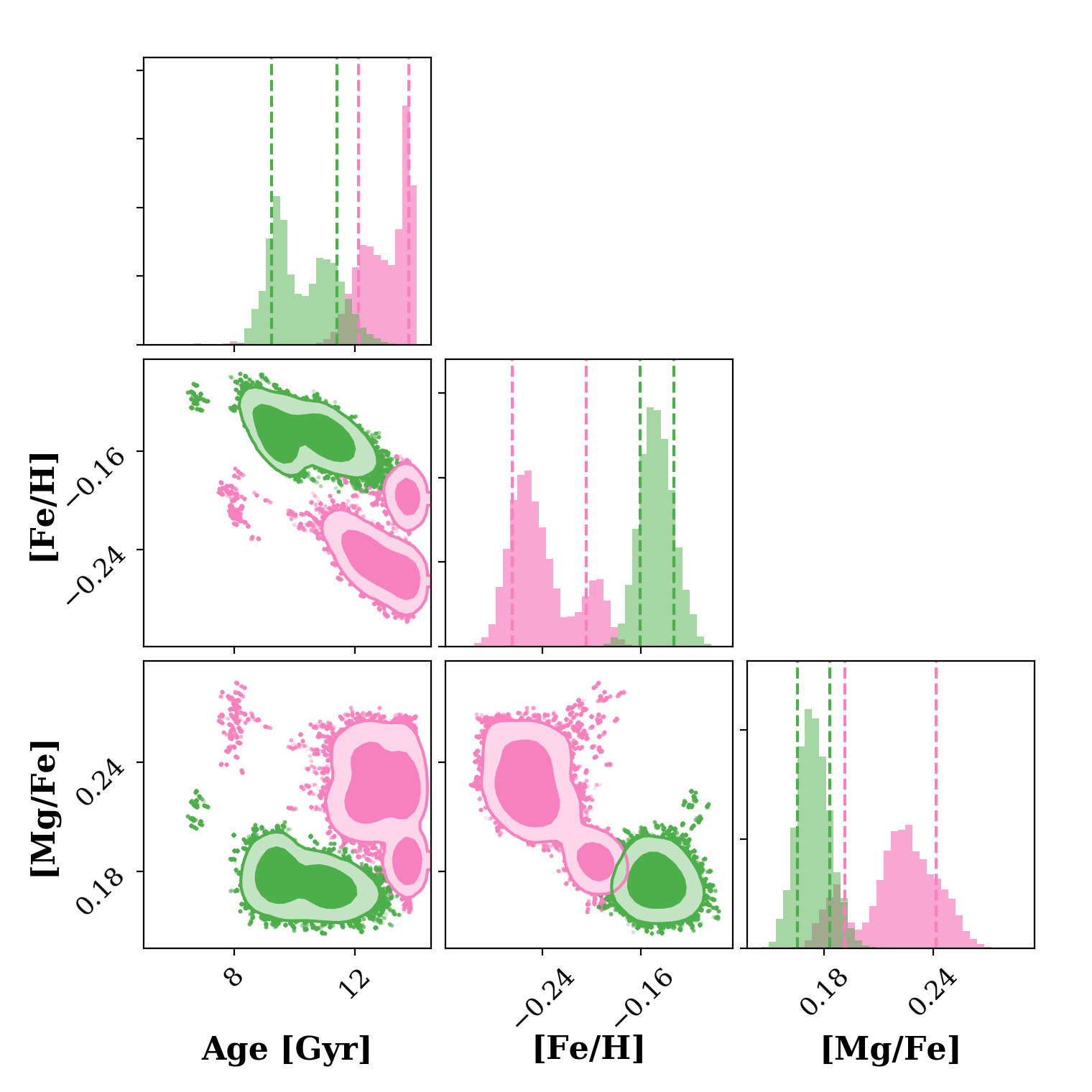}{.6\textwidth}{}}
		\caption{\textbf{Top:} Best fits and residuals for the $13<$logM$_h$[M$_{\odot}/h$]$<14$ bin ($R_e<R<1.5R_e$). \textbf{Bottom:} Posterior distributions. Details in Appendix {\ref{appendix1}}.}
		\label{figa3}
	\end{figure*}

\begin{figure*}[htp]
		\centering
		\gridline{\fig{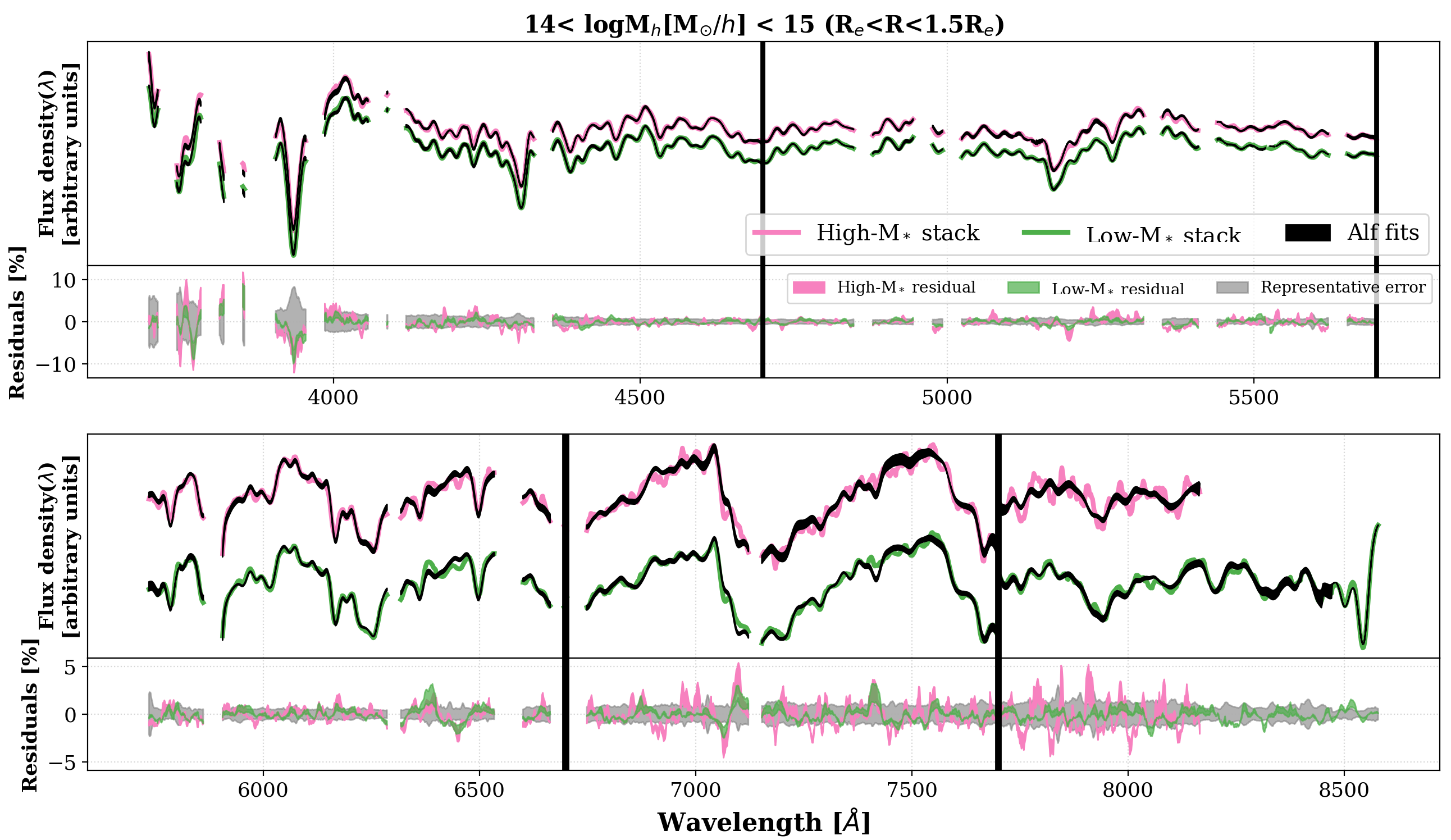}{\textwidth}{}}
		\gridline{\fig{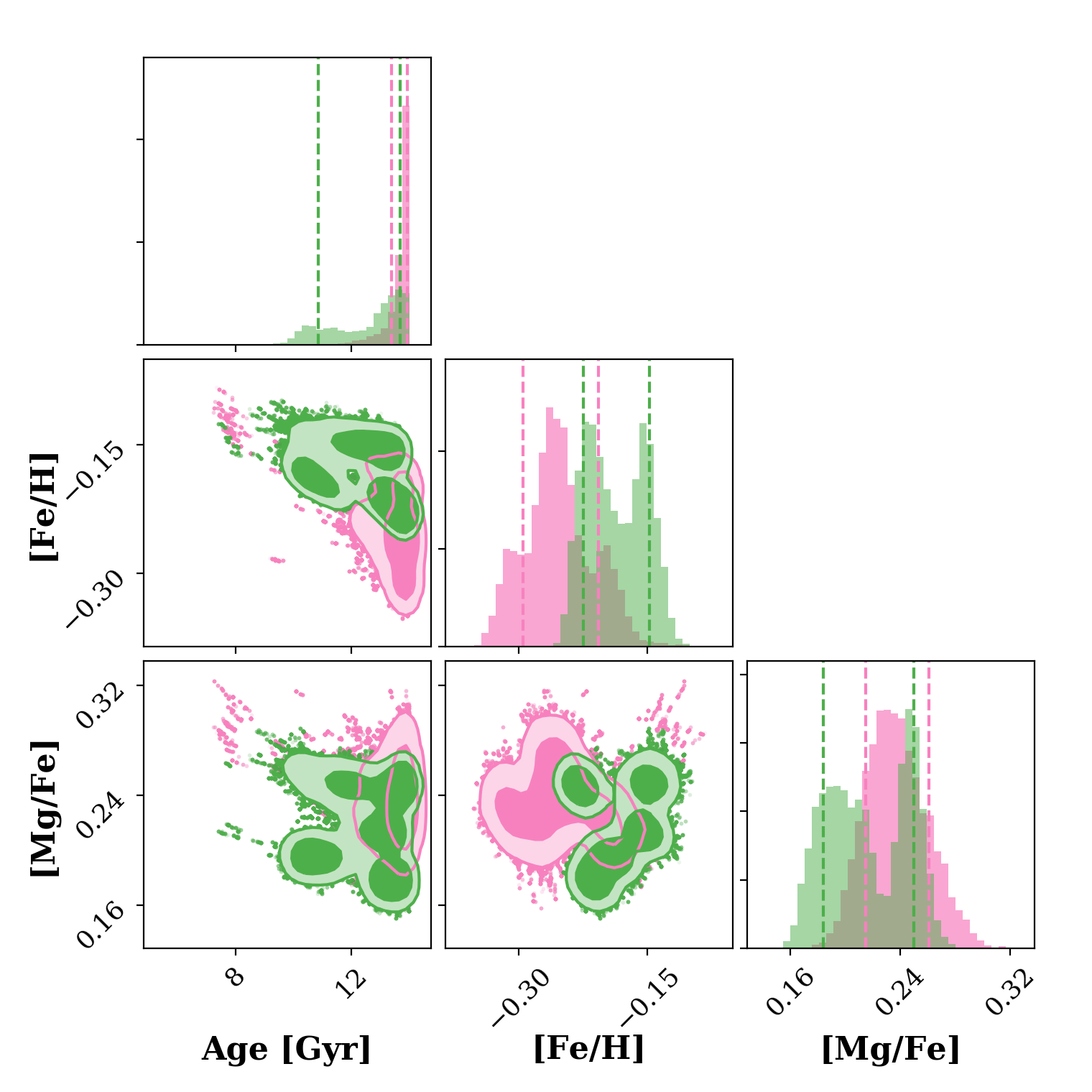}{.6\textwidth}{}}
		\caption{\textbf{Top:} Best fits and residuals for the $14<$logM$_h$[M$_{\odot}/h$]$<15$ bin ($R_e<R<1.5R_e$). \textbf{Bottom:} Posterior distributions. Details in Appendix {\ref{appendix1}}.}
		\label{figa4}
	\end{figure*}

	Figures \ref{figa1}, \ref{figa2}, \ref{figa3}, and \ref{figa4} show some of the fits and composite posterior distributions for high-M$_*$ (red) and low-M$_*$ (green) centrals. Halo mass increases with figure number. Top panels show the continuum-subtracted stacks and best fits. Residuals are plotted underneath. The errors shown in gray include the jitter and inflated error terms implemented by \texttt{alf}. Black vertical lines delimit the wavelength ranges used for continuum subtraction. 
	
	The bottom panels show the posterior distributions of mass-weighted ages, [Fe/H], and [Mg/Fe] derived with \texttt{alf} for the two corresponding spectra. The assumptions used to derive these quantities from raw \texttt{alf} outputs are described in Section \ref{3.5}. We fitted for a two-component SFH, stellar velocity dispersion, IMF, and the abundances of 19 elements. Posterior sampling was performed with emcee (\citealt{emcee}).
	
\newpage
\bibliography{mybib}{}

\begin{thebibliography}{}
\expandafter\ifx\csname natexlab\endcsname\relax\def\natexlab#1{#1}\fi
\providecommand{\url}[1]{\href{#1}{#1}}
\providecommand{\dodoi}[1]{doi:~\href{http://doi.org/#1}{\nolinkurl{#1}}}
\providecommand{\doeprint}[1]{\href{http://ascl.net/#1}{\nolinkurl{http://ascl.net/#1}}}
\providecommand{\doarXiv}[1]{\href{https://arxiv.org/abs/#1}{\nolinkurl{https://arxiv.org/abs/#1}}}

\bibitem[{{Adelman-McCarthy} {et~al.}(2006){Adelman-McCarthy}, {Ag{\"u}eros},
  {Allam}, {Anderson}, {Anderson}, {Annis}, {Bahcall}, {Baldry}, {Barentine},
  {Berlind}, {Bernardi}, {Blanton}, {Boroski}, {Brewington}, {Brinchmann},
  {Brinkmann}, {Brunner}, {Budav{\'a}ri}, {Carey}, {Carr}, {Castander},
  {Connolly}, {Csabai}, {Czarapata}, {Dalcanton}, {Doi}, {Dong}, {Eisenstein},
  {Evans}, {Fan}, {Finkbeiner}, {Friedman}, {Frieman}, {Fukugita}, {Gillespie},
  {Glazebrook}, {Gray}, {Grebel}, {Gunn}, {Gurbani}, {de Haas}, {Hall},
  {Harris}, {Harvanek}, {Hawley}, {Hayes}, {Hendry}, {Hennessy}, {Hindsley},
  {Hirata}, {Hogan}, {Hogg}, {Holmgren}, {Holtzman}, {Ichikawa}, {Ivezi{\'c}},
  {Jester}, {Johnston}, {Jorgensen}, {Juri{\'c}}, {Kent}, {Kleinman}, {Knapp},
  {Kniazev}, {Kron}, {Krzesinski}, {Kuropatkin}, {Lamb}, {Lampeitl}, {Lee},
  {Leger}, {Lin}, {Long}, {Loveday}, {Lupton}, {Margon},
  {Mart{\'\i}nez-Delgado}, {Mand elbaum}, {Matsubara}, {McGehee}, {McKay},
  {Meiksin}, {Munn}, {Nakajima}, {Nash}, {Neilsen}, {Newberg}, {Newman},
  {Nichol}, {Nicinski}, {Nieto-Santisteban}, {Nitta}, {O'Mullane}, {Okamura},
  {Owen}, {Padmanabhan}, {Pauls}, {Peoples}, {Pier}, {Pope}, {Pourbaix},
  {Quinn}, {Richards}, {Richmond}, {Rockosi}, {Schlegel}, {Schneider},
  {Schroeder}, {Scranton}, {Seljak}, {Sheldon}, {Shimasaku}, {Smith},
  {Smol{\v{c}}i{\'c}}, {Snedden}, {Stoughton}, {Strauss}, {SubbaRao}, {Szalay},
  {Szapudi}, {Szkody}, {Tegmark}, {Thakar}, {Tucker}, {Uomoto}, {Vanden Berk},
  {Vandenberg}, {Vogeley}, {Voges}, {Vogt}, {Walkowicz}, {Weinberg}, {West},
  {White}, {Xu}, {Yanny}, {Yocum}, {York}, {Zehavi}, {Zibetti}, \&
  {Zucker}}]{adelman2006}
{Adelman-McCarthy}, J.~K., {Ag{\"u}eros}, M.~A., {Allam}, S.~S., {et~al.} 2006,
  \apjs, 162, 38, \dodoi{10.1086/497917}

\bibitem[{{Aguado} {et~al.}(2019){Aguado}, {Ahumada}, {Almeida}, {Anderson},
  {Andrews}, {Anguiano}, {Aquino Ort{\'\i}z}, {Arag{\'o}n-Salamanca},
  {Argudo-Fern{\'a}ndez}, {Aubert}, {Avila-Reese}, {Badenes}, {Barboza
  Rembold}, {Barger}, {Barrera-Ballesteros}, {Bates}, {Bautista}, {Beaton},
  {Beers}, {Belfiore}, {Bernardi}, {Bershady}, {Beutler}, {Bird}, {Bizyaev},
  {Blanc}, {Blanton}, {Blomqvist}, {Bolton}, {Boquien}, {Borissova}, {Bovy},
  {Brand t}, {Brinkmann}, {Brownstein}, {Bundy}, {Burgasser}, {Byler}, {Cano
  Diaz}, {Cappellari}, {Carrera}, {Cervantes Sodi}, {Chen}, {Cherinka}, {Choi},
  {Chung}, {Coffey}, {Comerford}, {Comparat}, {Covey}, {da Silva Ilha}, {da
  Costa}, {Dai}, {Damke}, {Darling}, {Davies}, {Dawson}, {de Sainte Agathe},
  {Deconto Machado}, {Del Moro}, {De Lee}, {Diamond-Stanic}, {Dom{\'\i}nguez
  S{\'a}nchez}, {Donor}, {Drory}, {du Mas des Bourboux}, {Duckworth}, {Dwelly},
  {Ebelke}, {Emsellem}, {Escoffier}, {Fern{\'a}ndez-Trincado}, {Feuillet},
  {Fischer}, {Fleming}, {Fraser-McKelvie}, {Freischlad}, {Frinchaboy}, {Fu},
  {Galbany}, {Garcia-Dias}, {Garc{\'\i}a-Hern{\'a}ndez}, {Garma Oehmichen},
  {Geimba Maia}, {Gil-Mar{\'\i}n}, {Grabowski}, {Gu}, {Guo}, {Ha},
  {Harrington}, {Hasselquist}, {Hayes}, {Hearty}, {Hernandez Toledo}, {Hicks},
  {Hogg}, {Holley-Bockelmann}, {Holtzman}, {Hsieh}, {Hunt}, {Hwang},
  {Ibarra-Medel}, {Jimenez Angel}, {Johnson}, {Jones}, {J{\"o}nsson},
  {Kinemuchi}, {Kollmeier}, {Krawczyk}, {Kreckel}, {Kruk}, {Lacerna}, {Lan},
  {Lane}, {Law}, {Lee}, {Li}, {Lian}, {Lin}, {Lin}, {Lintott}, {Long},
  {Longa-Pe{\~n}a}, {Mackereth}, {de la Macorra}, {Majewski}, {Malanushenko},
  {Manchado}, {Maraston}, {Mariappan}, {Marinelli}, {Marques-Chaves},
  {Masseron}, {Masters}, {McDermid}, {Medina Pe{\~n}a}, {Meneses-Goytia},
  {Merloni}, {Merrifield}, {Meszaros}, {Minniti}, {Minsley}, {Muna}, {Myers},
  {Nair}, {Correa do Nascimento}, {Newman}, {Nitschelm}, {Olmstead}, {Oravetz},
  {Oravetz}, {Ortega Minakata}, {Pace}, {Padilla}, {Palicio}, {Pan}, {Pan},
  {Parikh}, {Parker}, {Peirani}, {Penny}, {Percival}, {Perez-Fournon},
  {Peterken}, {Pinsonneault}, {Prakash}, {Raddick}, {Raichoor}, {Riffel},
  {Riffel}, {Rix}, {Robin}, {Roman-Lopes}, {Rose}, {Ross}, {Rossi}, {Rowlands},
  {Rubin}, {S{\'a}nchez}, {S{\'a}nchez-Gallego}, {Sayres}, {Schaefer},
  {Schiavon}, {Schimoia}, {Schlafly}, {Schlegel}, {Schneider}, {Schultheis},
  {Seo}, {Shamsi}, {Shao}, {Shen}, {Shetty}, {Simonian}, {Smethurst}, {Sobeck},
  {Souter}, {Spindler}, {Stark}, {Stassun}, {Steinmetz}, {Storchi-Bergmann},
  {Stringfellow}, {Su{\'a}rez}, {Sun}, {Taghizadeh-Popp}, {Talbot}, {Tayar},
  {Thakar}, {Thomas}, {Tissera}, {Tojeiro}, {Troup}, {Unda-Sanzana},
  {Valenzuela}, {Vargas-Maga{\~n}a}, {V{\'a}zquez-Mata}, {Wake}, {Weaver},
  {Weijmans}, {Westfall}, {Wild}, {Wilson}, {Woods}, {Yan}, {Yang}, {Zamora},
  {Zasowski}, {Zhang}, {Zheng}, {Zheng}, {Zhu}, {Zinn}, \& {Zou}}]{dr15}
{Aguado}, D.~S., {Ahumada}, R., {Almeida}, A., {et~al.} 2019, \apjs, 240, 23,
  \dodoi{10.3847/1538-4365/aaf651}

\bibitem[{{Alam} {et~al.}(2015){Alam}, {Albareti}, {Allende Prieto}, {Anders},
  {Anderson}, {Anderton}, {Andrews}, {Armengaud}, {Aubourg}, {Bailey}, {Basu},
  {Bautista}, {Beaton}, {Beers}, {Bender}, {Berlind}, {Beutler}, {Bhardwaj},
  {Bird}, {Bizyaev}, {Blake}, {Blanton}, {Blomqvist}, {Bochanski}, {Bolton},
  {Bovy}, {Shelden Bradley}, {Brandt}, {Brauer}, {Brinkmann}, {Brown},
  {Brownstein}, {Burden}, {Burtin}, {Busca}, {Cai}, {Capozzi}, {Carnero
  Rosell}, {Carr}, {Carrera}, {Chambers}, {Chaplin}, {Chen}, {Chiappini},
  {Chojnowski}, {Chuang}, {Clerc}, {Comparat}, {Covey}, {Croft}, {Cuesta},
  {Cunha}, {da Costa}, {Da Rio}, {Davenport}, {Dawson}, {De Lee}, {Delubac},
  {Deshpande}, {Dhital}, {Dutra-Ferreira}, {Dwelly}, {Ealet}, {Ebelke},
  {Edmondson}, {Eisenstein}, {Ellsworth}, {Elsworth}, {Epstein}, {Eracleous},
  {Escoffier}, {Esposito}, {Evans}, {Fan}, {Fern{\'a}ndez-Alvar}, {Feuillet},
  {Filiz Ak}, {Finley}, {Finoguenov}, {Flaherty}, {Fleming}, {Font-Ribera},
  {Foster}, {Frinchaboy}, {Galbraith-Frew}, {Garc{\'\i}a},
  {Garc{\'\i}a-Hern{\'a}ndez}, {Garc{\'\i}a P{\'e}rez}, {Gaulme}, {Ge},
  {G{\'e}nova-Santos}, {Georgakakis}, {Ghezzi}, {Gillespie}, {Girardi},
  {Goddard}, {Gontcho}, {Gonz{\'a}lez Hern{\'a}ndez}, {Grebel}, {Green},
  {Grieb}, {Grieves}, {Gunn}, {Guo}, {Harding}, {Hasselquist}, {Hawley},
  {Hayden}, {Hearty}, {Hekker}, {Ho}, {Hogg}, {Holley-Bockelmann}, {Holtzman},
  {Honscheid}, {Huber}, {Huehnerhoff}, {Ivans}, {Jiang}, {Johnson},
  {Kinemuchi}, {Kirkby}, {Kitaura}, {Klaene}, {Knapp}, {Kneib}, {Koenig},
  {Lam}, {Lan}, {Lang}, {Laurent}, {Le Goff}, {Leauthaud}, {Lee}, {Lee},
  {Licquia}, {Liu}, {Long}, {L{\'o}pez-Corredoira}, {Lorenzo-Oliveira},
  {Lucatello}, {Lundgren}, {Lupton}, {Mack}, {Mahadevan}, {Maia}, {Majewski},
  {Malanushenko}, {Malanushenko}, {Manchado}, {Manera}, {Mao}, {Maraston},
  {Marchwinski}, {Margala}, {Martell}, {Martig}, {Masters}, {Mathur},
  {McBride}, {McGehee}, {McGreer}, {McMahon}, {M{\'e}nard}, {Menzel},
  {Merloni}, {M{\'e}sz{\'a}ros}, {Miller}, {Miralda-Escud{\'e}}, {Miyatake},
  {Montero-Dorta}, {More}, {Morganson}, {Morice-Atkinson}, {Morrison},
  {Mosser}, {Muna}, {Myers}, {Nandra}, {Newman}, {Neyrinck}, {Nguyen},
  {Nichol}, {Nidever}, {Noterdaeme}, {Nuza}, {O'Connell}, {O'Connell},
  {O'Connell}, {Ogando}, {Olmstead}, {Oravetz}, {Oravetz}, {Osumi}, {Owen},
  {Padgett}, {Padmanabhan}, {Paegert}, {Palanque-Delabrouille}, {Pan},
  {Parejko}, {P{\^a}ris}, {Park}, {Pattarakijwanich}, {Pellejero-Ibanez},
  {Pepper}, {Percival}, {P{\'e}rez-Fournon}, {P{\textasciiacute}rez-Ra`fols},
  {Petitjean}, {Pieri}, {Pinsonneault}, {Porto de Mello}, {Prada}, {Prakash},
  {Price-Whelan}, {Protopapas}, {Raddick}, {Rahman}, {Reid}, {Rich}, {Rix},
  {Robin}, {Rockosi}, {Rodrigues}, {Rodr{\'\i}guez-Torres}, {Roe}, {Ross},
  {Ross}, {Rossi}, {Ruan}, {Rubi{\~n}o-Mart{\'\i}n}, {Rykoff},
  {Salazar-Albornoz}, {Salvato}, {Samushia}, {S{\'a}nchez}, {Santiago},
  {Sayres}, {Schiavon}, {Schlegel}, {Schmidt}, {Schneider}, {Schultheis},
  {Schwope}, {Sc{\'o}ccola}, {Scott}, {Sellgren}, {Seo}, {Serenelli}, {Shane},
  {Shen}, {Shetrone}, {Shu}, {Silva Aguirre}, {Sivarani}, {Skrutskie},
  {Slosar}, {Smith}, {Sobreira}, {Souto}, {Stassun}, {Steinmetz}, {Stello},
  {Strauss}, {Streblyanska}, {Suzuki}, {Swanson}, {Tan}, {Tayar}, {Terrien},
  {Thakar}, {Thomas}, {Thomas}, {Thompson}, {Tinker}, {Tojeiro}, {Troup},
  {Vargas-Maga{\~n}a}, {Vazquez}, {Verde}, {Viel}, {Vogt}, {Wake}, {Wang},
  {Weaver}, {Weinberg}, {Weiner}, {White}, {Wilson}, {Wisniewski},
  {Wood-Vasey}, {Ye`che}, {York}, {Zakamska}, {Zamora}, {Zasowski}, {Zehavi},
  {Zhao}, {Zheng}, {Zhou}, {Zhou}, {Zou}, \& {Zhu}}]{dr12}
{Alam}, S., {Albareti}, F.~D., {Allende Prieto}, C., {et~al.} 2015, \apjs, 219,
  12, \dodoi{10.1088/0067-0049/219/1/12}

\bibitem[{{Allen} {et~al.}(2015){Allen}, {Croom}, {Konstantopoulos}, {Bryant},
  {Sharp}, {Cecil}, {Fogarty}, {Foster}, {Green}, {Ho}, {Owers}, {Schaefer},
  {Scott}, {Bauer}, {Baldry}, {Barnes}, {Bland-Hawthorn}, {Bloom}, {Brough},
  {Colless}, {Cortese}, {Couch}, {Drinkwater}, {Driver}, {Goodwin},
  {Gunawardhana}, {Hampton}, {Hopkins}, {Kewley}, {Lawrence}, {Leon-Saval},
  {Liske}, {L{\'o}pez-S{\'a}nchez}, {Lorente}, {McElroy}, {Medling}, {Mould},
  {Norberg}, {Parker}, {Power}, {Pracy}, {Richards}, {Robotham}, {Sweet},
  {Taylor}, {Thomas}, {Tonini}, \& {Walcher}}]{allen2015}
{Allen}, J.~T., {Croom}, S.~M., {Konstantopoulos}, I.~S., {et~al.} 2015,
  \mnras, 446, 1567, \dodoi{10.1093/mnras/stu2057}

\bibitem[{{Alton} {et~al.}(2018){Alton}, {Smith}, \& {Lucey}}]{alton2018}
{Alton}, P.~D., {Smith}, R.~J., \& {Lucey}, J.~R. 2018, \mnras, 478, 4464,
  \dodoi{10.1093/mnras/sty1242}

\bibitem[{{Argudo-Fern{\'a}ndez} {et~al.}(2015){Argudo-Fern{\'a}ndez},
  {Verley}, {Bergond}, {Duarte Puertas}, {Ramos Carmona}, {Sabater},
  {Fern{\'a}ndez Lorenzo}, {Espada}, {Sulentic}, {Ruiz}, \&
  {Leon}}]{argudo-fernandez2015}
{Argudo-Fern{\'a}ndez}, M., {Verley}, S., {Bergond}, G., {et~al.} 2015, \aap,
  578, A110, \dodoi{10.1051/0004-6361/201526016}

\bibitem[{{Arimoto} \& {Yoshii}(1987)}]{arimoto-yoshii1987}
{Arimoto}, N., \& {Yoshii}, Y. 1987, \aap, 173, 23

\bibitem[{{Barro} {et~al.}(2013){Barro}, {Faber}, {P{\'e}rez-Gonz{\'a}lez},
  {Koo}, {Williams}, {Kocevski}, {Trump}, {Mozena}, {McGrath}, {van der Wel},
  {Wuyts}, {Bell}, {Croton}, {Ceverino}, {Dekel}, {Ashby}, {Cheung},
  {Ferguson}, {Fontana}, {Fang}, {Giavalisco}, {Grogin}, {Guo}, {Hathi},
  {Hopkins}, {Huang}, {Koekemoer}, {Kartaltepe}, {Lee}, {Newman}, {Porter},
  {Primack}, {Ryan}, {Rosario}, {Somerville}, {Salvato}, \& {Hsu}}]{barro2013}
{Barro}, G., {Faber}, S.~M., {P{\'e}rez-Gonz{\'a}lez}, P.~G., {et~al.} 2013,
  \apj, 765, 104, \dodoi{10.1088/0004-637X/765/2/104}

\bibitem[{{Behroozi} {et~al.}(2019){Behroozi}, {Wechsler}, {Hearin}, \&
  {Conroy}}]{behroozi2019}
{Behroozi}, P., {Wechsler}, R.~H., {Hearin}, A.~P., \& {Conroy}, C. 2019,
  \mnras, 488, 3143, \dodoi{10.1093/mnras/stz1182}

\bibitem[{{Behroozi} {et~al.}(2013){Behroozi}, {Wechsler}, \&
  {Conroy}}]{behroozi2013}
{Behroozi}, P.~S., {Wechsler}, R.~H., \& {Conroy}, C. 2013, \apjl, 762, L31,
  \dodoi{10.1088/2041-8205/762/2/L31}

\bibitem[{{Belfiore} {et~al.}(2019){Belfiore}, {Westfall}, {Schaefer},
  {Cappellari}, {Ji}, {Bershady}, {Tremonti}, {Law}, {Yan}, {Bundy}, {Shetty},
  {Drory}, {Thomas}, {Emsellem}, \& {S{\'a}nchez}}]{belfiore2019}
{Belfiore}, F., {Westfall}, K.~B., {Schaefer}, A., {et~al.} 2019, \aj, 158,
  160, \dodoi{10.3847/1538-3881/ab3e4e}

\bibitem[{{Benson} {et~al.}(2000){Benson}, {Cole}, {Frenk}, {Baugh}, \&
  {Lacey}}]{benson2000}
{Benson}, A.~J., {Cole}, S., {Frenk}, C.~S., {Baugh}, C.~M., \& {Lacey}, C.~G.
  2000, \mnras, 311, 793, \dodoi{10.1046/j.1365-8711.2000.03101.x}

\bibitem[{{Berlind} {et~al.}(2006){Berlind}, {Kazin}, {Blanton}, {Pueblas},
  {Scoccimarro}, \& {Hogg}}]{berlind2006}
{Berlind}, A.~A., {Kazin}, E., {Blanton}, M.~R., {et~al.} 2006, arXiv e-prints,
  astro.
\newblock \doarXiv{astro-ph/0610524}

\bibitem[{{Bernardi} {et~al.}(2013){Bernardi}, {Meert}, {Sheth}, {Vikram},
  {Huertas-Company}, {Mei}, \& {Shankar}}]{bernardi2013}
{Bernardi}, M., {Meert}, A., {Sheth}, R.~K., {et~al.} 2013, \mnras, 436, 697,
  \dodoi{10.1093/mnras/stt1607}

\bibitem[{{Berti} {et~al.}(2017){Berti}, {Coil}, {Behroozi}, {Eisenstein},
  {Bray}, {Cool}, \& {Moustakas}}]{berti2017}
{Berti}, A.~M., {Coil}, A.~L., {Behroozi}, P.~S., {et~al.} 2017, \apj, 834, 87,
  \dodoi{10.3847/1538-4357/834/1/87}

\bibitem[{{Bezanson} {et~al.}(2009){Bezanson}, {van Dokkum}, {Tal},
  {Marchesini}, {Kriek}, {Franx}, \& {Coppi}}]{bezanson2009}
{Bezanson}, R., {van Dokkum}, P.~G., {Tal}, T., {et~al.} 2009, \apj, 697, 1290,
  \dodoi{10.1088/0004-637X/697/2/1290}

\bibitem[{{Blanton} {et~al.}(2011){Blanton}, {Kazin}, {Muna}, {Weaver}, \&
  {Price-Whelan}}]{blanton2011}
{Blanton}, M.~R., {Kazin}, E., {Muna}, D., {Weaver}, B.~A., \& {Price-Whelan},
  A. 2011, \aj, 142, 31, \dodoi{10.1088/0004-6256/142/1/31}

\bibitem[{{Blanton} \& {Roweis}(2007)}]{blanton2007}
{Blanton}, M.~R., \& {Roweis}, S. 2007, \aj, 133, 734, \dodoi{10.1086/510127}

\bibitem[{{Blanton} {et~al.}(2005){Blanton}, {Schlegel}, {Strauss},
  {Brinkmann}, {Finkbeiner}, {Fukugita}, {Gunn}, {Hogg}, {Ivezi{\'c}}, {Knapp},
  {Lupton}, {Munn}, {Schneider}, {Tegmark}, \& {Zehavi}}]{blanton2005b}
{Blanton}, M.~R., {Schlegel}, D.~J., {Strauss}, M.~A., {et~al.} 2005, \aj, 129,
  2562, \dodoi{10.1086/429803}

\bibitem[{{Blanton} {et~al.}(2017){Blanton}, {Bershady}, {Abolfathi},
  {Albareti}, {Allende Prieto}, {Almeida}, {Alonso-Garc{\'{\i}}a}, {Anders},
  {Anderson}, {Andrews}, \& et~al.}]{blanton2017}
{Blanton}, M.~R., {Bershady}, M.~A., {Abolfathi}, B., {et~al.} 2017, \aj, 154,
  28, \dodoi{10.3847/1538-3881/aa7567}

\bibitem[{{Bluck} {et~al.}(2020){Bluck}, {Maiolino}, {S{\'a}nchez}, {Ellison},
  {Thorp}, {Piotrowska}, {Teimoorinia}, \& {Bundy}}]{bluck2020}
{Bluck}, A. F.~L., {Maiolino}, R., {S{\'a}nchez}, S.~F., {et~al.} 2020, \mnras,
  492, 96, \dodoi{10.1093/mnras/stz3264}

\bibitem[{{Booth} \& {Schaye}(2010)}]{booth-schaye2010}
{Booth}, C.~M., \& {Schaye}, J. 2010, \mnras, 405, L1,
  \dodoi{10.1111/j.1745-3933.2010.00832.x}

\bibitem[{{Bradshaw} {et~al.}(2020){Bradshaw}, {Leauthaud}, {Hearin}, {Huang},
  \& {Behroozi}}]{bradshaw2020}
{Bradshaw}, C., {Leauthaud}, A., {Hearin}, A., {Huang}, S., \& {Behroozi}, P.
  2020, \mnras, 493, 337, \dodoi{10.1093/mnras/staa081}

\bibitem[{{Bressan} {et~al.}(1994){Bressan}, {Chiosi}, \&
  {Fagotto}}]{bressan1994}
{Bressan}, A., {Chiosi}, C., \& {Fagotto}, F. 1994, \apjs, 94, 63,
  \dodoi{10.1086/192073}

\bibitem[{{Buitrago} {et~al.}(2008){Buitrago}, {Trujillo}, {Conselice},
  {Bouwens}, {Dickinson}, \& {Yan}}]{buitrago2008}
{Buitrago}, F., {Trujillo}, I., {Conselice}, C.~J., {et~al.} 2008, \apjl, 687,
  L61, \dodoi{10.1086/592836}

\bibitem[{{Bundy} {et~al.}(2015){Bundy}, {Bershady}, {Law}, {Yan}, {Drory},
  {MacDonald}, {Wake}, {Cherinka}, {S{\'a}nchez-Gallego}, {Weijmans}, {Thomas},
  {Tremonti}, {Masters}, {Coccato}, {Diamond-Stanic}, {Arag{\'o}n-Salamanca},
  {Avila-Reese}, {Badenes}, {Falc{\'o}n-Barroso}, {Belfiore}, {Bizyaev},
  {Blanc}, {Bland-Hawthorn}, {Blanton}, {Brownstein}, {Byler}, {Cappellari},
  {Conroy}, {Dutton}, {Emsellem}, {Etherington}, {Frinchaboy}, {Fu}, {Gunn},
  {Harding}, {Johnston}, {Kauffmann}, {Kinemuchi}, {Klaene}, {Knapen},
  {Leauthaud}, {Li}, {Lin}, {Maiolino}, {Malanushenko}, {Malanushenko}, {Mao},
  {Maraston}, {McDermid}, {Merrifield}, {Nichol}, {Oravetz}, {Pan}, {Parejko},
  {Sanchez}, {Schlegel}, {Simmons}, {Steele}, {Steinmetz}, {Thanjavur},
  {Thompson}, {Tinker}, {van den Bosch}, {Westfall}, {Wilkinson}, {Wright},
  {Xiao}, \& {Zhang}}]{bundy2015}
{Bundy}, K., {Bershady}, M.~A., {Law}, D.~R., {et~al.} 2015, \apj, 798, 7,
  \dodoi{10.1088/0004-637X/798/1/7}

\bibitem[{{Calderon} {et~al.}(2018){Calderon}, {Berlind}, \&
  {Sinha}}]{calderon2018}
{Calderon}, V.~F., {Berlind}, A.~A., \& {Sinha}, M. 2018, \mnras, 480, 2031,
  \dodoi{10.1093/mnras/sty2000}

\bibitem[{{Campbell} {et~al.}(2015){Campbell}, {van den Bosch}, {Hearin},
  {Padmanabhan}, {Berlind}, {Mo}, {Tinker}, \& {Yang}}]{campbell2015}
{Campbell}, D., {van den Bosch}, F.~C., {Hearin}, A., {et~al.} 2015, \mnras,
  452, 444, \dodoi{10.1093/mnras/stv1091}

\bibitem[{{Cappellari}(2017)}]{cappellari2017}
{Cappellari}, M. 2017, \mnras, 466, 798, \dodoi{10.1093/mnras/stw3020}

\bibitem[{{Cappellari} \& {Emsellem}(2004)}]{cappellari-emsellem2004}
{Cappellari}, M., \& {Emsellem}, E. 2004, \pasp, 116, 138,
  \dodoi{10.1086/381875}

\bibitem[{{Cappellari} {et~al.}(2013){Cappellari}, {McDermid}, {Alatalo},
  {Blitz}, {Bois}, {Bournaud}, {Bureau}, {Crocker}, {Davies}, {Davis}, {de
  Zeeuw}, {Duc}, {Emsellem}, {Khochfar}, {Krajnovi{\'c}}, {Kuntschner},
  {Morganti}, {Naab}, {Oosterloo}, {Sarzi}, {Scott}, {Serra}, {Weijmans}, \&
  {Young}}]{cappellari2013b}
{Cappellari}, M., {McDermid}, R.~M., {Alatalo}, K., {et~al.} 2013, \mnras, 432,
  1862, \dodoi{10.1093/mnras/stt644}

\bibitem[{{Cassata} {et~al.}(2010){Cassata}, {Giavalisco}, {Guo}, {Ferguson},
  {Koekemoer}, {Renzini}, {Fontana}, {Salimbeni}, {Dickinson}, {Casertano},
  {Conselice}, {Grogin}, {Lotz}, {Papovich}, {Lucas}, {Straughn}, {Gardner}, \&
  {Moustakas}}]{cassata2010}
{Cassata}, P., {Giavalisco}, M., {Guo}, Y., {et~al.} 2010, \apjl, 714, L79,
  \dodoi{10.1088/2041-8205/714/1/L79}

\bibitem[{{Cassata} {et~al.}(2011){Cassata}, {Giavalisco}, {Guo}, {Renzini},
  {Ferguson}, {Koekemoer}, {Salimbeni}, {Scarlata}, {Grogin}, {Conselice},
  {Dahlen}, {Lotz}, {Dickinson}, \& {Lin}}]{cassata2011b}
---. 2011, \apj, 743, 96, \dodoi{10.1088/0004-637X/743/1/96}

\bibitem[{{Chen} {et~al.}(2020){Chen}, {Faber}, {Koo}, {Somerville}, {Primack},
  {Dekel}, {Rodr{\'\i}guez-Puebla}, {Guo}, {Barro}, {Kocevski}, {van der Wel},
  {Woo}, {Bell}, {Fang}, {Ferguson}, {Giavalisco}, {Huertas-Company}, {Jiang},
  {Kassin}, {Lin}, {Liu}, {Luo}, {Luo}, {Pacifici}, {Pandya}, {Salim}, {Shu},
  {Tacchella}, {Terrazas}, \& {Yesuf}}]{chen2020}
{Chen}, Z., {Faber}, S.~M., {Koo}, D.~C., {et~al.} 2020, \apj, 897, 102,
  \dodoi{10.3847/1538-4357/ab9633}

\bibitem[{{Cherinka} {et~al.}(2019){Cherinka}, {Andrews},
  {S{\'a}nchez-Gallego}, {Brownstein}, {Argudo-Fern{\'a}ndez}, {Blanton},
  {Bundy}, {Jones}, {Masters}, {Law}, {Rowlands}, {Weijmans}, {Westfall}, \&
  {Yan}}]{cherinka2017}
{Cherinka}, B., {Andrews}, B.~H., {S{\'a}nchez-Gallego}, J., {et~al.} 2019,
  \aj, 158, 74, \dodoi{10.3847/1538-3881/ab2634}

\bibitem[{{Choi} {et~al.}(2016){Choi}, {Dotter}, {Conroy}, {Cantiello},
  {Paxton}, \& {Johnson}}]{choi2016}
{Choi}, J., {Dotter}, A., {Conroy}, C., {et~al.} 2016, \apj, 823, 102,
  \dodoi{10.3847/0004-637X/823/2/102}

\bibitem[{{Cid Fernandes} {et~al.}(2005){Cid Fernandes}, {Mateus}, {Sodr{\'e}},
  {Stasi{\'n}ska}, \& {Gomes}}]{cidfernandes2005}
{Cid Fernandes}, R., {Mateus}, A., {Sodr{\'e}}, L., {Stasi{\'n}ska}, G., \&
  {Gomes}, J.~M. 2005, \mnras, 358, 363,
  \dodoi{10.1111/j.1365-2966.2005.08752.x}

\bibitem[{{Cimatti} {et~al.}(2008){Cimatti}, {Cassata}, {Pozzetti}, {Kurk},
  {Mignoli}, {Renzini}, {Daddi}, {Bolzonella}, {Brusa}, {Rodighiero},
  {Dickinson}, {Franceschini}, {Zamorani}, {Berta}, {Rosati}, \&
  {Halliday}}]{cimatti2008}
{Cimatti}, A., {Cassata}, P., {Pozzetti}, L., {et~al.} 2008, \aap, 482, 21,
  \dodoi{10.1051/0004-6361:20078739}

\bibitem[{{Conroy}(2013)}]{conroy2013}
{Conroy}, C. 2013, \araa, 51, 393, \dodoi{10.1146/annurev-astro-082812-141017}

\bibitem[{{Conroy} {et~al.}(2014){Conroy}, {Graves}, \& {van
  Dokkum}}]{conroy2014}
{Conroy}, C., {Graves}, G.~J., \& {van Dokkum}, P.~G. 2014, \apj, 780, 33,
  \dodoi{10.1088/0004-637X/780/1/33}

\bibitem[{{Conroy} \& {Gunn}(2010)}]{conroy-gunn2010}
{Conroy}, C., \& {Gunn}, J.~E. 2010, \apj, 712, 833,
  \dodoi{10.1088/0004-637X/712/2/833}

\bibitem[{{Conroy} {et~al.}(2009){Conroy}, {Gunn}, \& {White}}]{conroy2009}
{Conroy}, C., {Gunn}, J.~E., \& {White}, M. 2009, \apj, 699, 486,
  \dodoi{10.1088/0004-637X/699/1/486}

\bibitem[{{Conroy} \& {van Dokkum}(2012)}]{conroy2012a}
{Conroy}, C., \& {van Dokkum}, P. 2012, \apj, 747, 69,
  \dodoi{10.1088/0004-637X/747/1/69}

\bibitem[{{Conroy} {et~al.}(2018){Conroy}, {Villaume}, {van Dokkum}, \&
  {Lind}}]{conroy2018}
{Conroy}, C., {Villaume}, A., {van Dokkum}, P.~G., \& {Lind}, K. 2018, \apj,
  854, 139, \dodoi{10.3847/1538-4357/aaab49}

\bibitem[{{Contini} {et~al.}(2019){Contini}, {Gu}, {Kang}, {Rhee}, \&
  {Yi}}]{contini2019}
{Contini}, E., {Gu}, Q., {Kang}, X., {Rhee}, J., \& {Yi}, S.~K. 2019, \apj,
  882, 167, \dodoi{10.3847/1538-4357/ab3b03}

\bibitem[{{Contreras} {et~al.}(2021){Contreras}, {Angulo}, \&
  {Zennaro}}]{contreras2021}
{Contreras}, S., {Angulo}, R.~E., \& {Zennaro}, M. 2021, \mnras,
  \dodoi{10.1093/mnras/stab1170}

\bibitem[{{Contreras} {et~al.}(2019){Contreras}, {Zehavi}, {Padilla}, {Baugh},
  {Jim{\'e}nez}, \& {Lacerna}}]{contreras2019}
{Contreras}, S., {Zehavi}, I., {Padilla}, N., {et~al.} 2019, \mnras, 484, 1133,
  \dodoi{10.1093/mnras/stz018}

\bibitem[{{Cook} {et~al.}(2016){Cook}, {Conroy}, {Pillepich},
  {Rodriguez-Gomez}, \& {Hernquist}}]{cook2016}
{Cook}, B.~A., {Conroy}, C., {Pillepich}, A., {Rodriguez-Gomez}, V., \&
  {Hernquist}, L. 2016, \apj, 833, 158, \dodoi{10.3847/1538-4357/833/2/158}

\bibitem[{{Cooper} {et~al.}(2015){Cooper}, {Parry}, {Lowing}, {Cole}, \&
  {Frenk}}]{cooper2015}
{Cooper}, A.~P., {Parry}, O.~H., {Lowing}, B., {Cole}, S., \& {Frenk}, C. 2015,
  \mnras, 454, 3185, \dodoi{10.1093/mnras/stv2057}

\bibitem[{{Correa} {et~al.}(2015){Correa}, {Wyithe}, {Schaye}, \&
  {Duffy}}]{correa2015}
{Correa}, C.~A., {Wyithe}, J. S.~B., {Schaye}, J., \& {Duffy}, A.~R. 2015,
  \mnras, 450, 1521, \dodoi{10.1093/mnras/stv697}

\bibitem[{{Crain} {et~al.}(2015){Crain}, {Schaye}, {Bower}, {Furlong},
  {Schaller}, {Theuns}, {Dalla Vecchia}, {Frenk}, {McCarthy}, {Helly},
  {Jenkins}, {Rosas-Guevara}, {White}, \& {Trayford}}]{crain2015}
{Crain}, R.~A., {Schaye}, J., {Bower}, R.~G., {et~al.} 2015, \mnras, 450, 1937,
  \dodoi{10.1093/mnras/stv725}

\bibitem[{{Crook} {et~al.}(2007){Crook}, {Huchra}, {Martimbeau}, {Masters},
  {Jarrett}, \& {Macri}}]{crook2007}
{Crook}, A.~C., {Huchra}, J.~P., {Martimbeau}, N., {et~al.} 2007, \apj, 655,
  790, \dodoi{10.1086/510201}

\bibitem[{{Croton} {et~al.}(2007){Croton}, {Gao}, \& {White}}]{croton2007}
{Croton}, D.~J., {Gao}, L., \& {White}, S. D.~M. 2007, \mnras, 374, 1303,
  \dodoi{10.1111/j.1365-2966.2006.11230.x}

\bibitem[{{Daddi} {et~al.}(2005){Daddi}, {Renzini}, {Pirzkal}, {Cimatti},
  {Malhotra}, {Stiavelli}, {Xu}, {Pasquali}, {Rhoads}, {Brusa}, {di Serego
  Alighieri}, {Ferguson}, {Koekemoer}, {Moustakas}, {Panagia}, \&
  {Windhorst}}]{daddi2005}
{Daddi}, E., {Renzini}, A., {Pirzkal}, N., {et~al.} 2005, \apj, 626, 680,
  \dodoi{10.1086/430104}

\bibitem[{{Damjanov} {et~al.}(2009){Damjanov}, {McCarthy}, {Abraham},
  {Glazebrook}, {Yan}, {Mentuch}, {Le Borgne}, {Savaglio}, {Crampton},
  {Murowinski}, {Juneau}, {Carlberg}, {J{\o}rgensen}, {Roth}, {Chen}, \&
  {Marzke}}]{damjanov2009}
{Damjanov}, I., {McCarthy}, P.~J., {Abraham}, R.~G., {et~al.} 2009, \apj, 695,
  101, \dodoi{10.1088/0004-637X/695/1/101}

\bibitem[{{Damjanov} {et~al.}(2011){Damjanov}, {Abraham}, {Glazebrook},
  {McCarthy}, {Caris}, {Carlberg}, {Chen}, {Crampton}, {Green}, {J{\o}rgensen},
  {Juneau}, {Le Borgne}, {Marzke}, {Mentuch}, {Murowinski}, {Roth}, {Savaglio},
  \& {Yan}}]{damjanov2011}
{Damjanov}, I., {Abraham}, R.~G., {Glazebrook}, K., {et~al.} 2011, \apjl, 739,
  L44, \dodoi{10.1088/2041-8205/739/2/L44}

\bibitem[{{Dekel} \& {Birnboim}(2006)}]{dekel2006}
{Dekel}, A., \& {Birnboim}, Y. 2006, \mnras, 368, 2,
  \dodoi{10.1111/j.1365-2966.2006.10145.x}

\bibitem[{{Dekel} \& {Burkert}(2014)}]{dekel-burkert2014}
{Dekel}, A., \& {Burkert}, A. 2014, \mnras, 438, 1870,
  \dodoi{10.1093/mnras/stt2331}

\bibitem[{{Dey} {et~al.}(2019){Dey}, {Schlegel}, {Lang}, {Blum}, {Burleigh},
  {Fan}, {Findlay}, {Finkbeiner}, {Herrera}, {Juneau}, {Landriau}, {Levi},
  {McGreer}, {Meisner}, {Myers}, {Moustakas}, {Nugent}, {Patej}, {Schlafly},
  {Walker}, {Valdes}, {Weaver}, {Y{\`e}che}, {Zou}, {Zhou}, {Abareshi},
  {Abbott}, {Abolfathi}, {Aguilera}, {Alam}, {Allen}, {Alvarez}, {Annis},
  {Ansarinejad}, {Aubert}, {Beechert}, {Bell}, {BenZvi}, {Beutler}, {Bielby},
  {Bolton}, {Brice{\~n}o}, {Buckley-Geer}, {Butler}, {Calamida}, {Carlberg},
  {Carter}, {Casas}, {Castander}, {Choi}, {Comparat}, {Cukanovaite}, {Delubac},
  {DeVries}, {Dey}, {Dhungana}, {Dickinson}, {Ding}, {Donaldson}, {Duan},
  {Duckworth}, {Eftekharzadeh}, {Eisenstein}, {Etourneau}, {Fagrelius},
  {Farihi}, {Fitzpatrick}, {Font-Ribera}, {Fulmer}, {G{\"a}nsicke},
  {Gaztanaga}, {George}, {Gerdes}, {Gontcho}, {Gorgoni}, {Green}, {Guy},
  {Harmer}, {Hernandez}, {Honscheid}, {Huang}, {James}, {Jannuzi}, {Jiang},
  {Joyce}, {Karcher}, {Karkar}, {Kehoe}, {Kneib}, {Kueter-Young}, {Lan},
  {Lauer}, {Le Guillou}, {Le Van Suu}, {Lee}, {Lesser}, {Perreault Levasseur},
  {Li}, {Mann}, {Marshall}, {Mart{\'\i}nez-V{\'a}zquez}, {Martini}, {du Mas des
  Bourboux}, {McManus}, {Meier}, {M{\'e}nard}, {Metcalfe},
  {Mu{\~n}oz-Guti{\'e}rrez}, {Najita}, {Napier}, {Narayan}, {Newman}, {Nie},
  {Nord}, {Norman}, {Olsen}, {Paat}, {Palanque-Delabrouille}, {Peng},
  {Poppett}, {Poremba}, {Prakash}, {Rabinowitz}, {Raichoor}, {Rezaie},
  {Robertson}, {Roe}, {Ross}, {Ross}, {Rudnick}, {Safonova}, {Saha},
  {S{\'a}nchez}, {Savary}, {Schweiker}, {Scott}, {Seo}, {Shan}, {Silva},
  {Slepian}, {Soto}, {Sprayberry}, {Staten}, {Stillman}, {Stupak}, {Summers},
  {Sien Tie}, {Tirado}, {Vargas-Maga{\~n}a}, {Vivas}, {Wechsler}, {Williams},
  {Yang}, {Yang}, {Yapici}, {Zaritsky}, {Zenteno}, {Zhang}, {Zhang}, {Zhou}, \&
  {Zhou}}]{dey2019}
{Dey}, A., {Schlegel}, D.~J., {Lang}, D., {et~al.} 2019, \aj, 157, 168,
  \dodoi{10.3847/1538-3881/ab089d}

\bibitem[{{Dotter}(2016)}]{dotter2016}
{Dotter}, A. 2016, \apjs, 222, 8, \dodoi{10.3847/0067-0049/222/1/8}

\bibitem[{{Dressler}(1980)}]{dressler1980}
{Dressler}, A. 1980, \apj, 236, 351, \dodoi{10.1086/157753}

\bibitem[{{Drory} {et~al.}(2015){Drory}, {MacDonald}, {Bershady}, {Bundy},
  {Gunn}, {Law}, {Smith}, {Stoll}, {Tremonti}, {Wake}, {Yan}, {Weijmans},
  {Byler}, {Cherinka}, {Cope}, {Eigenbrot}, {Harding}, {Holder}, {Huehnerhoff},
  {Jaehnig}, {Jansen}, {Klaene}, {Paat}, {Percival}, \& {Sayres}}]{drory2015}
{Drory}, N., {MacDonald}, N., {Bershady}, M.~A., {et~al.} 2015, \aj, 149, 77,
  \dodoi{10.1088/0004-6256/149/2/77}

\bibitem[{{Dunkley} {et~al.}(2009){Dunkley}, {Komatsu}, {Nolta}, {Spergel},
  {Larson}, {Hinshaw}, {Page}, {Bennett}, {Gold}, {Jarosik}, {Weiland},
  {Halpern}, {Hill}, {Kogut}, {Limon}, {Meyer}, {Tucker}, {Wollack}, \&
  {Wright}}]{dunkley2009}
{Dunkley}, J., {Komatsu}, E., {Nolta}, M.~R., {et~al.} 2009, \apjs, 180, 306,
  \dodoi{10.1088/0067-0049/180/2/306}

\bibitem[{{Eggen} {et~al.}(1962){Eggen}, {Lynden-Bell}, \&
  {Sandage}}]{eggen1962}
{Eggen}, O.~J., {Lynden-Bell}, D., \& {Sandage}, A.~R. 1962, \apj, 136, 748,
  \dodoi{10.1086/147433}

\bibitem[{{El-Badry} {et~al.}(2016){El-Badry}, {Wetzel}, {Geha}, {Hopkins},
  {Kere{\v s}}, {Chan}, \& {Faucher-Gigu{\`e}re}}]{el-badry2016}
{El-Badry}, K., {Wetzel}, A., {Geha}, M., {et~al.} 2016, \apj, 820, 131,
  \dodoi{10.3847/0004-637X/820/2/131}

\bibitem[{{Estrada-Carpenter} {et~al.}(2020){Estrada-Carpenter}, {Papovich},
  {Momcheva}, {Brammer}, {Simons}, {Bridge}, {Cleri}, {Ferguson},
  {Finkelstein}, {Giavalisco}, {Jung}, {Matharu}, {Trump}, \&
  {Weiner}}]{estrada-carpenter2020}
{Estrada-Carpenter}, V., {Papovich}, C., {Momcheva}, I., {et~al.} 2020, \apj,
  898, 171, \dodoi{10.3847/1538-4357/aba004}

\bibitem[{{Faber} {et~al.}(1985){Faber}, {Friel}, {Burstein}, \&
  {Gaskell}}]{faber1985}
{Faber}, S.~M., {Friel}, E.~D., {Burstein}, D., \& {Gaskell}, C.~M. 1985,
  \apjs, 57, 711, \dodoi{10.1086/191024}

\bibitem[{{Faber} \& {Jackson}(1976{\natexlab{a}})}]{faber1976}
{Faber}, S.~M., \& {Jackson}, R.~E. 1976{\natexlab{a}}, \apj, 204, 668,
  \dodoi{10.1086/154215}

\bibitem[{{Faber} \& {Jackson}(1976{\natexlab{b}})}]{faber-jackson1976}
---. 1976{\natexlab{b}}, \apj, 204, 668, \dodoi{10.1086/154215}

\bibitem[{{Fern{\'a}ndez-Trincado} {et~al.}(2020){Fern{\'a}ndez-Trincado},
  {Beers}, \& {Minniti}}]{fernandez-trincado2020}
{Fern{\'a}ndez-Trincado}, J.~G., {Beers}, T.~C., \& {Minniti}, D. 2020, \aap,
  644, A83, \dodoi{10.1051/0004-6361/202039434}

\bibitem[{{Fern{\'a}ndez-Trincado} {et~al.}(2019){Fern{\'a}ndez-Trincado},
  {Beers}, {Placco}, {Moreno}, {Alves-Brito}, {Minniti}, {Tang},
  {P{\'e}rez-Villegas}, {Reyl{\'e}}, {Robin}, \&
  {Villanova}}]{fernandez-trincado2019}
{Fern{\'a}ndez-Trincado}, J.~G., {Beers}, T.~C., {Placco}, V.~M., {et~al.}
  2019, \apjl, 886, L8, \dodoi{10.3847/2041-8213/ab5286}

\bibitem[{{Foreman-Mackey} {et~al.}(2013){Foreman-Mackey}, {Hogg}, {Lang}, \&
  {Goodman}}]{emcee}
{Foreman-Mackey}, D., {Hogg}, D.~W., {Lang}, D., \& {Goodman}, J. 2013, \pasp,
  125, 306, \dodoi{10.1086/670067}

\bibitem[{{Franx} {et~al.}(2008){Franx}, {van Dokkum}, {F{\"o}rster Schreiber},
  {Wuyts}, {Labb{\'e}}, \& {Toft}}]{franx2008}
{Franx}, M., {van Dokkum}, P.~G., {F{\"o}rster Schreiber}, N.~M., {et~al.}
  2008, \apj, 688, 770, \dodoi{10.1086/592431}

\bibitem[{{Furlong} {et~al.}(2017){Furlong}, {Bower}, {Crain}, {Schaye},
  {Theuns}, {Trayford}, {Qu}, {Schaller}, {Berthet}, \& {Helly}}]{furlong2017}
{Furlong}, M., {Bower}, R.~G., {Crain}, R.~A., {et~al.} 2017, \mnras, 465, 722,
  \dodoi{10.1093/mnras/stw2740}

\bibitem[{{Gallazzi} {et~al.}(2005){Gallazzi}, {Charlot}, {Brinchmann},
  {White}, \& {Tremonti}}]{gallazzi2005}
{Gallazzi}, A., {Charlot}, S., {Brinchmann}, J., {White}, S.~D.~M., \&
  {Tremonti}, C.~A. 2005, \mnras, 362, 41,
  \dodoi{10.1111/j.1365-2966.2005.09321.x}

\bibitem[{{Gao} {et~al.}(2005){Gao}, {Springel}, \& {White}}]{gao2005}
{Gao}, L., {Springel}, V., \& {White}, S.~D.~M. 2005, \mnras, 363, L66,
  \dodoi{10.1111/j.1745-3933.2005.00084.x}

\bibitem[{{Garc{\'\i}a-Benito} {et~al.}(2017){Garc{\'\i}a-Benito},
  {Gonz{\'a}lez Delgado}, {P{\'e}rez}, {Cid Fernandes}, {Cortijo-Ferrero},
  {L{\'o}pez Fern{\'a}ndez}, {de Amorim}, {Lacerda}, {Vale Asari}, \&
  {S{\'a}nchez}}]{garcia-benito2017}
{Garc{\'\i}a-Benito}, R., {Gonz{\'a}lez Delgado}, R.~M., {P{\'e}rez}, E.,
  {et~al.} 2017, \aap, 608, A27, \dodoi{10.1051/0004-6361/201731357}

\bibitem[{{Girelli} {et~al.}(2020){Girelli}, {Pozzetti}, {Bolzonella},
  {Giocoli}, {Marulli}, \& {Baldi}}]{girelli2020}
{Girelli}, G., {Pozzetti}, L., {Bolzonella}, M., {et~al.} 2020, \aap, 634,
  A135, \dodoi{10.1051/0004-6361/201936329}

\bibitem[{{Goddard} {et~al.}(2017{\natexlab{a}}){Goddard}, {Thomas},
  {Maraston}, {Westfall}, {Etherington}, {Riffel}, {Mallmann}, {Zheng},
  {Argudo-Fern{\'a}ndez}, {Lian}, {Bershady}, {Bundy}, {Drory}, {Law}, {Yan},
  {Wake}, {Weijmans}, {Bizyaev}, {Brownstein}, {Lane}, {Maiolino}, {Masters},
  {Merrifield}, {Nitschelm}, {Pan}, {Roman-Lopes}, {Storchi-Bergmann}, \&
  {Schneider}}]{goddard2017a}
{Goddard}, D., {Thomas}, D., {Maraston}, C., {et~al.} 2017{\natexlab{a}},
  \mnras, 466, 4731, \dodoi{10.1093/mnras/stw3371}

\bibitem[{{Goddard} {et~al.}(2017{\natexlab{b}}){Goddard}, {Thomas},
  {Maraston}, {Westfall}, {Etherington}, {Riffel}, {Mallmann}, {Zheng},
  {Argudo-Fern{\'a}ndez}, {Bershady}, {Bundy}, {Drory}, {Law}, {Yan}, {Wake},
  {Weijmans}, {Bizyaev}, {Brownstein}, {Lane}, {Maiolino}, {Masters},
  {Merrifield}, {Nitschelm}, {Pan}, {Roman-Lopes}, \&
  {Storchi-Bergmann}}]{goddard2017b}
---. 2017{\natexlab{b}}, \mnras, 465, 688, \dodoi{10.1093/mnras/stw2719}

\bibitem[{{Gonz{\'a}lez Delgado} {et~al.}(2014){Gonz{\'a}lez Delgado}, {Cid
  Fernandes}, {Garc{\'{\i}}a-Benito}, {P{\'e}rez}, {de Amorim},
  {Cortijo-Ferrero}, {Lacerda}, {L{\'o}pez Fern{\'a}ndez}, {S{\'a}nchez}, {Vale
  Asari}, {Alves}, {Bland-Hawthorn}, {Galbany}, {Gallazzi}, {Husemann},
  {Bekeraite}, {Jungwiert}, {L{\'o}pez-S{\'a}nchez}, {de Lorenzo-C{\'a}ceres},
  {Marino}, {Mast}, {Moll{\'a}}, {del Olmo}, {S{\'a}nchez-Bl{\'a}zquez}, {van
  de Ven}, {V{\'{\i}}lchez}, {Walcher}, {Wisotzki}, {Ziegler}, \&
  {Collaboration920}}]{gonzalez-delgado2014}
{Gonz{\'a}lez Delgado}, R.~M., {Cid Fernandes}, R., {Garc{\'{\i}}a-Benito}, R.,
  {et~al.} 2014, \apjl, 791, L16, \dodoi{10.1088/2041-8205/791/1/L16}

\bibitem[{{Graham} {et~al.}(2018){Graham}, {Cappellari}, {Li}, {Mao},
  {Bershady}, {Bizyaev}, {Brinkmann}, {Brownstein}, {Bundy}, {Drory}, {Law},
  {Pan}, {Thomas}, {Wake}, {Weijmans}, {Westfall}, \& {Yan}}]{graham2018}
{Graham}, M.~T., {Cappellari}, M., {Li}, H., {et~al.} 2018, \mnras, 477, 4711,
  \dodoi{10.1093/mnras/sty504}

\bibitem[{{Greene} {et~al.}(2015){Greene}, {Janish}, {Ma}, {McConnell},
  {Blakeslee}, {Thomas}, \& {Murphy}}]{greene2015}
{Greene}, J.~E., {Janish}, R., {Ma}, C.-P., {et~al.} 2015, \apj, 807, 11,
  \dodoi{10.1088/0004-637X/807/1/11}

\bibitem[{{Greene} {et~al.}(2013){Greene}, {Murphy}, {Graves}, {Gunn},
  {Raskutti}, {Comerford}, \& {Gebhardt}}]{greene2013}
{Greene}, J.~E., {Murphy}, J.~D., {Graves}, G.~J., {et~al.} 2013, \apj, 776,
  64, \dodoi{10.1088/0004-637X/776/2/64}

\bibitem[{{Gunn} {et~al.}(2006){Gunn}, {Siegmund}, {Mannery}, {Owen}, {Hull},
  {Leger}, {Carey}, {Knapp}, {York}, {Boroski}, {Kent}, {Lupton}, {Rockosi},
  {Evans}, {Waddell}, {Anderson}, {Annis}, {Barentine}, {Bartoszek}, {Bastian},
  {Bracker}, {Brewington}, {Briegel}, {Brinkmann}, {Brown}, {Carr},
  {Czarapata}, {Drennan}, {Dombeck}, {Federwitz}, {Gillespie}, {Gonzales},
  {Hansen}, {Harvanek}, {Hayes}, {Jordan}, {Kinney}, {Klaene}, {Kleinman},
  {Kron}, {Kresinski}, {Lee}, {Limmongkol}, {Lindenmeyer}, {Long}, {Loomis},
  {McGehee}, {Mantsch}, {Neilsen}, {Neswold}, {Newman}, {Nitta}, {Peoples},
  {Pier}, {Prieto}, {Prosapio}, {Rivetta}, {Schneider}, {Snedden}, \&
  {Wang}}]{gunn2006}
{Gunn}, J.~E., {Siegmund}, W.~A., {Mannery}, E.~J., {et~al.} 2006, \aj, 131,
  2332, \dodoi{10.1086/500975}

\bibitem[{{Hearin} {et~al.}(2016{\natexlab{a}}){Hearin}, {Behroozi}, \& {van
  den Bosch}}]{hearin2016b}
{Hearin}, A.~P., {Behroozi}, P.~S., \& {van den Bosch}, F.~C.
  2016{\natexlab{a}}, \mnras, 461, 2135, \dodoi{10.1093/mnras/stw1462}

\bibitem[{{Hearin} {et~al.}(2016{\natexlab{b}}){Hearin}, {Zentner}, {van den
  Bosch}, {Campbell}, \& {Tollerud}}]{hearin2016}
{Hearin}, A.~P., {Zentner}, A.~R., {van den Bosch}, F.~C., {Campbell}, D., \&
  {Tollerud}, E. 2016{\natexlab{b}}, \mnras, 460, 2552,
  \dodoi{10.1093/mnras/stw840}

\bibitem[{{Hopkins} {et~al.}(2010{\natexlab{a}}){Hopkins}, {Bundy},
  {Hernquist}, {Wuyts}, \& {Cox}}]{hopkins2010}
{Hopkins}, P.~F., {Bundy}, K., {Hernquist}, L., {Wuyts}, S., \& {Cox}, T.~J.
  2010{\natexlab{a}}, \mnras, 401, 1099,
  \dodoi{10.1111/j.1365-2966.2009.15699.x}

\bibitem[{{Hopkins} {et~al.}(2010{\natexlab{b}}){Hopkins}, {Bundy}, {Croton},
  {Hernquist}, {Keres}, {Khochfar}, {Stewart}, {Wetzel}, \&
  {Younger}}]{hopkins2010b}
{Hopkins}, P.~F., {Bundy}, K., {Croton}, D., {et~al.} 2010{\natexlab{b}}, \apj,
  715, 202, \dodoi{10.1088/0004-637X/715/1/202}

\bibitem[{{Huang} {et~al.}(2013{\natexlab{a}}){Huang}, {Ho}, {Peng}, {Li}, \&
  {Barth}}]{huang2013a}
{Huang}, S., {Ho}, L.~C., {Peng}, C.~Y., {Li}, Z.-Y., \& {Barth}, A.~J.
  2013{\natexlab{a}}, \apj, 766, 47, \dodoi{10.1088/0004-637X/766/1/47}

\bibitem[{{Huang} {et~al.}(2013{\natexlab{b}}){Huang}, {Ho}, {Peng}, {Li}, \&
  {Barth}}]{huang2013b}
---. 2013{\natexlab{b}}, \apjl, 768, L28, \dodoi{10.1088/2041-8205/768/2/L28}

\bibitem[{{Huang} {et~al.}(2018){Huang}, {Leauthaud}, {Greene}, {Bundy}, {Lin},
  {Tanaka}, {Miyazaki}, \& {Komiyama}}]{huang2018}
{Huang}, S., {Leauthaud}, A., {Greene}, J.~E., {et~al.} 2018, \mnras, 475,
  3348, \dodoi{10.1093/mnras/stx3200}

\bibitem[{{Huang} {et~al.}(2020){Huang}, {Leauthaud}, {Hearin}, {Behroozi},
  {Bradshaw}, {Ardila}, {Speagle}, {Tenneti}, {Bundy}, {Greene}, {Sif{\'o}n},
  \& {Bahcall}}]{huang2020}
{Huang}, S., {Leauthaud}, A., {Hearin}, A., {et~al.} 2020, \mnras, 492, 3685,
  \dodoi{10.1093/mnras/stz3314}

\bibitem[{{Johansson} {et~al.}(2012{\natexlab{a}}){Johansson}, {Thomas}, \&
  {Maraston}}]{johansson2012b}
{Johansson}, J., {Thomas}, D., \& {Maraston}, C. 2012{\natexlab{a}}, \mnras,
  421, 1908, \dodoi{10.1111/j.1365-2966.2011.20316.x}

\bibitem[{{Johansson} {et~al.}(2012{\natexlab{b}}){Johansson}, {Naab}, \&
  {Ostriker}}]{johansson2012}
{Johansson}, P.~H., {Naab}, T., \& {Ostriker}, J.~P. 2012{\natexlab{b}}, \apj,
  754, 115, \dodoi{10.1088/0004-637X/754/2/115}

\bibitem[{{Johnson} {et~al.}(2020){Johnson}, {Leja}, {Conroy}, \&
  {Speagle}}]{johnson2020}
{Johnson}, B.~D., {Leja}, J., {Conroy}, C., \& {Speagle}, J.~S. 2020, arXiv
  e-prints, arXiv:2012.01426.
\newblock \doarXiv{2012.01426}

\bibitem[{{Kauffmann} {et~al.}(2004){Kauffmann}, {White}, {Heckman},
  {M{\'e}nard}, {Brinchmann}, {Charlot}, {Tremonti}, \&
  {Brinkmann}}]{kauffmann2004}
{Kauffmann}, G., {White}, S. D.~M., {Heckman}, T.~M., {et~al.} 2004, \mnras,
  353, 713, \dodoi{10.1111/j.1365-2966.2004.08117.x}

\bibitem[{{Kirby} {et~al.}(2008){Kirby}, {Guhathakurta}, \&
  {Sneden}}]{kirby2008}
{Kirby}, E.~N., {Guhathakurta}, P., \& {Sneden}, C. 2008, \apj, 682, 1217,
  \dodoi{10.1086/589627}

\bibitem[{{Klypin} {et~al.}(2016){Klypin}, {Yepes}, {Gottl{\"o}ber}, {Prada},
  \& {He{\ss}}}]{klypin2016}
{Klypin}, A., {Yepes}, G., {Gottl{\"o}ber}, S., {Prada}, F., \& {He{\ss}}, S.
  2016, \mnras, 457, 4340, \dodoi{10.1093/mnras/stw248}

\bibitem[{{Kobayashi}(2004)}]{kobayashi2004}
{Kobayashi}, C. 2004, \mnras, 347, 740,
  \dodoi{10.1111/j.1365-2966.2004.07258.x}

\bibitem[{{Kriek} \& {Conroy}(2013)}]{kriek-conroy2013}
{Kriek}, M., \& {Conroy}, C. 2013, \apjl, 775, L16,
  \dodoi{10.1088/2041-8205/775/1/L16}

\bibitem[{{Kroupa}(2001)}]{kroupa2001}
{Kroupa}, P. 2001, \mnras, 322, 231, \dodoi{10.1046/j.1365-8711.2001.04022.x}

\bibitem[{{Kulier} {et~al.}(2019){Kulier}, {Padilla}, {Schaye}, {Crain},
  {Schaller}, {Bower}, {Theuns}, \& {Paillas}}]{kulier2019}
{Kulier}, A., {Padilla}, N., {Schaye}, J., {et~al.} 2019, \mnras, 482, 3261,
  \dodoi{10.1093/mnras/sty2914}

\bibitem[{{Kurucz}(2018)}]{kurucz2018}
{Kurucz}, R.~L. 2018, in Astronomical Society of the Pacific Conference Series,
  Vol. 515, Workshop on Astrophysical Opacities, 47

\bibitem[{{La Barbera} {et~al.}(2014){La Barbera}, {Pasquali}, {Ferreras},
  {Gallazzi}, {de Carvalho}, \& {de la Rosa}}]{labarbera2014}
{La Barbera}, F., {Pasquali}, A., {Ferreras}, I., {et~al.} 2014, \mnras, 445,
  1977, \dodoi{10.1093/mnras/stu1626}

\bibitem[{{Lacerna} {et~al.}(2020){Lacerna}, {Ibarra-Medel}, {Avila-Reese},
  {Hern{\'a}ndez-Toledo}, {V{\'a}zquez-Mata}, \& {S{\'a}nchez}}]{lacerna2020}
{Lacerna}, I., {Ibarra-Medel}, H., {Avila-Reese}, V., {et~al.} 2020, \aap, 644,
  A117, \dodoi{10.1051/0004-6361/202037503}

\bibitem[{{Lacerna} {et~al.}(2014){Lacerna}, {Padilla}, \&
  {Stasyszyn}}]{lacerna2014}
{Lacerna}, I., {Padilla}, N., \& {Stasyszyn}, F. 2014, \mnras, 443, 3107,
  \dodoi{10.1093/mnras/stu1318}

\bibitem[{{Larson}(1974)}]{larson1974}
{Larson}, R.~B. 1974, \mnras, 169, 229, \dodoi{10.1093/mnras/169.2.229}

\bibitem[{{Law} {et~al.}(2015){Law}, {Yan}, {Bershady}, {Bundy}, {Cherinka},
  {Drory}, {MacDonald}, {S{\'a}nchez-Gallego}, {Wake}, {Weijmans}, {Blanton},
  {Klaene}, {Moran}, {Sanchez}, \& {Zhang}}]{law2015}
{Law}, D.~R., {Yan}, R., {Bershady}, M.~A., {et~al.} 2015, \aj, 150, 19,
  \dodoi{10.1088/0004-6256/150/1/19}

\bibitem[{{Law} {et~al.}(2016){Law}, {Cherinka}, {Yan}, {Andrews}, {Bershady},
  {Bizyaev}, {Blanc}, {Blanton}, {Bolton}, {Brownstein}, {Bundy}, {Chen},
  {Drory}, {D'Souza}, {Fu}, {Jones}, {Kauffmann}, {MacDonald}, {Masters},
  {Newman}, {Parejko}, {S{\'a}nchez-Gallego}, {S{\'a}nchez}, {Schlegel},
  {Thomas}, {Wake}, {Weijmans}, {Westfall}, \& {Zhang}}]{law2016}
{Law}, D.~R., {Cherinka}, B., {Yan}, R., {et~al.} 2016, \aj, 152, 83,
  \dodoi{10.3847/0004-6256/152/4/83}

\bibitem[{{Law} {et~al.}(2021){Law}, {Westfall}, {Bershady}, {Cappellari},
  {Yan}, {Belfiore}, {Bizyaev}, {Brownstein}, {Chen}, {Cherinka}, {Drory},
  {Lazarz}, \& {Shetty}}]{law2021}
{Law}, D.~R., {Westfall}, K.~B., {Bershady}, M.~A., {et~al.} 2021, \aj, 161,
  52, \dodoi{10.3847/1538-3881/abcaa2}

\bibitem[{{Leauthaud} {et~al.}(2017){Leauthaud}, {Saito}, {Hilbert},
  {Barreira}, {More}, {White}, {Alam}, {Behroozi}, {Bundy}, {Coupon}, {Erben},
  {Heymans}, {Hildebrandt}, {Mandelbaum}, {Miller}, {Moraes}, {Pereira},
  {Rodr{\'\i}guez-Torres}, {Schmidt}, {Shan}, {Viel}, \&
  {Villaescusa-Navarro}}]{leauthaud2017}
{Leauthaud}, A., {Saito}, S., {Hilbert}, S., {et~al.} 2017, \mnras, 467, 3024,
  \dodoi{10.1093/mnras/stx258}

\bibitem[{{Lehmann} {et~al.}(2017){Lehmann}, {Mao}, {Becker}, {Skillman}, \&
  {Wechsler}}]{lehmann2017}
{Lehmann}, B.~V., {Mao}, Y.-Y., {Becker}, M.~R., {Skillman}, S.~W., \&
  {Wechsler}, R.~H. 2017, \apj, 834, 37, \dodoi{10.3847/1538-4357/834/1/37}

\bibitem[{{Leja} {et~al.}(2019){Leja}, {Carnall}, {Johnson}, {Conroy}, \&
  {Speagle}}]{leja2019}
{Leja}, J., {Carnall}, A.~C., {Johnson}, B.~D., {Conroy}, C., \& {Speagle},
  J.~S. 2019, \apj, 876, 3, \dodoi{10.3847/1538-4357/ab133c}

\bibitem[{{Leja} {et~al.}(2017){Leja}, {Johnson}, {Conroy}, {van Dokkum}, \&
  {Byler}}]{leja2017}
{Leja}, J., {Johnson}, B.~D., {Conroy}, C., {van Dokkum}, P.~G., \& {Byler}, N.
  2017, \apj, 837, 170, \dodoi{10.3847/1538-4357/aa5ffe}

\bibitem[{{Li} {et~al.}(2008){Li}, {Mo}, \& {Gao}}]{li2008}
{Li}, Y., {Mo}, H.~J., \& {Gao}, L. 2008, \mnras, 389, 1419,
  \dodoi{10.1111/j.1365-2966.2008.13667.x}

\bibitem[{{Lin} {et~al.}(2016){Lin}, {Mandelbaum}, {Huang}, {Huang}, {Dalal},
  {Diemer}, {Jian}, \& {Kravtsov}}]{lin2016}
{Lin}, Y.-T., {Mandelbaum}, R., {Huang}, Y.-H., {et~al.} 2016, \apj, 819, 119,
  \dodoi{10.3847/0004-637X/819/2/119}

\bibitem[{{Mansfield} \& {Kravtsov}(2020)}]{mansfield2020}
{Mansfield}, P., \& {Kravtsov}, A.~V. 2020, \mnras, 493, 4763,
  \dodoi{10.1093/mnras/staa430}

\bibitem[{{Mao} {et~al.}(2017){Mao}, {Zentner}, \& {Wechsler}}]{mao2017}
{Mao}, Y.-Y., {Zentner}, A.~R., \& {Wechsler}, R.~H. 2017, ArXiv e-prints.
\newblock \doarXiv{1705.03888}

\bibitem[{{Matteucci}(1994)}]{matteucci1994}
{Matteucci}, F. 1994, \aap, 288, 57

\bibitem[{{Matthee} {et~al.}(2017){Matthee}, {Schaye}, {Crain}, {Schaller},
  {Bower}, \& {Theuns}}]{matthee2017}
{Matthee}, J., {Schaye}, J., {Crain}, R.~A., {et~al.} 2017, \mnras, 465, 2381,
  \dodoi{10.1093/mnras/stw2884}

\bibitem[{{McDermid} {et~al.}(2015){McDermid}, {Alatalo}, {Blitz}, {Bournaud},
  {Bureau}, {Cappellari}, {Crocker}, {Davies}, {Davis}, {de Zeeuw}, {Duc},
  {Emsellem}, {Khochfar}, {Krajnovi{\'c}}, {Kuntschner}, {Morganti}, {Naab},
  {Oosterloo}, {Sarzi}, {Scott}, {Serra}, {Weijmans}, \&
  {Young}}]{mcdermid2015}
{McDermid}, R.~M., {Alatalo}, K., {Blitz}, L., {et~al.} 2015, \mnras, 448,
  3484, \dodoi{10.1093/mnras/stv105}

\bibitem[{{Minchev} {et~al.}(2012){Minchev}, {Famaey}, {Quillen}, {Di Matteo},
  {Combes}, {Vlaji{\'c}}, {Erwin}, \& {Bland-Hawthorn}}]{minchev2012}
{Minchev}, I., {Famaey}, B., {Quillen}, A.~C., {et~al.} 2012, \aap, 548, A126,
  \dodoi{10.1051/0004-6361/201219198}

\bibitem[{{Montero-Dorta} {et~al.}(2021){Montero-Dorta}, {Chaves-Montero},
  {Artale}, \& {Favole}}]{montero-dorta2021}
{Montero-Dorta}, A.~D., {Chaves-Montero}, J., {Artale}, M.~C., \& {Favole}, G.
  2021, arXiv e-prints, arXiv:2105.05274.
\newblock \doarXiv{2105.05274}

\bibitem[{{Montero-Dorta} {et~al.}(2017){Montero-Dorta}, {P{\'e}rez}, {Prada},
  {Rodr{\'\i}guez-Torres}, {Favole}, {Klypin}, {Cid Fernandes}, {Gonz{\'a}lez
  Delgado}, {Dom{\'\i}nguez}, {Bolton}, {Garc{\'\i}a-Benito}, {Jullo}, \&
  {Niemiec}}]{montero-dorta2017}
{Montero-Dorta}, A.~D., {P{\'e}rez}, E., {Prada}, F., {et~al.} 2017, \apjl,
  848, L2, \dodoi{10.3847/2041-8213/aa8cc5}

\bibitem[{{Montero-Dorta} {et~al.}(2020){Montero-Dorta}, {Artale}, {Abramo},
  {Tucci}, {Padilla}, {Sato-Polito}, {Lacerna}, {Rodriguez}, \&
  {Angulo}}]{montero-dorta2020}
{Montero-Dorta}, A.~D., {Artale}, M.~C., {Abramo}, L.~R., {et~al.} 2020,
  \mnras, 496, 1182, \dodoi{10.1093/mnras/staa1624}

\bibitem[{{Moster} {et~al.}(2013){Moster}, {Naab}, \& {White}}]{moster2013}
{Moster}, B.~P., {Naab}, T., \& {White}, S.~D.~M. 2013, \mnras, 428, 3121,
  \dodoi{10.1093/mnras/sts261}

\bibitem[{{Moster} {et~al.}(2010){Moster}, {Somerville}, {Maulbetsch}, {van den
  Bosch}, {Macci{\`o}}, {Naab}, \& {Oser}}]{moster2010}
{Moster}, B.~P., {Somerville}, R.~S., {Maulbetsch}, C., {et~al.} 2010, \apj,
  710, 903, \dodoi{10.1088/0004-637X/710/2/903}

\bibitem[{{Naab} {et~al.}(2014){Naab}, {Oser}, {Emsellem}, {Cappellari},
  {Krajnovi{\'c}}, {McDermid}, {Alatalo}, {Bayet}, {Blitz}, {Bois}, {Bournaud},
  {Bureau}, {Crocker}, {Davies}, {Davis}, {de Zeeuw}, {Duc}, {Hirschmann},
  {Johansson}, {Khochfar}, {Kuntschner}, {Morganti}, {Oosterloo}, {Sarzi},
  {Scott}, {Serra}, {van de Ven}, {Weijmans}, \& {Young}}]{naab2014}
{Naab}, T., {Oser}, L., {Emsellem}, E., {et~al.} 2014, \mnras, 444, 3357,
  \dodoi{10.1093/mnras/stt1919}

\bibitem[{{Nelson} {et~al.}(2015){Nelson}, {Pillepich}, {Genel},
  {Vogelsberger}, {Springel}, {Torrey}, {Rodriguez-Gomez}, {Sijacki}, {Snyder},
  {Griffen}, {Marinacci}, {Blecha}, {Sales}, {Xu}, \& {Hernquist}}]{nelson2015}
{Nelson}, D., {Pillepich}, A., {Genel}, S., {et~al.} 2015, Astronomy and
  Computing, 13, 12, \dodoi{10.1016/j.ascom.2015.09.003}

\bibitem[{{Newman} {et~al.}(2010){Newman}, {Ellis}, {Treu}, \&
  {Bundy}}]{newman2010}
{Newman}, A.~B., {Ellis}, R.~S., {Treu}, T., \& {Bundy}, K. 2010, \apjl, 717,
  L103, \dodoi{10.1088/2041-8205/717/2/L103}

\bibitem[{{Oke} \& {Gunn}(1983)}]{oke1983}
{Oke}, J.~B., \& {Gunn}, J.~E. 1983, \apj, 266, 713, \dodoi{10.1086/160817}

\bibitem[{{Oser} {et~al.}(2012){Oser}, {Naab}, {Ostriker}, \&
  {Johansson}}]{oser2012}
{Oser}, L., {Naab}, T., {Ostriker}, J.~P., \& {Johansson}, P.~H. 2012, \apj,
  744, 63, \dodoi{10.1088/0004-637X/744/1/63}

\bibitem[{{Oser} {et~al.}(2010){Oser}, {Ostriker}, {Naab}, {Johansson}, \&
  {Burkert}}]{oser2010}
{Oser}, L., {Ostriker}, J.~P., {Naab}, T., {Johansson}, P.~H., \& {Burkert}, A.
  2010, \apj, 725, 2312, \dodoi{10.1088/0004-637X/725/2/2312}

\bibitem[{{Oyarz{\'u}n} {et~al.}(2017){Oyarz{\'u}n}, {Blanc}, {Gonz{\'a}lez},
  {Mateo}, \& {Bailey}}]{oyarzun2017}
{Oyarz{\'u}n}, G.~A., {Blanc}, G.~A., {Gonz{\'a}lez}, V., {Mateo}, M., \&
  {Bailey}, John~I., I. 2017, \apj, 843, 133, \dodoi{10.3847/1538-4357/aa7552}

\bibitem[{{Oyarz{\'u}n} {et~al.}(2019){Oyarz{\'u}n}, {Bundy}, {Westfall},
  {Belfiore}, {Thomas}, {Maraston}, {Lian}, {Arag{\'o}n-Salamanca}, {Zheng},
  {Gonzalez-Perez}, {Law}, {Drory}, \& {Andrews}}]{oyarzun2019}
{Oyarz{\'u}n}, G.~A., {Bundy}, K., {Westfall}, K.~B., {et~al.} 2019, \apj, 880,
  111, \dodoi{10.3847/1538-4357/ab297c}

\bibitem[{{Parikh} {et~al.}(2021){Parikh}, {Thomas}, {Maraston}, {Westfall},
  {Andrews}, {Boardman}, {Drory}, \& {Oyarzun}}]{parikh2021}
{Parikh}, T., {Thomas}, D., {Maraston}, C., {et~al.} 2021, arXiv e-prints,
  arXiv:2102.06703.
\newblock \doarXiv{2102.06703}

\bibitem[{{Parikh} {et~al.}(2018){Parikh}, {Thomas}, {Maraston}, {Westfall},
  {Goddard}, {Lian}, {Meneses-Goytia}, {Jones}, {Vaughan}, {Andrews},
  {Bershady}, {Bizyaev}, {Brinkmann}, {Brownstein}, {Bundy}, {Drory},
  {Emsellem}, {Law}, {Newman}, {Roman-Lopes}, {Wake}, {Yan}, \&
  {Zheng}}]{parikh2018}
---. 2018, \mnras, 477, 3954, \dodoi{10.1093/mnras/sty785}

\bibitem[{{Parikh} {et~al.}(2019){Parikh}, {Thomas}, {Maraston}, {Westfall},
  {Lian}, {Fraser-McKelvie}, {Andrews}, {Drory}, \&
  {Meneses-Goytia}}]{parikh2019}
---. 2019, \mnras, 483, 3420, \dodoi{10.1093/mnras/sty3339}

\bibitem[{{Pasquali} {et~al.}(2010){Pasquali}, {Gallazzi}, {Fontanot}, {van den
  Bosch}, {De Lucia}, {Mo}, \& {Yang}}]{pasquali2010}
{Pasquali}, A., {Gallazzi}, A., {Fontanot}, F., {et~al.} 2010, \mnras, 407,
  937, \dodoi{10.1111/j.1365-2966.2010.17074.x}

\bibitem[{{Peng} {et~al.}(2012){Peng}, {Lilly}, {Renzini}, \&
  {Carollo}}]{peng2012}
{Peng}, Y.-j., {Lilly}, S.~J., {Renzini}, A., \& {Carollo}, M. 2012, \apj, 757,
  4, \dodoi{10.1088/0004-637X/757/1/4}

\bibitem[{{Peng} {et~al.}(2010){Peng}, {Lilly}, {Kova{\v{c}}}, {Bolzonella},
  {Pozzetti}, {Renzini}, {Zamorani}, {Ilbert}, {Knobel}, {Iovino}, {Maier},
  {Cucciati}, {Tasca}, {Carollo}, {Silverman}, {Kampczyk}, {de Ravel},
  {Sanders}, {Scoville}, {Contini}, {Mainieri}, {Scodeggio}, {Kneib}, {Le
  F{\`e}vre}, {Bardelli}, {Bongiorno}, {Caputi}, {Coppa}, {de la Torre},
  {Franzetti}, {Garilli}, {Lamareille}, {Le Borgne}, {Le Brun}, {Mignoli},
  {Perez Montero}, {Pello}, {Ricciardelli}, {Tanaka}, {Tresse}, {Vergani},
  {Welikala}, {Zucca}, {Oesch}, {Abbas}, {Barnes}, {Bordoloi}, {Bottini},
  {Cappi}, {Cassata}, {Cimatti}, {Fumana}, {Hasinger}, {Koekemoer},
  {Leauthaud}, {Maccagni}, {Marinoni}, {McCracken}, {Memeo}, {Meneux}, {Nair},
  {Porciani}, {Presotto}, \& {Scaramella}}]{peng2010b}
{Peng}, Y.-j., {Lilly}, S.~J., {Kova{\v{c}}}, K., {et~al.} 2010, \apj, 721,
  193, \dodoi{10.1088/0004-637X/721/1/193}

\bibitem[{{Posti} \& {Fall}(2021)}]{posti-fall2021}
{Posti}, L., \& {Fall}, S.~M. 2021, arXiv e-prints, arXiv:2102.11282.
\newblock \doarXiv{2102.11282}

\bibitem[{{Renzini}(2006)}]{renzini2006}
{Renzini}, A. 2006, \araa, 44, 141,
  \dodoi{10.1146/annurev.astro.44.051905.092450}

\bibitem[{{Rodriguez-Gomez} {et~al.}(2016){Rodriguez-Gomez}, {Pillepich},
  {Sales}, {Genel}, {Vogelsberger}, {Zhu}, {Wellons}, {Nelson}, {Torrey},
  {Springel}, {Ma}, \& {Hernquist}}]{rodriguez-gomez2016}
{Rodriguez-Gomez}, V., {Pillepich}, A., {Sales}, L.~V., {et~al.} 2016, \mnras,
  458, 2371, \dodoi{10.1093/mnras/stw456}

\bibitem[{{Rodr{\'\i}guez-Puebla} {et~al.}(2017){Rodr{\'\i}guez-Puebla},
  {Primack}, {Avila-Reese}, \& {Faber}}]{rodriguez-puebla2017}
{Rodr{\'\i}guez-Puebla}, A., {Primack}, J.~R., {Avila-Reese}, V., \& {Faber},
  S.~M. 2017, \mnras, 470, 651, \dodoi{10.1093/mnras/stx1172}

\bibitem[{{Rosani} {et~al.}(2018){Rosani}, {Pasquali}, {La Barbera},
  {Ferreras}, \& {Vazdekis}}]{rosani2018}
{Rosani}, G., {Pasquali}, A., {La Barbera}, F., {Ferreras}, I., \& {Vazdekis},
  A. 2018, \mnras, 476, 5233, \dodoi{10.1093/mnras/sty528}

\bibitem[{{Salpeter}(1955)}]{salpeter1955}
{Salpeter}, E.~E. 1955, \apj, 121, 161, \dodoi{10.1086/145971}

\bibitem[{{S{\'a}nchez} {et~al.}(2016){S{\'a}nchez}, {P{\'e}rez},
  {S{\'a}nchez-Bl{\'a}zquez}, {Garc{\'{\i}}a-Benito}, {Ibarra-Mede},
  {Gonz{\'a}lez}, {Rosales-Ortega}, {S{\'a}nchez-Menguiano}, {Ascasibar},
  {Bitsakis}, {Law}, {Cano-D{\'{\i}}az}, {L{\'o}pez-Cob{\'a}}, {Marino}, {Gil
  de Paz}, {L{\'o}pez-S{\'a}nchez}, {Barrera-Ballesteros}, {Galbany}, {Mast},
  {Abril-Melgarejo}, \& {Roman-Lopes}}]{sanchez2016}
{S{\'a}nchez}, S.~F., {P{\'e}rez}, E., {S{\'a}nchez-Bl{\'a}zquez}, P., {et~al.}
  2016, \rmxaa, 52, 171.
\newblock \doarXiv{1602.01830}

\bibitem[{{S{\'a}nchez-Bl{\'a}zquez} {et~al.}(2006){S{\'a}nchez-Bl{\'a}zquez},
  {Peletier}, {Jim{\'e}nez-Vicente}, {Cardiel}, {Cenarro},
  {Falc{\'o}n-Barroso}, {Gorgas}, {Selam}, \&
  {Vazdekis}}]{sanchez-blazquez2006}
{S{\'a}nchez-Bl{\'a}zquez}, P., {Peletier}, R.~F., {Jim{\'e}nez-Vicente}, J.,
  {et~al.} 2006, \mnras, 371, 703, \dodoi{10.1111/j.1365-2966.2006.10699.x}

\bibitem[{{Santucci} {et~al.}(2020){Santucci}, {Brough}, {Scott}, {Montes},
  {Owers}, {van de Sande}, {Bland -Hawthorn}, {Bryant}, {Croom}, {Ferreras},
  {Lawrence}, {L{\'o}pez-S{\'a}nchez}, \& {Richards}}]{santucci2020}
{Santucci}, G., {Brough}, S., {Scott}, N., {et~al.} 2020, arXiv e-prints,
  arXiv:2005.00541.
\newblock \doarXiv{2005.00541}

\bibitem[{{Schaye} {et~al.}(2015){Schaye}, {Crain}, {Bower}, {Furlong},
  {Schaller}, {Theuns}, {Dalla Vecchia}, {Frenk}, {McCarthy}, {Helly},
  {Jenkins}, {Rosas-Guevara}, {White}, {Baes}, {Booth}, {Camps}, {Navarro},
  {Qu}, {Rahmati}, {Sawala}, {Thomas}, \& {Trayford}}]{schaye2015}
{Schaye}, J., {Crain}, R.~A., {Bower}, R.~G., {et~al.} 2015, \mnras, 446, 521,
  \dodoi{10.1093/mnras/stu2058}

\bibitem[{{Schiavon}(2007)}]{schiavon2007}
{Schiavon}, R.~P. 2007, \apjs, 171, 146, \dodoi{10.1086/511753}

\bibitem[{{Scholz-D{\'\i}az} {et~al.}(2022){Scholz-D{\'\i}az},
  {Mart{\'\i}n-Navarro}, \& {Falc{\'o}n-Barroso}}]{scholz-diaz2022}
{Scholz-D{\'\i}az}, L., {Mart{\'\i}n-Navarro}, I., \& {Falc{\'o}n-Barroso}, J.
  2022, \mnras, 511, 4900, \dodoi{10.1093/mnras/stac361}

\bibitem[{{Scott} {et~al.}(2017){Scott}, {Brough}, {Croom}, {Davies}, {van de
  Sande}, {Allen}, {Bland-Hawthorn}, {Bryant}, {Cortese}, {D'Eugenio},
  {Federrath}, {Ferreras}, {Goodwin}, {Groves}, {Konstantopoulos}, {Lawrence},
  {Medling}, {Moffett}, {Owers}, {Richards}, {Robotham}, {Tonini}, \&
  {Yi}}]{scott2017}
{Scott}, N., {Brough}, S., {Croom}, S.~M., {et~al.} 2017, \mnras, 472, 2833,
  \dodoi{10.1093/mnras/stx2166}

\bibitem[{{Smee} {et~al.}(2013){Smee}, {Gunn}, {Uomoto}, {Roe}, {Schlegel},
  {Rockosi}, {Carr}, {Leger}, {Dawson}, {Olmstead}, {Brinkmann}, {Owen},
  {Barkhouser}, {Honscheid}, {Harding}, {Long}, {Lupton}, {Loomis}, {Anderson},
  {Annis}, {Bernardi}, {Bhardwaj}, {Bizyaev}, {Bolton}, {Brewington}, {Briggs},
  {Burles}, {Burns}, {Castander}, {Connolly}, {Davenport}, {Ebelke}, {Epps},
  {Feldman}, {Friedman}, {Frieman}, {Heckman}, {Hull}, {Knapp}, {Lawrence},
  {Loveday}, {Mannery}, {Malanushenko}, {Malanushenko}, {Merrelli}, {Muna},
  {Newman}, {Nichol}, {Oravetz}, {Pan}, {Pope}, {Ricketts}, {Shelden},
  {Sandford}, {Siegmund}, {Simmons}, {Smith}, {Snedden}, {Schneider},
  {SubbaRao}, {Tremonti}, {Waddell}, \& {York}}]{smee2013}
{Smee}, S.~A., {Gunn}, J.~E., {Uomoto}, A., {et~al.} 2013, \aj, 146, 32,
  \dodoi{10.1088/0004-6256/146/2/32}

\bibitem[{{Somerville} \& {Dav{\'e}}(2015)}]{somerville2015}
{Somerville}, R.~S., \& {Dav{\'e}}, R. 2015, \araa, 53, 51,
  \dodoi{10.1146/annurev-astro-082812-140951}

\bibitem[{{Speagle}(2019)}]{speagle2019}
{Speagle}, J.~S. 2019, arXiv e-prints.
\newblock \doarXiv{1904.02180}

\bibitem[{{Spergel} {et~al.}(2007){Spergel}, {Bean}, {Dor{\'e}}, {Nolta},
  {Bennett}, {Dunkley}, {Hinshaw}, {Jarosik}, {Komatsu}, {Page}, {Peiris},
  {Verde}, {Halpern}, {Hill}, {Kogut}, {Limon}, {Meyer}, {Odegard}, {Tucker},
  {Weiland }, {Wollack}, \& {Wright}}]{spergel2007}
{Spergel}, D.~N., {Bean}, R., {Dor{\'e}}, O., {et~al.} 2007, \apjs, 170, 377,
  \dodoi{10.1086/513700}

\bibitem[{{Szomoru} {et~al.}(2013){Szomoru}, {Franx}, {van Dokkum}, {Trenti},
  {Illingworth}, {Labb{\'e}}, \& {Oesch}}]{szomoru2013}
{Szomoru}, D., {Franx}, M., {van Dokkum}, P.~G., {et~al.} 2013, \apj, 763, 73,
  \dodoi{10.1088/0004-637X/763/2/73}

\bibitem[{Tange(2018)}]{tange2018}
Tange, O. 2018, GNU Parallel 2018 (Ole Tange), \dodoi{10.5281/zenodo.1146014}.
\newblock \url{https://doi.org/10.5281/zenodo.1146014}

\bibitem[{{Taylor} \& {Kobayashi}(2017)}]{taylor2017}
{Taylor}, P., \& {Kobayashi}, C. 2017, \mnras, 471, 3856,
  \dodoi{10.1093/mnras/stx1860}

\bibitem[{{Thomas} {et~al.}(2005){Thomas}, {Maraston}, {Bender}, \& {Mendes de
  Oliveira}}]{thomas2005}
{Thomas}, D., {Maraston}, C., {Bender}, R., \& {Mendes de Oliveira}, C. 2005,
  \apj, 621, 673, \dodoi{10.1086/426932}

\bibitem[{{Thomas} {et~al.}(2010){Thomas}, {Maraston}, {Schawinski}, {Sarzi},
  \& {Silk}}]{thomas2010}
{Thomas}, D., {Maraston}, C., {Schawinski}, K., {Sarzi}, M., \& {Silk}, J.
  2010, \mnras, 404, 1775, \dodoi{10.1111/j.1365-2966.2010.16427.x}

\bibitem[{{Tinker}(2020{\natexlab{a}})}]{tinker2020a}
{Tinker}, J.~L. 2020{\natexlab{a}}, arXiv e-prints, arXiv:2007.12200.
\newblock \doarXiv{2007.12200}

\bibitem[{{Tinker}(2020{\natexlab{b}})}]{tinker2020b}
---. 2020{\natexlab{b}}, arXiv e-prints, arXiv:2010.02946.
\newblock \doarXiv{2010.02946}

\bibitem[{{Tinker} {et~al.}(2018){Tinker}, {Hahn}, {Mao}, {Wetzel}, \&
  {Conroy}}]{tinker2018}
{Tinker}, J.~L., {Hahn}, C., {Mao}, Y.-Y., {Wetzel}, A.~R., \& {Conroy}, C.
  2018, \mnras, 477, 935, \dodoi{10.1093/mnras/sty666}

\bibitem[{{Tissera} {et~al.}(2014){Tissera}, {Beers}, {Carollo}, \&
  {Scannapieco}}]{tissera2014}
{Tissera}, P.~B., {Beers}, T.~C., {Carollo}, D., \& {Scannapieco}, C. 2014,
  \mnras, 439, 3128, \dodoi{10.1093/mnras/stu181}

\bibitem[{{Tissera} {et~al.}(2013){Tissera}, {Scannapieco}, {Beers}, \&
  {Carollo}}]{tissera2013}
{Tissera}, P.~B., {Scannapieco}, C., {Beers}, T.~C., \& {Carollo}, D. 2013,
  \mnras, 432, 3391, \dodoi{10.1093/mnras/stt691}

\bibitem[{{Toft} {et~al.}(2007){Toft}, {van Dokkum}, {Franx}, {Labbe},
  {F{\"o}rster Schreiber}, {Wuyts}, {Webb}, {Rudnick}, {Zirm}, {Kriek}, {van
  der Werf}, {Blakeslee}, {Illingworth}, {Rix}, {Papovich}, \&
  {Moorwood}}]{toft2007}
{Toft}, S., {van Dokkum}, P., {Franx}, M., {et~al.} 2007, \apj, 671, 285,
  \dodoi{10.1086/521810}

\bibitem[{{Toomre}(1977)}]{toomre1977}
{Toomre}, A. 1977, in Evolution of Galaxies and Stellar Populations, ed. B.~M.
  {Tinsley} \& D.~C. {Larson}, Richard B.~Gehret, 401

\bibitem[{{Tripicco} \& {Bell}(1995)}]{tripicco-bell1995}
{Tripicco}, M.~J., \& {Bell}, R.~A. 1995, \aj, 110, 3035,
  \dodoi{10.1086/117744}

\bibitem[{{Trujillo} {et~al.}(2007){Trujillo}, {Conselice}, {Bundy}, {Cooper},
  {Eisenhardt}, \& {Ellis}}]{trujillo2007}
{Trujillo}, I., {Conselice}, C.~J., {Bundy}, K., {et~al.} 2007, \mnras, 382,
  109, \dodoi{10.1111/j.1365-2966.2007.12388.x}

\bibitem[{{Trujillo} {et~al.}(2006{\natexlab{a}}){Trujillo}, {Feulner},
  {Goranova}, {Hopp}, {Longhetti}, {Saracco}, {Bender}, {Braito}, {Della Ceca},
  {Drory}, {Mannucci}, \& {Severgnini}}]{trujillo2006b}
{Trujillo}, I., {Feulner}, G., {Goranova}, Y., {et~al.} 2006{\natexlab{a}},
  \mnras, 373, L36, \dodoi{10.1111/j.1745-3933.2006.00238.x}

\bibitem[{{Trujillo} {et~al.}(2006{\natexlab{b}}){Trujillo}, {F{\"o}rster
  Schreiber}, {Rudnick}, {Barden}, {Franx}, {Rix}, {Caldwell}, {McIntosh},
  {Toft}, {H{\"a}ussler}, {Zirm}, {van Dokkum}, {Labb{\'e}}, {Moorwood},
  {R{\"o}ttgering}, {van der Wel}, {van der Werf}, \& {van
  Starkenburg}}]{trujillo2006a}
{Trujillo}, I., {F{\"o}rster Schreiber}, N.~M., {Rudnick}, G., {et~al.}
  2006{\natexlab{b}}, \apj, 650, 18, \dodoi{10.1086/506464}

\bibitem[{{van den Bosch} {et~al.}(2007){van den Bosch}, {Yang}, {Mo},
  {Weinmann}, {Macci{\`o}}, {More}, {Cacciato}, {Skibba}, \&
  {Kang}}]{vandenbosch2007}
{van den Bosch}, F.~C., {Yang}, X., {Mo}, H.~J., {et~al.} 2007, \mnras, 376,
  841, \dodoi{10.1111/j.1365-2966.2007.11493.x}

\bibitem[{{van der Wel} {et~al.}(2008){van der Wel}, {Holden}, {Zirm}, {Franx},
  {Rettura}, {Illingworth}, \& {Ford}}]{vanderwel2008}
{van der Wel}, A., {Holden}, B.~P., {Zirm}, A.~W., {et~al.} 2008, \apj, 688,
  48, \dodoi{10.1086/592267}

\bibitem[{{van der Wel} {et~al.}(2011){van der Wel}, {Rix}, {Wuyts}, {McGrath},
  {Koekemoer}, {Bell}, {Holden}, {Robaina}, \& {McIntosh}}]{vanderwel2011}
{van der Wel}, A., {Rix}, H.-W., {Wuyts}, S., {et~al.} 2011, \apj, 730, 38,
  \dodoi{10.1088/0004-637X/730/1/38}

\bibitem[{{van der Wel} {et~al.}(2014){van der Wel}, {Franx}, {van Dokkum},
  {Skelton}, {Momcheva}, {Whitaker}, {Brammer}, {Bell}, {Rix}, {Wuyts},
  {Ferguson}, {Holden}, {Barro}, {Koekemoer}, {Chang}, {McGrath},
  {H{\"a}ussler}, {Dekel}, {Behroozi}, {Fumagalli}, {Leja}, {Lundgren},
  {Maseda}, {Nelson}, {Wake}, {Patel}, {Labb{\'e}}, {Faber}, {Grogin}, \&
  {Kocevski}}]{vanderwel2014}
{van der Wel}, A., {Franx}, M., {van Dokkum}, P.~G., {et~al.} 2014, \apj, 788,
  28, \dodoi{10.1088/0004-637X/788/1/28}

\bibitem[{{van Dokkum} {et~al.}(2017){van Dokkum}, {Conroy}, {Villaume},
  {Brodie}, \& {Romanowsky}}]{vandokkum2017}
{van Dokkum}, P., {Conroy}, C., {Villaume}, A., {Brodie}, J., \& {Romanowsky},
  A.~J. 2017, \apj, 841, 68, \dodoi{10.3847/1538-4357/aa7135}

\bibitem[{{van Dokkum} {et~al.}(2009){van Dokkum}, {Kriek}, \&
  {Franx}}]{vandokkum2009}
{van Dokkum}, P.~G., {Kriek}, M., \& {Franx}, M. 2009, \nat, 460, 717,
  \dodoi{10.1038/nature08220}

\bibitem[{{van Dokkum} {et~al.}(2008){van Dokkum}, {Franx}, {Kriek}, {Holden},
  {Illingworth}, {Magee}, {Bouwens}, {Marchesini}, {Quadri}, {Rudnick},
  {Taylor}, \& {Toft}}]{vandokkum2008}
{van Dokkum}, P.~G., {Franx}, M., {Kriek}, M., {et~al.} 2008, \apjl, 677, L5,
  \dodoi{10.1086/587874}

\bibitem[{{van Dokkum} {et~al.}(2010){van Dokkum}, {Whitaker}, {Brammer},
  {Franx}, {Kriek}, {Labb{\'e}}, {Marchesini}, {Quadri}, {Bezanson},
  {Illingworth}, {Muzzin}, {Rudnick}, {Tal}, \& {Wake}}]{vandokkum2010}
{van Dokkum}, P.~G., {Whitaker}, K.~E., {Brammer}, G., {et~al.} 2010, \apj,
  709, 1018, \dodoi{10.1088/0004-637X/709/2/1018}

\bibitem[{{Vazdekis} {et~al.}(2010){Vazdekis}, {S{\'a}nchez-Bl{\'a}zquez},
  {Falc{\'o}n-Barroso}, {Cenarro}, {Beasley}, {Cardiel}, {Gorgas}, \&
  {Peletier}}]{vazdekis2010}
{Vazdekis}, A., {S{\'a}nchez-Bl{\'a}zquez}, P., {Falc{\'o}n-Barroso}, J.,
  {et~al.} 2010, \mnras, 404, 1639, \dodoi{10.1111/j.1365-2966.2010.16407.x}

\bibitem[{{Villaume} {et~al.}(2017){Villaume}, {Conroy}, {Johnson}, {Rayner},
  {Mann}, \& {van Dokkum}}]{villaume2017}
{Villaume}, A., {Conroy}, C., {Johnson}, B., {et~al.} 2017, \apjs, 230, 23,
  \dodoi{10.3847/1538-4365/aa72ed}

\bibitem[{{Vogelsberger} {et~al.}(2014){Vogelsberger}, {Genel}, {Springel},
  {Torrey}, {Sijacki}, {Xu}, {Snyder}, {Nelson}, \&
  {Hernquist}}]{vogelsberger2014a}
{Vogelsberger}, M., {Genel}, S., {Springel}, V., {et~al.} 2014, \mnras, 444,
  1518, \dodoi{10.1093/mnras/stu1536}

\bibitem[{{Wake} {et~al.}(2012){Wake}, {van Dokkum}, \& {Franx}}]{wake2012}
{Wake}, D.~A., {van Dokkum}, P.~G., \& {Franx}, M. 2012, \apjl, 751, L44,
  \dodoi{10.1088/2041-8205/751/2/L44}

\bibitem[{{Wake} {et~al.}(2017){Wake}, {Bundy}, {Diamond-Stanic}, {Yan},
  {Blanton}, {Bershady}, {S{\'a}nchez-Gallego}, {Drory}, {Jones}, {Kauffmann},
  {Law}, {Li}, {MacDonald}, {Masters}, {Thomas}, {Tinker}, {Weijmans}, \&
  {Brownstein}}]{wake2017}
{Wake}, D.~A., {Bundy}, K., {Diamond-Stanic}, A.~M., {et~al.} 2017, \aj, 154,
  86, \dodoi{10.3847/1538-3881/aa7ecc}

\bibitem[{{Wang} {et~al.}(2009){Wang}, {Mo}, {Jing}, {Guo}, {van den Bosch}, \&
  {Yang}}]{wang2009}
{Wang}, H., {Mo}, H.~J., {Jing}, Y.~P., {et~al.} 2009, \mnras, 394, 398,
  \dodoi{10.1111/j.1365-2966.2008.14301.x}

\bibitem[{{Wang} {et~al.}(2012){Wang}, {Mo}, {Yang}, \& {van den
  Bosch}}]{wang2012}
{Wang}, H., {Mo}, H.~J., {Yang}, X., \& {van den Bosch}, F.~C. 2012, \mnras,
  420, 1809, \dodoi{10.1111/j.1365-2966.2011.20174.x}

\bibitem[{{Wang} {et~al.}(2016){Wang}, {Mo}, {Yang}, {Zhang}, {Shi}, {Jing},
  {Liu}, {Li}, {Kang}, \& {Gao}}]{wang2016}
{Wang}, H., {Mo}, H.~J., {Yang}, X., {et~al.} 2016, \apj, 831, 164,
  \dodoi{10.3847/0004-637X/831/2/164}

\bibitem[{{Wechsler} {et~al.}(2002){Wechsler}, {Bullock}, {Primack},
  {Kravtsov}, \& {Dekel}}]{wechsler2002}
{Wechsler}, R.~H., {Bullock}, J.~S., {Primack}, J.~R., {Kravtsov}, A.~V., \&
  {Dekel}, A. 2002, \apj, 568, 52, \dodoi{10.1086/338765}

\bibitem[{{Wechsler} \& {Tinker}(2018)}]{wechsler2018}
{Wechsler}, R.~H., \& {Tinker}, J.~L. 2018, \araa, 56, 435,
  \dodoi{10.1146/annurev-astro-081817-051756}

\bibitem[{{Wechsler} {et~al.}(2006){Wechsler}, {Zentner}, {Bullock},
  {Kravtsov}, \& {Allgood}}]{wechsler2006}
{Wechsler}, R.~H., {Zentner}, A.~R., {Bullock}, J.~S., {Kravtsov}, A.~V., \&
  {Allgood}, B. 2006, \apj, 652, 71, \dodoi{10.1086/507120}

\bibitem[{{Weinmann} {et~al.}(2006){Weinmann}, {van den Bosch}, {Yang}, \&
  {Mo}}]{weinmann2006}
{Weinmann}, S.~M., {van den Bosch}, F.~C., {Yang}, X., \& {Mo}, H.~J. 2006,
  \mnras, 366, 2, \dodoi{10.1111/j.1365-2966.2005.09865.x}

\bibitem[{{Wellons} {et~al.}(2015){Wellons}, {Torrey}, {Ma}, {Rodriguez-Gomez},
  {Vogelsberger}, {Kriek}, {van Dokkum}, {Nelson}, {Genel}, {Pillepich},
  {Springel}, {Sijacki}, {Snyder}, {Nelson}, {Sales}, \&
  {Hernquist}}]{wellons2015b}
{Wellons}, S., {Torrey}, P., {Ma}, C.-P., {et~al.} 2015, \mnras, 449, 361,
  \dodoi{10.1093/mnras/stv303}

\bibitem[{{Wellons} {et~al.}(2016){Wellons}, {Torrey}, {Ma}, {Rodriguez-Gomez},
  {Pillepich}, {Nelson}, {Genel}, {Vogelsberger}, \& {Hernquist}}]{wellons2015}
---. 2016, \mnras, 456, 1030, \dodoi{10.1093/mnras/stv2738}

\bibitem[{{Westfall} {et~al.}(2019){Westfall}, {Cappellari}, {Bershady},
  {Bundy}, {Belfiore}, {Ji}, {Law}, {Schaefer}, {Shetty}, {Tremonti}, {Yan},
  {Andrews}, {Brownstein}, {Cherinka}, {Coccato}, {Drory}, {Maraston},
  {Parikh}, {S{\'a}nchez-Gallego}, {Thomas}, {Weijmans}, {Barrera-Ballesteros},
  {Du}, {Goddard}, {Li}, {Masters}, {Ibarra Medel}, {S{\'a}nchez}, {Yang},
  {Zheng}, \& {Zhou}}]{westfall2019}
{Westfall}, K.~B., {Cappellari}, M., {Bershady}, M.~A., {et~al.} 2019, \aj,
  158, 231, \dodoi{10.3847/1538-3881/ab44a2}

\bibitem[{{Whitaker} {et~al.}(2012){Whitaker}, {Kriek}, {van Dokkum},
  {Bezanson}, {Brammer}, {Franx}, \& {Labb{\'e}}}]{whitaker2012b}
{Whitaker}, K.~E., {Kriek}, M., {van Dokkum}, P.~G., {et~al.} 2012, \apj, 745,
  179, \dodoi{10.1088/0004-637X/745/2/179}

\bibitem[{{White} \& {Rees}(1978)}]{white-rees1978}
{White}, S.~D.~M., \& {Rees}, M.~J. 1978, \mnras, 183, 341,
  \dodoi{10.1093/mnras/183.3.341}

\bibitem[{{Worthey} {et~al.}(1994){Worthey}, {Faber}, {Gonzalez}, \&
  {Burstein}}]{worthey1994}
{Worthey}, G., {Faber}, S.~M., {Gonzalez}, J.~J., \& {Burstein}, D. 1994,
  \apjs, 94, 687, \dodoi{10.1086/192087}

\bibitem[{{Xu} \& {Zheng}(2020)}]{xu2020}
{Xu}, X., \& {Zheng}, Z. 2020, \mnras, 492, 2739, \dodoi{10.1093/mnras/staa009}

\bibitem[{{Yan} {et~al.}(2016{\natexlab{a}}){Yan}, {Bundy}, {Law}, {Bershady},
  {Andrews}, {Cherinka}, {Diamond-Stanic}, {Drory}, {MacDonald},
  {S{\'a}nchez-Gallego}, {Thomas}, {Wake}, {Weijmans}, {Westfall}, {Zhang},
  {Arag{\'o}n-Salamanca}, {Belfiore}, {Bizyaev}, {Blanc}, {Blanton},
  {Brownstein}, {Cappellari}, {D'Souza}, {Emsellem}, {Fu}, {Gaulme}, {Graham},
  {Goddard}, {Gunn}, {Harding}, {Jones}, {Kinemuchi}, {Li}, {Li}, {Maiolino},
  {Mao}, {Maraston}, {Masters}, {Merrifield}, {Oravetz}, {Pan}, {Parejko},
  {Sanchez}, {Schlegel}, {Simmons}, {Thanjavur}, {Tinker}, {Tremonti}, {van den
  Bosch}, \& {Zheng}}]{yan2016b}
{Yan}, R., {Bundy}, K., {Law}, D.~R., {et~al.} 2016{\natexlab{a}}, \aj, 152,
  197, \dodoi{10.3847/0004-6256/152/6/197}

\bibitem[{{Yan} {et~al.}(2016{\natexlab{b}}){Yan}, {Tremonti}, {Bershady},
  {Law}, {Schlegel}, {Bundy}, {Drory}, {MacDonald}, {Bizyaev}, {Blanc},
  {Blanton}, {Cherinka}, {Eigenbrot}, {Gunn}, {Harding}, {Hogg},
  {S{\'a}nchez-Gallego}, {S{\'a}nchez}, {Wake}, {Weijmans}, {Xiao}, \&
  {Zhang}}]{yan2016a}
{Yan}, R., {Tremonti}, C., {Bershady}, M.~A., {et~al.} 2016{\natexlab{b}}, \aj,
  151, 8, \dodoi{10.3847/0004-6256/151/1/8}

\bibitem[{{Yang} {et~al.}(2005){Yang}, {Mo}, {van den Bosch}, \&
  {Jing}}]{yang2005}
{Yang}, X., {Mo}, H.~J., {van den Bosch}, F.~C., \& {Jing}, Y.~P. 2005, \mnras,
  356, 1293, \dodoi{10.1111/j.1365-2966.2005.08560.x}

\bibitem[{{Yang} {et~al.}(2007){Yang}, {Mo}, {van den Bosch}, {Pasquali}, {Li},
  \& {Barden}}]{yang2007}
{Yang}, X., {Mo}, H.~J., {van den Bosch}, F.~C., {et~al.} 2007, \apj, 671, 153,
  \dodoi{10.1086/522027}

\bibitem[{{York} {et~al.}(2000){York}, {Adelman}, {Anderson}, {Anderson},
  {Annis}, {Bahcall}, {Bakken}, {Barkhouser}, {Bastian}, {Berman}, {Boroski},
  {Bracker}, {Briegel}, {Briggs}, {Brinkmann}, {Brunner}, {Burles}, {Carey},
  {Carr}, {Castander}, {Chen}, {Colestock}, {Connolly}, {Crocker}, {Csabai},
  {Czarapata}, {Davis}, {Doi}, {Dombeck}, {Eisenstein}, {Ellman}, {Elms},
  {Evans}, {Fan}, {Federwitz}, {Fiscelli}, {Friedman}, {Frieman}, {Fukugita},
  {Gillespie}, {Gunn}, {Gurbani}, {de Haas}, {Haldeman}, {Harris}, {Hayes},
  {Heckman}, {Hennessy}, {Hindsley}, {Holm}, {Holmgren}, {Huang}, {Hull},
  {Husby}, {Ichikawa}, {Ichikawa}, {Ivezi{\'c}}, {Kent}, {Kim}, {Kinney},
  {Klaene}, {Kleinman}, {Kleinman}, {Knapp}, {Korienek}, {Kron}, {Kunszt},
  {Lamb}, {Lee}, {Leger}, {Limmongkol}, {Lindenmeyer}, {Long}, {Loomis},
  {Loveday}, {Lucinio}, {Lupton}, {MacKinnon}, {Mannery}, {Mantsch}, {Margon},
  {McGehee}, {McKay}, {Meiksin}, {Merelli}, {Monet}, {Munn}, {Narayanan},
  {Nash}, {Neilsen}, {Neswold}, {Newberg}, {Nichol}, {Nicinski}, {Nonino},
  {Okada}, {Okamura}, {Ostriker}, {Owen}, {Pauls}, {Peoples}, {Peterson},
  {Petravick}, {Pier}, {Pope}, {Pordes}, {Prosapio}, {Rechenmacher}, {Quinn},
  {Richards}, {Richmond}, {Rivetta}, {Rockosi}, {Ruthmansdorfer}, {Sandford},
  {Schlegel}, {Schneider}, {Sekiguchi}, {Sergey}, {Shimasaku}, {Siegmund},
  {Smee}, {Smith}, {Snedden}, {Stone}, {Stoughton}, {Strauss}, {Stubbs},
  {SubbaRao}, {Szalay}, {Szapudi}, {Szokoly}, {Thakar}, {Tremonti}, {Tucker},
  {Uomoto}, {Vanden Berk}, {Vogeley}, {Waddell}, {Wang}, {Watanabe},
  {Weinberg}, {Yanny}, {Yasuda}, \& {SDSS Collaboration}}]{york2000}
{York}, D.~G., {Adelman}, J., {Anderson}, Jr., J.~E., {et~al.} 2000, \aj, 120,
  1579, \dodoi{10.1086/301513}

\bibitem[{{Zehavi} {et~al.}(2018){Zehavi}, {Contreras}, {Padilla}, {Smith},
  {Baugh}, \& {Norberg}}]{zehavi2018}
{Zehavi}, I., {Contreras}, S., {Padilla}, N., {et~al.} 2018, \apj, 853, 84,
  \dodoi{10.3847/1538-4357/aaa54a}

\bibitem[{{Zentner} {et~al.}(2019){Zentner}, {Hearin}, {van den Bosch},
  {Lange}, \& {Villarreal}}]{zentner2019}
{Zentner}, A.~R., {Hearin}, A., {van den Bosch}, F.~C., {Lange}, J.~U., \&
  {Villarreal}, A. 2019, \mnras, 485, 1196, \dodoi{10.1093/mnras/stz470}

\bibitem[{{Zentner} {et~al.}(2014){Zentner}, {Hearin}, \& {van den
  Bosch}}]{zentner2014}
{Zentner}, A.~R., {Hearin}, A.~P., \& {van den Bosch}, F.~C. 2014, \mnras, 443,
  3044, \dodoi{10.1093/mnras/stu1383}

\bibitem[{{Zhao} {et~al.}(2009){Zhao}, {Jing}, {Mo}, \&
  {B{\"o}rner}}]{zhao2009}
{Zhao}, D.~H., {Jing}, Y.~P., {Mo}, H.~J., \& {B{\"o}rner}, G. 2009, \apj, 707,
  354, \dodoi{10.1088/0004-637X/707/1/354}

\bibitem[{{Zheng} {et~al.}(2017){Zheng}, {Wang}, {Ge}, {Mao}, {Li}, {Li}, {Mo},
  {Goddard}, {Bundy}, {Li}, {Nair}, {Lin}, {Long}, {Riffel}, {Thomas},
  {Masters}, {Bizyaev}, {Brownstein}, {Zhang}, {Law}, {Drory}, {Roman Lopes},
  \& {Malanushenko}}]{zheng2017}
{Zheng}, Z., {Wang}, H., {Ge}, J., {et~al.} 2017, \mnras, 465, 4572,
  \dodoi{10.1093/mnras/stw3030}

\bibitem[{{Zheng} {et~al.}(2019){Zheng}, {Li}, {Mao}, {Wang}, {Liu}, {Mo},
  {Yuan}, {Maraston}, {Thomas}, {Yan}, {Bundy}, {Long}, {Parikh},
  {Oyarz{\'u}n}, {Bizyaev}, \& {Lacerna}}]{zheng2019}
{Zheng}, Z., {Li}, C., {Mao}, S., {et~al.} 2019, \apj, 873, 63,
  \dodoi{10.3847/1538-4357/ab03d2}

\bibitem[{{Zirm} {et~al.}(2007){Zirm}, {van der Wel}, {Franx}, {Labb{\'e}},
  {Trujillo}, {van Dokkum}, {Toft}, {Daddi}, {Rudnick}, {Rix},
  {R{\"o}ttgering}, \& {van der Werf}}]{zirm2007}
{Zirm}, A.~W., {van der Wel}, A., {Franx}, M., {et~al.} 2007, \apj, 656, 66,
  \dodoi{10.1086/510713}

\bibitem[{{Zolotov} {et~al.}(2009){Zolotov}, {Willman}, {Brooks}, {Governato},
  {Brook}, {Hogg}, {Quinn}, \& {Stinson}}]{zolotov2009}
{Zolotov}, A., {Willman}, B., {Brooks}, A.~M., {et~al.} 2009, \apj, 702, 1058,
  \dodoi{10.1088/0004-637X/702/2/1058}

\bibitem[{{Zolotov} {et~al.}(2015){Zolotov}, {Dekel}, {Mandelker}, {Tweed},
  {Inoue}, {DeGraf}, {Ceverino}, {Primack}, {Barro}, \& {Faber}}]{zolotov2015}
{Zolotov}, A., {Dekel}, A., {Mandelker}, N., {et~al.} 2015, \mnras, 450, 2327,
  \dodoi{10.1093/mnras/stv740}

\bibitem[{{Zu} \& {Mandelbaum}(2016)}]{zu2016}
{Zu}, Y., \& {Mandelbaum}, R. 2016, \mnras, 457, 4360,
  \dodoi{10.1093/mnras/stw221}

\end{thebibliography}
\bibliographystyle{aasjournal}



\end{document}